\documentclass[longauth]{aa}  

\usepackage{comment}
\usepackage{subfig}
\usepackage{soul}       
\usepackage{xcolor}
\usepackage{hyperref}

\newcount\Comments  
\newcommand{\kibitz}[2]{\ifnum\Comments=1\textcolor{#1}{#2}\fi}

\usepackage{graphicx}
\usepackage{txfonts}
\usepackage{gensymb}
\usepackage{orcidlink}

\begin{document}

\title{Gamma-ray signature of superluminous supernovae: \textit{Fermi}-LAT GeV detection of SN 2017egm and evidence of a central engine }

\author{
F.~Acero*$^{(1,2)}$ \and 
A.~Acharyya$^{(3)}$ \and 
A.~Adelfio$^{(4)}$ \and 
M.~Ajello$^{(5)}$ \and 
E.~Aviano$^{(6,7)}$ \and 
L.~Baldini$^{(8,9)}$ \and 
J.~Ballet$^{(1)}$ \and 
C.~Bartolini$^{(10,11)}$ \and 
D.~Bastieri$^{(12,13,14)}$ \and 
J.~Becerra~Gonzalez$^{(15)}$
% ...
}

\institute{
\inst{1}~Universit\'e Paris-Saclay, Universit\'e Paris Cit\'e, CEA, CNRS, AIM, F-91191 Gif-sur-Yvette Cedex, France\\ 
\inst{2}~FSLAC IRL 2009, CNRS/IAC, La Laguna, Tenerife, Spain\\ 
\inst{3}~Center for Cosmology and Particle Physics Phenomenology, University of Southern Denmark, Campusvej 55, DK-5230 Odense M, Denmark\\ 
\inst{4}~Istituto Nazionale di Fisica Nucleare, Sezione di Perugia, I-06123 Perugia, Italy\\ 
\inst{5}~Department of Physics and Astronomy, Clemson University, Kinard Lab of Physics, Clemson, SC 29634-0978, USA\\ 
\inst{6}~Dipartimento di Fisica, Universit\`a di Trieste, I-34127 Trieste, Italy\\ 
\inst{7}~Istituto Nazionale di Fisica Nucleare, Sezione di Trieste, I-34127 Trieste, Italy\\ 
\inst{8}~Universit\`a di Pisa, Dipartimento di Fisica E. Fermi, I-56127 Pisa, Italy\\ 
\inst{9}~Istituto Nazionale di Fisica Nucleare, Sezione di Pisa, I-56127 Pisa, Italy\\ 
\inst{10}~Istituto Nazionale di Fisica Nucleare, Sezione di Bari, I-70126 Bari, Italy\\ 
\inst{11}~Universit\`a degli studi di Trento, via Calepina 14, 38122 Trento, Italy\\ 
\inst{12}~Istituto Nazionale di Fisica Nucleare, Sezione di Padova, I-35131 Padova, Italy\\ 
\inst{13}~Dipartimento di Fisica e Astronomia ``G. Galilei'', Universit\`a di Padova, Via F. Marzolo, 8, I-35131 Padova, Italy\\ 
\inst{14}~Center for Space Studies and Activities ``G. Colombo", University of Padova, Via Venezia 15, I-35131 Padova, Italy\\ 
\inst{15}~Instituto de Astrof\'isica de Canarias and Universidad de La Laguna, Dpto. Astrof\'isica, 38200 La Laguna, Tenerife, Spain\\ 
\inst{16}~Dipartimento di Fisica ``M. Merlin" dell'Universit\`a e del Politecnico di Bari, via Amendola 173, I-70126 Bari, Italy\\ 
\inst{17}~Istituto Nazionale di Fisica Nucleare, Sezione di Torino, I-10125 Torino, Italy\\ 
\inst{18}~Dipartimento di Fisica, Universit\`a degli Studi di Torino, I-10125 Torino, Italy\\ 
\inst{19}~Laboratoire Leprince-Ringuet, CNRS/IN2P3, \'Ecole polytechnique, Institut Polytechnique de Paris, 91120 Palaiseau, France\\ 
\inst{20}~Deutsches Elektronen Synchrotron DESY, D-15738 Zeuthen, Germany\\ 
\inst{21}~Institut f\"ur Theoretische Physik and Astrophysik, Universit\"at W\"urzburg, D-97074 W\"urzburg, Germany\\ 
\inst{22}~W. W. Hansen Experimental Physics Laboratory, Kavli Institute for Particle Astrophysics and Cosmology, Department of Physics and SLAC National Accelerator Laboratory, Stanford University, Stanford, CA 94305, USA\\ 
\inst{23}~INAF-Istituto di Astrofisica Spaziale e Fisica Cosmica Milano, via E. Bassini 15, I-20133 Milano, Italy\\ 
\inst{24}~Istituto Nazionale di Fisica Nucleare, Sezione di Roma ``Tor Vergata", I-00133 Roma, Italy\\ 
\inst{25}~Space Science Data Center - Agenzia Spaziale Italiana, Via del Politecnico, snc, I-00133, Roma, Italy\\ 
\inst{26}~Dipartimento di Fisica, Universit\`a La Sapienza, Piazzale A. Moro, 2, I-00185 Roma, Italy\\ 
\inst{27}~Dipartimento di Fisica, Universit\`a degli Studi di Perugia, I-06123 Perugia, Italy\\ 
\inst{28}~Italian Space Agency, Via del Politecnico snc, 00133 Roma, Italy\\ 
\inst{29}~Space Science Division, Naval Research Laboratory, Washington, DC 20375-5352, USA\\ 
\inst{30}~Friedrich-Alexander Universit\"at Erlangen-N\"urnberg, Erlangen Centre for Astroparticle Physics, Erwin-Rommel-Str. 1, 91058 Erlangen, Germany\\ 
\inst{31}~Friedrich-Alexander-Universit\"at, Erlangen-N\"urnberg, Schlossplatz 4, 91054 Erlangen, Germany\\ 
\inst{32}~INAF Istituto di Radioastronomia, I-40129 Bologna, Italy\\ 
\inst{33}~Grupo de Altas Energ\'ias, Universidad Complutense de Madrid, E-28040 Madrid, Spain\\ 
\inst{34}~Universit\'e Bordeaux, CNRS, LP2I Bordeaux, UMR 5797, F-33170 Gradignan, France\\ 
\inst{35}~Universit\`a di Pisa and Istituto Nazionale di Fisica Nucleare, Sezione di Pisa I-56127 Pisa, Italy\\ 
\inst{36}~Ruhr University Bochum, Faculty of Physics and Astronomy, Astronomical Institute (AIRUB), 44780 Bochum, Germany\\ 
\inst{37}~Department of Physical Sciences, Hiroshima University, Higashi-Hiroshima, Hiroshima 739-8526, Japan\\ 
\inst{38}~Dipartimento di Fisica, Universit\`a di Roma ``Tor Vergata", I-00133 Roma, Italy\\ 
\inst{39}~Dipartimento di Fisica e Geologia, Universit\`a degli Studi di Perugia, via Pascoli snc, I-06123 Perugia, Italy\\ 
\inst{40}~Department of Fundamental Physics, University of Salamanca, Plaza de la Merced s/n, E-37008 Salamanca, Spain\\ 
\inst{41}~Universit\'e Paris Cit\'e, Universit\'e Paris-Saclay, CEA, CNRS, AIM, F-91191 Gif-sur-Yvette, France\\ 
\inst{42}~The George Washington University, Department of Physics, 725 21st St, NW, Washington, DC 20052, USA\\ 
\inst{43}~Astrophysics Science Division, NASA Goddard Space Flight Center, Greenbelt, MD 20771, USA\\ 
\inst{44}~University of North Florida, Department of Physics, 1 UNF Drive, Jacksonville, FL 32224 , USA\\ 
\inst{45}~Yunnan Observatories, Chinese Academy of Sciences, Kunming 650216, China\\ 
\inst{46}~NASA Postdoctoral Program Fellow, USA\\ 
\inst{47}~Department of Astronomy, University of Science and Technology of China, Hefei 230026, China\\ 
\inst{48}~School of Astronomy and Space Science, University of Science and Technology of China, Hefei 230026, China\\ 
\inst{49}~Institute of Astrophysics, Foundation for Research and Technology-Hellas, Heraklion, GR-70013, Greece\\ 
\inst{50}~The Aerospace Corporation, 14745 Lee Rd, Chantilly, VA 20151, USA\\ 
\inst{51}~Institute of Space Sciences (ICE, CSIC), Campus UAB, Carrer de Magrans s/n, E-08193 Barcelona, Spain\\
\inst{52}~Center for Space Science and Technology, University of Maryland Baltimore County, 1000 Hilltop Circle, Baltimore, MD 21250, USA\\ 
\inst{53}~Hiroshima Astrophysical Science Center, Hiroshima University, Higashi-Hiroshima, Hiroshima 739-8526, Japan\\ 
\inst{54}~Vatican Observatory, Castel Gandolfo, V-00120, Vatican City State\\ 
\inst{55}~Department of physics and Astronomy, Louisiana State University, Baton Rouge, LA 70803, USA\\ 
\inst{56}~INAF-Astronomical Observatory of Padova, Vicolo dell'Osservatorio 5, I-35122 Padova, Italy\\ 
\inst{57}~Institut f\"ur Astro- und Teilchenphysik, Leopold-Franzens-Universit\"at Innsbruck, A-6020 Innsbruck, Austria\\ 
\inst{58}~Instituto de F\'isica Te\'orica UAM/CSIC, Universidad Aut\'onoma de Madrid, E-28049 Madrid, Spain\\ 
\inst{59}~Departamento de F\'isica Te\'orica, Universidad Aut\'onoma de Madrid, 28049 Madrid, Spain\\ 
\inst{60}~Santa Cruz Institute for Particle Physics, Department of Physics and Department of Astronomy and Astrophysics, University of California at Santa Cruz, Santa Cruz, CA 95064, USA\\ 
\inst{61}~NYCB Real-Time Computing Inc., Lattingtown, NY 11560-1025, USA\\ 
\inst{62}~Purdue University Northwest, Hammond, IN 46323, USA\\ 
\inst{63}~Nagoya University, Institute for Space-Earth Environmental Research, Furo-cho, Chikusa-ku, Nagoya 464-8601, Japan\\ 
\inst{64}~Kobayashi-Maskawa Institute for the Origin of Particles and the Universe, Nagoya University, Furo-cho, Chikusa-ku, Nagoya, Japan\\ 
\inst{65}~Department of Astronomy, University of Maryland, College Park, MD 20742, USA\\ 
\inst{66}~Institute of Space Sciences (ICE, CSIC), Campus UAB, Carrer de Magrans s/n, E-08193 Barcelona, Spain and Institut d'Estudis Espacials de Catalunya (IEEC), E-08034 Barcelona, Spain and Instituci\'o Catalana de Recerca i Estudis Avan\c{c}ats (ICREA), E-08010 Barcelona, Spain\\ 
\inst{67}~Praxis Inc., Alexandria, VA 22303, resident at Naval Research Laboratory, Washington, DC 20375, USA\\ 
\inst{68}~Center for Astrophysics and Cosmology, University of Nova Gorica, Nova Gorica, Slovenia\\ 
\inst{69}~Institute of Space Sciences (ICE, CSIC), Campus UAB, Carrer de Magrans s/n, E-08193 Barcelona, Spain; and Institut d'Estudis Espacials de Catalunya (IEEC), E-08034 Barcelona, Spain\\ 
\inst{70}~Department of Physics and Columbia Astrophysics Laboratory, Columbia University, New York, NY 10027, USA\\
\inst{71}~Center for Computational Astrophysics, Flatiron Institute, 162 5th Ave, New York, NY 10010, USA\\
\inst{72}~The Oskar Klein Centre, Department of Astronomy, Stockholm University, Albanova University Center, SE 106 91 Stockholm, Sweden\\ 
\inst{73}~Astrophysics Division, National Centre for Nuclear Research, Pasteura 7, 02-093 Warsaw, Poland\\
\inst{74}~Tartu Observatory, University of Tartu, T\~oravere, 61602 Tartumaa, Estonia\\
Corresponding authors :\\
\email{fabio.acero@cnrs.fr} \\
\email{gmarti@ice.csic.es} \\
}

   \date{Received December 12, 2025; accepted March 7, 2026}


  \abstract
   {Superluminous supernovae (SLSNe) are a rare class of transients with peak luminosities 10–100 times greater than those of standard core-collapse supernovae (SNe). The mechanisms powering their extreme brightness remain debated, with circumstellar medium (CSM) interaction, or energy injection from a central engine like a magnetar wind nebula being the most plausible scenarios.
   While the optical properties of SLSNe are extensively studied, their $\gamma$-ray signatures remain poorly constrained.
   }
   {To further constrain the underlying mechanism, we carried out
   a systematic search for giga-electronvolt $\gamma$-ray emission using the \textit{Fermi} Large Area Telescope (LAT) from a sample of nearby hydrogen-poor (Type I) and hydrogen-rich (Type II) SLSNe over the past 16 years. Our objective is to test predictions from CSM and magnetar models, and to assess the prospects for future detections with the Cherenkov Telescope Array Observatory (CTAO).
   }
   {For the six targets of this sample, we  studied the time variability  of a putative $\gamma$-ray signal at the optical position of the SLSN on a six-month timescale, and in the case of SN 2017egm, we further investigated variability on 15-day intervals and applied a Bayesian block algorithm to characterize the time variability of the signal. We then compared  the temporal evolution and spectral properties to the predictions from a magnetar and CSM interaction model.}
   {Among the sample, only SN 2017egm shows significant $\gamma$-ray emission, with likelihood test statistic (TS) values of 26-33 (i.e., $> 5\sigma$) depending on the adopted time window. The signal arises between 50 and 160 days after explosion and is well described by a power-law spectrum with index $\Gamma=2.17\pm0.23$. The emission is consistent both in terms of its light curve and its spectrum, with predictions from magnetar models requiring either low nebular magnetization or faster spin-down than dipole losses. The CSM shell interaction scenario can reproduce the observed flux level but not the observed timing of the $\gamma$-ray signal. In addition, the observed ratio, $L_{\gamma}/L_{\rm opt}\sim 1$, is inconsistent with theoretical expectations and not in line with ratio measurements in other interacting CSM-dominated objects (e.g., novae or SNe) where this ratio is less than $10^{-2}$.
   }
   {Our study strongly suggests that a central engine like a magnetar plays a key role in this SLSN and could explain the bulk of the optical and $\gamma$-ray light curves properties.    In order to explain the observed late-time bumps in the optical light curve of SN 2017egm, we require either: a hybrid picture combining magnetar and multiple CSM shells for the optical bumps or a pure magnetar model with infalling matter on an accretion disk. Finally, simulations of 50 hours of CTAO observations indicate that a SN 2017egm-like event would be detectable up to $\sim140$ Mpc in the magnetar model but not in the CSM model due to strong $\gamma$-$\gamma$ absorption.}

   \keywords{   supernovae: general –  supernovae: SN 2017egm - Stars: magnetars - circumstellar matter - Gamma rays: general   }

\authorrunning{Fermi-LAT Collaboration et al.}
\titlerunning{Gamma-ray signature of superluminous supernovae with \textit{Fermi}-LAT}

   \maketitle

\section{Introduction} \label{sec:intro}
Superluminous supernovae (SLSNe) are a recently recognized class of astronomical transients whose optical luminosities exceed those of typical core-collapse supernovae (SNe) by a factor of 10 -- 100. Reaching peak magnitude typically 40 days after the explosion, some SLSNe have reached a \textit{u} band absolute magnitude brighter than $-22$ mag, which translates to luminosities in the range of $10^{44}$--$10^{45}$ erg s$^{-1}$. 
\citet{Quimby11} noted that the late-time decay rates of these particular SNe are inconsistent with a radioactive decay scenario and require a late deposition of a large amount of energy to explain their light curves.
As for the early classification of SNe, SLSNe are divided into two main categories based on their spectral properties; namely, the absence or presence of hydrogen lines, corresponding, respectively, to  type I (hydrogen-poor) and type II SLSNe (hydrogen-rich). 
Type I SLSNe are the most studied class and there are now over 260 SLSNe reported in the catalog of \citet{Gomez24}.

What makes SLSNe so different from regular core-collapse SNe is still being debated. There are mainly four different energy sources being considered to explain the high peak luminosity of SLSNe: ejecta fallback accretion onto a black hole, radioactive decay of $^{56}$Ni, circumstellar interaction, and a magnetar wind nebula \citep[see, e.g.,][for a review]{Moriya18-review,Gal-Yam19,Inserra19,Nicholl21,Moriya24}. 
In the last of these scenarios, a neutron star with extreme properties (millisecond period and $\sim10^{14}$--$10^{15}$ G magnetic field; a magnetar) will convert its rotational energy into a wind of electrons and positrons. This magnetar wind nebula will produce high-energy nonthermal radiation (via inverse Compton (IC) and synchrotron) that upon interaction with the surrounding ejecta will be reprocessed into lower-energy photons (optical/UV) and power the SLSNe.
In theoretical models aiming to predict the optical light curves from engine-powered SLSN \citep[e.g.,][]{Dessart12,Chatzopoulos13,Metzger15}, it is generally assumed that the thermalization efficiency is 100\% (all high-energy radiation is reprocessed by the ejecta).
While this is certainly a viable assumption in the early phase where the ejecta are opaque, this efficiency will depart from 100\%  as the ejecta expands and some $\gamma$-rays start leaking.
Therefore models predicting the light curve of escaping $\gamma$-rays need to treat this efficiency carefully in a time-dependent manner, requiring the computation of the diffusion and escape of the photons through the ejecta \citep[for more details, see][]{Vurm21}.
Note that while the aforementioned model is focused on a magnetar as central engine, most of the general principles outlined in the model remain valid for other central engines capable of injecting a pool of nonthermal radiation.
An alternative scenario is the circumstellar medium (CSM) interaction, which could also account for most thermal observables \citep[e.g., see][]{Chatzopoulos13}. It proposes that the extreme luminosity arises from the conversion of the SN ejecta's kinetic energy into radiation.
When the fast-moving ejecta collides with a dense shell of CSM, a strong shock is created that transforms kinetic energy into thermal energy radiating mostly in the X-ray band.
This high-energy emission is absorbed by the optically thick CSM and then reemitted as optical/UV light.
The slow diffusion of the photons through the CSM broadens the event's duration, accounting for the observed long-lasting light curves.
Observations of some type I SLSNe such as SN 2017egm have revealed irregularities in their light curves, including post-peak bumps, which require more complex multiple CSM shell scenarios \citep{Lin23}. 
Other scenarios such as radioactive decay or accretion could play a role in some objects but are for now not thought to be the dominant mechanism. 
For example, based on their SLSNe catalog, \citet{Gomez24} concluded that only a small fraction of the sample can accommodate a significant radioactive decay component. For the accretion scenario, \citet{Moriya18}  studied the required accreted mass to power the SLSNe light curve and concluded that in most cases the mass is greater than the ejecta mass.

The search for SLSNe in $\gamma$-rays is more recent and only upper limits were reported up to the claimed $\gamma$-ray detection of SN~2017egm by \citet[see our Sects. \ref{sec:intro} and \ref{sect:comparison}]{Li24}.
Using \textit{Fermi}-LAT data, \citet{Renault18} presented individual and stacked analysis of a sample of 45 SLSNe and only upper limits were reported. However their samples ranged from 2008 to 2015 and the nearest object tested (CSS140222) was at a distance of 145 Mpc with relatively poor multiwavelength coverage. All other sources investigated lie at a redshift of $z>0.1$ ($d> 400$ Mpc). Naturally SN~2017egm was not present in the sample.

Regarding Type IIn (i.e., with narrow emission hydrogen lines, conveying strong CSM interaction), individual studies \citep[e.g., SN 2009ip,][]{Margutti2014} or systematic searches have been carried out, as in \citet{Ackermann2015} on 147 SNe Type IIn evolving in dense CSM in a one year time window. No significant excess was detected, including the most promising source SN 2010jl \citep[at 49 Mpc; see also][]{Murase19}.
Another $\gamma$-ray study was carried out  by \citet{Prokhorov21}, who did a systematic search with \textit{Fermi}-LAT from a large sample of more than 55000 SNe candidates from the open SN catalog (not only SLSNe).
Their analysis used a sliding time window technique in an aperture photometry mode that allowed for the exploration of a large number of candidates but that is less sensitive than a full forward folding likelihood analysis such as the one that we present in this work for a small sample of SLSNe.
A few candidate detections were reported, including variable emission spatially coincident with the SN iPTF 14hls.
However, a detailed study by \citet{Yuan18} reported that the $\gamma$-ray emission lies within 0.045$\degree$ of a blazar, which offers an alternative scenario for the high-energy signal.

\citet{Acharyya23} present a $\gamma$-ray follow-up with \textit{Fermi}-LAT and VERITAS telescopes of the nearby SLSNe SN 2015bn and SN 2017egm, and only upper limits are reported. We note that in a 6 month time bin for SN 2017egm with \textit{Fermi}-LAT data an interesting hint of a signal (test statistic TS of 10.1) is reported but not considered further.

\begin{table*}
\centering                          
\caption{Sample of nearby ($<$ 200 Mpc) SLSNe used in this study based on the curated list from \citet{Gomez24} and \citet{Pessi25}.}  
\begin{tabular}{l c c c c c c c c c}
\hline\hline                 
Name & Type & RA  & Dec  &  Discovery date & $z$ & Distance   &  Peak Mag  & Peak date & \textit{Fermi}-LAT TS \\
   &  &  deg  &  deg  &    &    &  Mpc  &   & MJD  &  \\
\hline
SN 2017egm  & I  & 154.773 & 46.453 & 2017-05-23  & 0.0307 & 135 & $-20.9$ &  57926.4 & 25.7 \\
SN 2018bsz  & I  &242.412 & -32.062 & 2018-05-09 & 0.0267 & 111 & $-20.5$ & 58267.5 &  0 \\
SN 2019ieh  & I  &250.545 & 6.984 & 2019-06-23  & 0.032 & 140 & $-19.4$ & 58671.6 & 5.9\\
SN 2020wnt  & I  &56.658 & 43.229 & 2020-10-11  & 0.032 & 140 & $-20.6$ & 59213.7 & 0\\
SN 2021adxl & II &177.028 & -12.644 & 2021-11-03 & 0.018 & 79 & $-20.2$ & 59521.0 & 2.2\\
SN 2022mma  & II  &219.756 & 15.986 & 2022-06-10 & 0.038 & 167 & $-20.4$ & 59790.7 & 0 \\
\hline
\vspace{-5mm}
\end{tabular}           
\tablefoot{The peak absolute  magnitude is given in the $r$ band. The TS of a putative $\gamma$-ray signal was evaluated with a \textit{Fermi}-LAT analysis at the SN optical position and integrated over the first year post-SN explosion assuming a spectral photon index of $\Gamma=2$. See Sect. \ref{sect:results} for more details.}
\label{tab:sample}
\end{table*}

   \begin{figure*}
   \centering
   \includegraphics[width=0.75\textwidth]{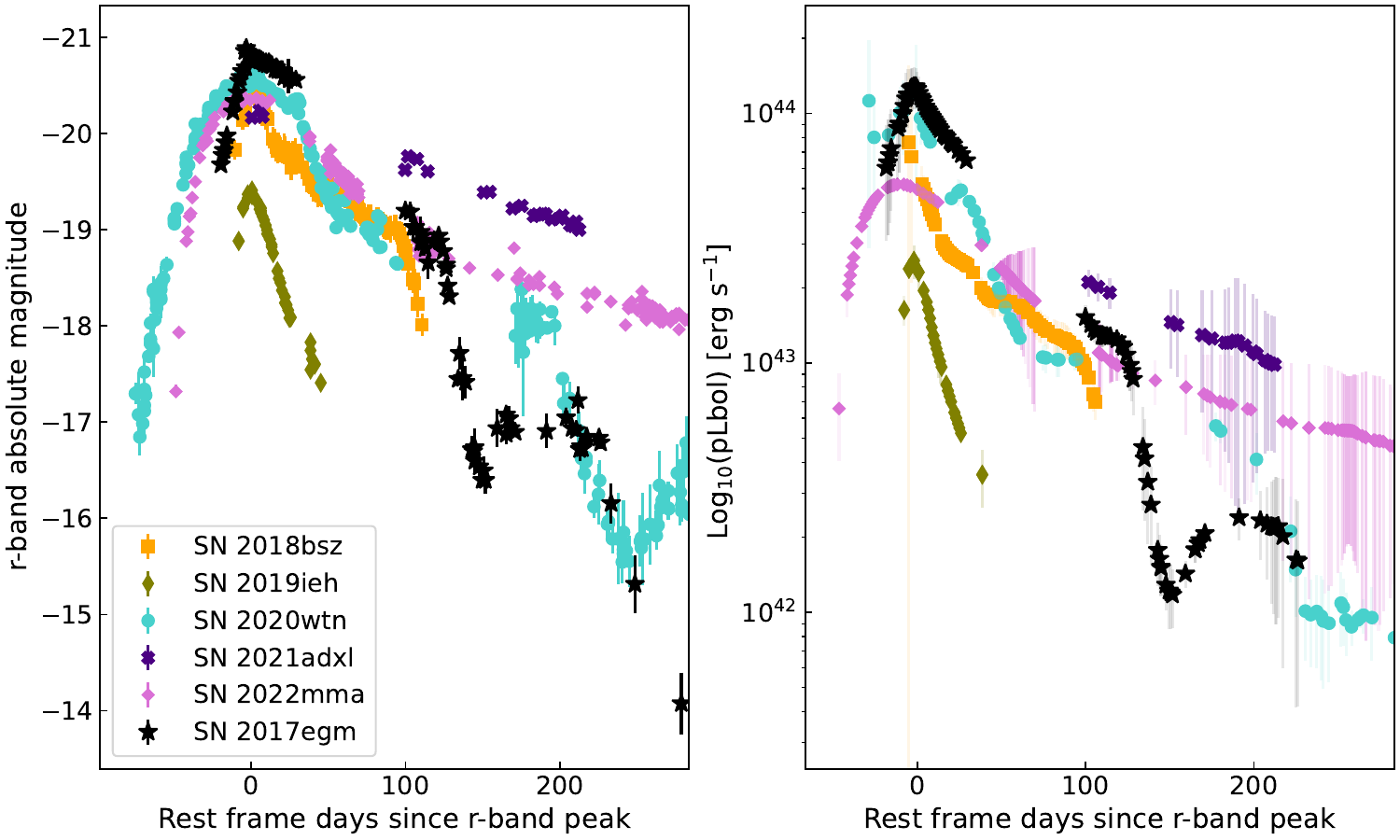}

   \caption{Comparison of the $r$ band absolute magnitude (left) and the pseudo-bolometric (using the $gri$ bands; right) luminosity  light curves for the objects of our samples.
   The light curves have been aligned to $r$ band peak; see Sect. \ref{sect:optical} for more details. Fewer time bins appear in the luminosity panel because of the unfulfilled requirement of simultaneous $g-r-i$ observations in some bins. }
              \label{fig:opticalLC}%
    \end{figure*}

SN 2017egm was first reported by the \textit{Gaia} telescope as Gaia17biu on May 23, 2017.
At a redshift of $z=0.03$ (135 Mpc\footnote{Throughout the paper, we assume a flat $\Lambda$CDM cosmology with $H_0=67.66$\,km\,s$^{-1}$\,Mpc$^{-1}$, $\Omega_{\Lambda}$=0.6889 and $\Omega_{M}$=0.3111 \citep{Planck18}.}), it was the closest type I SLSNe at the time of detection\footnote{Type I SN 2018bsz at a distance of 111 Mpc is only slightly closer and the Type II SN 2021adxl is at 79 Mpc.}. 
In the spectral analysis, the source revealed no H emission (Type I SLSN) and He I emission lines classifying the source as member of the small but growing class of helium-rich SLSN-Ib \citep{Zhu23}.
The host galaxy of SN 2017egm (NGC 3191) is a massive, metal-rich spiral galaxy, in contrast with most Type I SLSNe, which are mostly found in metal-poor dwarf galaxies \citep{Nicholl17}.
While the early-time light curve of the first months can be reproduced with only a magnetar model in \citet{Nicholl17}, the late-time observations revealed multiple bumps starting 200 days after the SN that cannot be explained by either a magnetar model alone or a one-zone CSM interaction \citep{Zhu23,Lin23}. 
Although temporal variation in the properties of the magnetar and in the opacity could be invoked to modulate the observed flux in a magnetar-dominated model \citep[see][]{Hosseinzadeh22},
the scenario favored by \citet{Lin23} to explain the bumpy features is the multiple CSM shell interaction plus radioactive decay model.
This multiple-shell scenario could result from  pulsational mass ejections from the SN progenitor due to the electron-positron pair instability  mechanism that ejects large quantities of matter without totally disrupting the progenitor \citep{Woosley07,Chatzopoulos12,Woosley17,Renzo20}.

At high radio frequencies (5-34 GHz), non-detections were reported in the 34--74 day time interval after the explosion \citep{Coppejans18}.
In the X-ray band, \textit{Chandra} follow-up observations obtained throughout the first year post-explosion were reported in \citet{Zhu23} and yielded the most stringent upper limits of any SLSN-I to date.
At higher energies, the recently reported detection of giga-electronvolt $\gamma$-ray emission from SN 2017egm by \citet{Li24} with \textit{Fermi}-LAT adds another ingredient to the understanding of this complex system.

This triggered our interest to revisit a sample of nearby ($<$ 200 Mpc) sources based on the curated lists of Type I SLSNe by \citet{Gomez24} and type II SLSNe by \citet{Pessi25}.
In Sect. \ref{sample} we define a list of interesting SLSNe candidates for $\gamma$-ray follow-up with \textit{Fermi}-LAT and present the light curves for the 16 yrs of data available with \textit{Fermi}-LAT.
In Sect. \ref{sect:results} we carry out a detailed analysis of the most interesting target SN 2017egm and compare our results with those of \citet{Li24}, while we confront the temporal and spectral properties with the models of \citet{Vurm21} in Sect. \ref{sect:models}. Predictions for the detection of such a target at tera-electronvolt $\gamma$-rays with the Cherenkov telescope CTAO are then explored in Sect. \ref{sect:cta} and alternative multiwavelength counterparts to the $\gamma$-ray signal are discussed in Sect. \ref{sect:mwl}.

\section{Sample selection and data analysis}
\label{sample}

\subsection{SLSN sample selection}
The classification of a transient source as a SLSN is non trivial and can evolve over months to year timescales after the SN discovery as more data become available. In particular, a critical parameter to claim the SN to be luminous or super-luminous is the distance to the source. Depending on  the quality of the spectral data obtained, the association with a host galaxy and the redshift estimation can evolve with time but not always end up being updated in online databases such as WISeREP\footnote{\url{https://www.wiserep.org/} , also see \cite{Wiserep}.} or TNS\footnote{\url{https://www.wis-tns.org/}}. 
In addition, some sources can be later reclassified in another category such as tidal disruption events or other classes as discussed  in Appendix D of \citet{Gomez24} and Sect. 6 about contaminants in \citet{Pessi25}.
Therefore, we decided to base our $\gamma$-ray follow-up on curated lists of events to ensure a robust distance estimation and classification. We selected the Type I SLSNe (H-poor) catalog of \citet{Gomez24}, which provides a verified sample of 262 SLSNe reported through the end of 2022.
For type II SLSNe (H-rich), we used the catalog of \citet{Pessi25}, which comprise  a sample of 107 objects. Combining both catalogs with a distance threshold of 200 Mpc, we end up with the list of 6 targets presented in Table \ref{tab:sample}. Such a threshold is motivated by the \textit{Fermi}-LAT sensitivity: a SLSN at 200~Mpc with a $\gamma$-ray luminosity similar to that in the optical band ($\sim10^{44}$ erg s$^{-1}$) would produce an energy flux comparable to the LAT $2\sigma$ sensitivity level for 1 month of exposure \citep[obtained by scaling the upper limit derived in the SN~2023ixf \textit{Fermi}-LAT study;][]{Marti-Devesa24}.

\subsection{Optical properties of the sample}
\label{sect:optical}

 To compare the optical properties of our sample in a self-consistent way, we retrieved the publicly available light curves of each object from the literature and public archives to reconstruct the r band absolute magnitudes and the pseudo-bolometric luminosities for each SLSN with a set of common bands. For this, we used data from various sources: \citet{Zhu23} for SN 2017egm; \citet{Chen21} for SN 2018bsz; \citet{Gomez24} for SN 2019ieh; \citet{Tinyanont23} for SN 2020wtn; \citet{Brennan24} and \citet{Pessi25} for SN 2021adxl; and \citet{Pessi25} for SN 2022mma. Data from the ZTF data release were also used\footnote{\url{https://irsa.ipac.caltech.edu/data/ZTF/docs/releases/ztf_release_notes_latest}}.
 SN 2019ieh is technically a luminous SN I \citep{Gomez22} but we include it here as it is part of the sample of SLSN I of \citet{Gomez24}.

We note that the only photometric bands that are commonly available for all objects of our sample are $g$, $r$ and $i$. Thus, we constructed pseudo-bolometric light curves considering only these bands following the second method presented by \citet{Pessi25}, which considers the integration of the spectral energy distribution (SED) over the interpolated $gri$ band light curves with respect to $r$ band peak, considering extrapolations to both the UV and NIR by fitting a blackbody to consider the missing flux. In the UV flux, we extrapolate the blackbody fit from 0 Å to our $g$ band and in the NIR flux, $i$ band to 25000 Å. To compare the light curves, we aligned them to their  $r$ band peak epoch as reported in
\citet{Chen21} for SN 2018bsz (MJD 58267.5),  \citet{Gomez24} for SN 2019ieh (MJD 58671.6), \citet{Tinyanont23} for SN 2020wnt (59213.75 MJD), \citet{Brennan24} for SN 2021adxl\footnote{Note that the peak of SN 2021adxl has not been observed and the peak is considered to be the first available observation.} (MJD 59521), \citet{Pessi25} for SN 2022mma (MJD 59790.76).
For SN 2017egm the reported peak has been calculated on the $g$ band by \citet{Bose2018}. Using Gaussian process interpolation we obtained the $r$ band peak epoch of SN 2017egm to be MJD $57926.4\pm1$. 

Our pseudo-bolometric light curves differ from those available in the literature for each object, either because different cosmology parameters are considered or additional bands are included. Although the pseudo-bolometric luminosities obtained in this work only provide lower limits, they allow us to perform comparisons in a self-consistent way. We also consider the $r$ band alone as it is often used as a proxy for bolometric luminosity. The $r$ band absolute magnitude is calculated as $M_{\rm \lambda} = m_{\rm \lambda} - \mu - A_{\rm \lambda} - K_{\rm corr}$ where $\mu$ is the distance modulus, $A_{\rm \lambda}$ is the Milky Way extinction in the corresponding wavelength and $K_{\rm corr}$ is the K-correction approximation $K_{\rm corr} = -2.5 \times log(1+z)$. We do not consider host extinction for any of the events as they have been deemed negligible or has not been considered by the different authors analyzing these events.
With the aforementioned procedure, the $r$ band light curve and the pseudo-bolometric luminosity light curves are presented in Fig. \ref{fig:opticalLC} and the peak value in Table \ref{tab:sample}.
We can see that SN 2017egm is the most luminous event in the considered sample, closely followed by SN 2020wnt and SN 2018bsz.

Given the diversity on the different light curves
shape and spectral features, individual analysis conclude that most events should have at least some degree
of CSM interaction.
This does not exclude the possibility of a central engine in some cases.
Gamma-ray emission in the giga-electronvolt to tera-electronvolt range is predicted for both the CSM interaction powered and magnetar powered scenarios, although no clear association  has been published \cite[see ][and references therein]{Renault18} until the recent case of SN 2017egm that is discussed in this work.

\subsection{Fermi-LAT data analysis}
\label{sect:Fermi-analysis}

   \begin{figure*}
   \centering
   \includegraphics[width=0.33\textwidth]{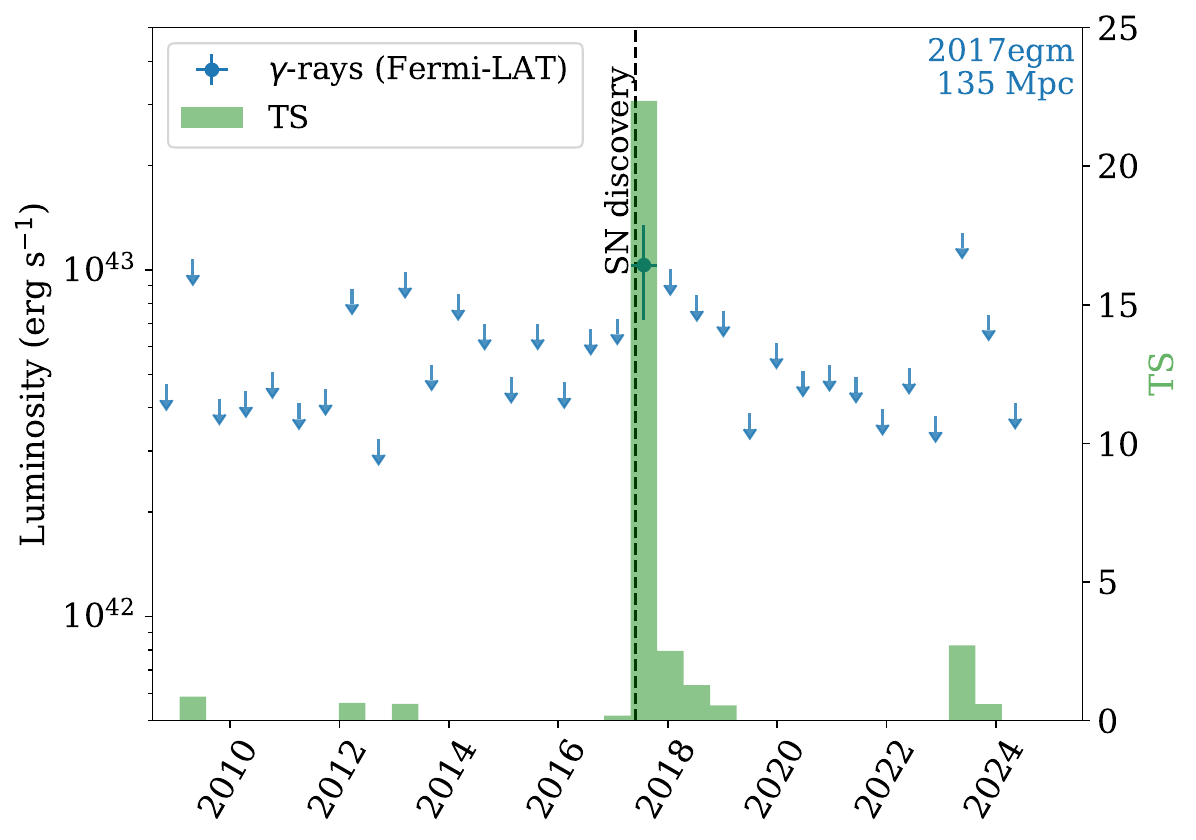} 
   \includegraphics[width=0.33\textwidth]{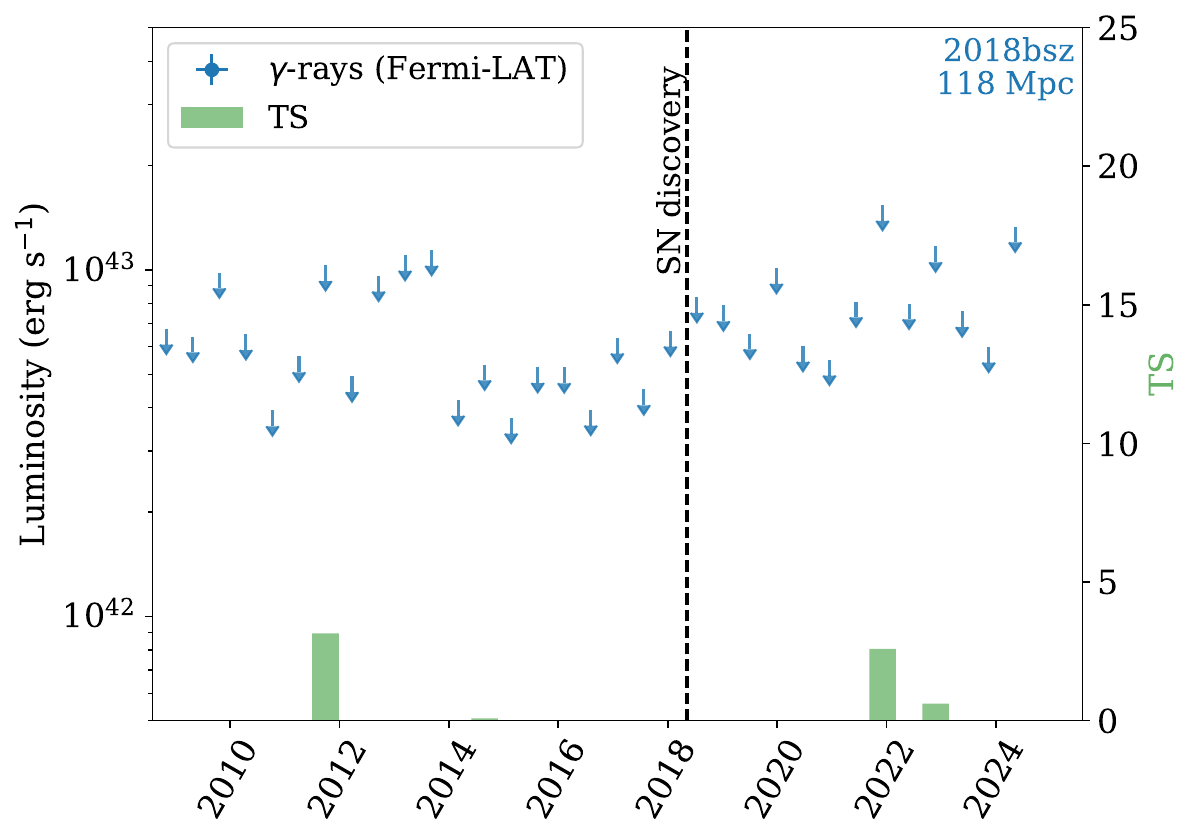} 
   \includegraphics[width=0.33\textwidth]{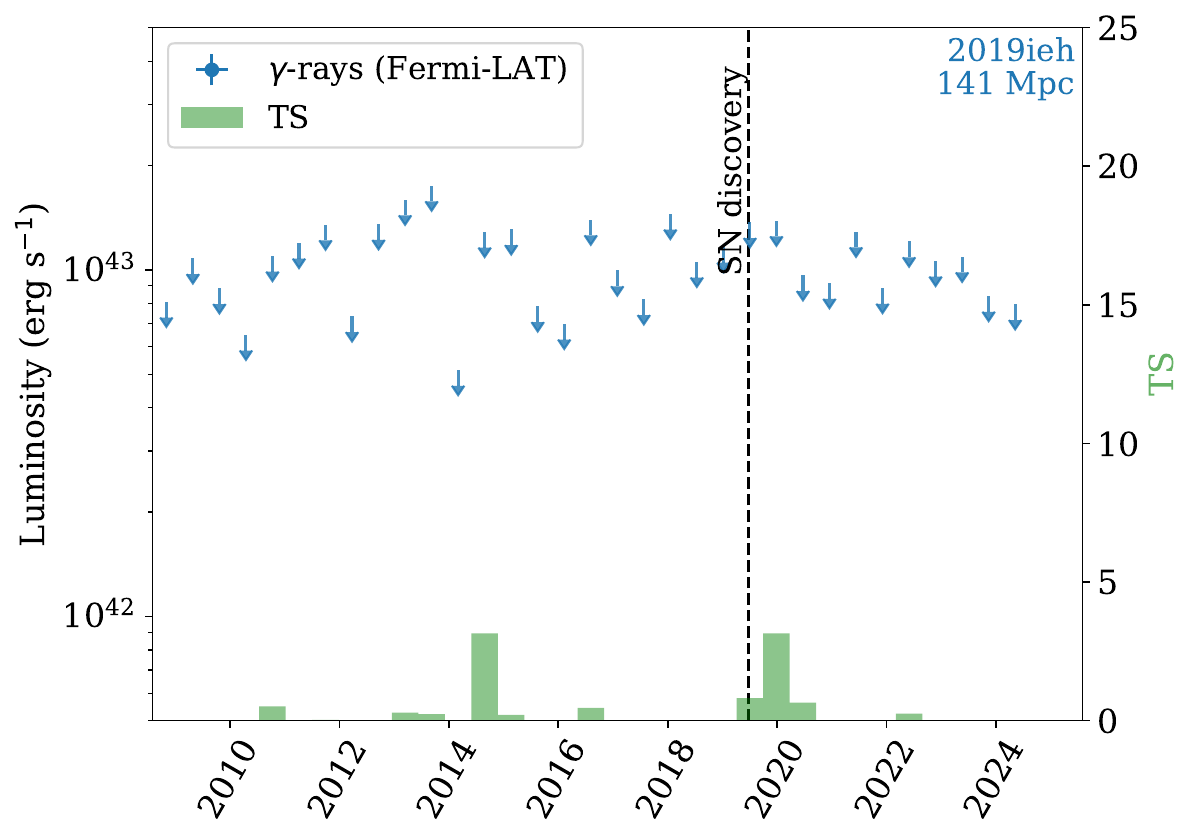} \\
   \includegraphics[width=0.33\textwidth]{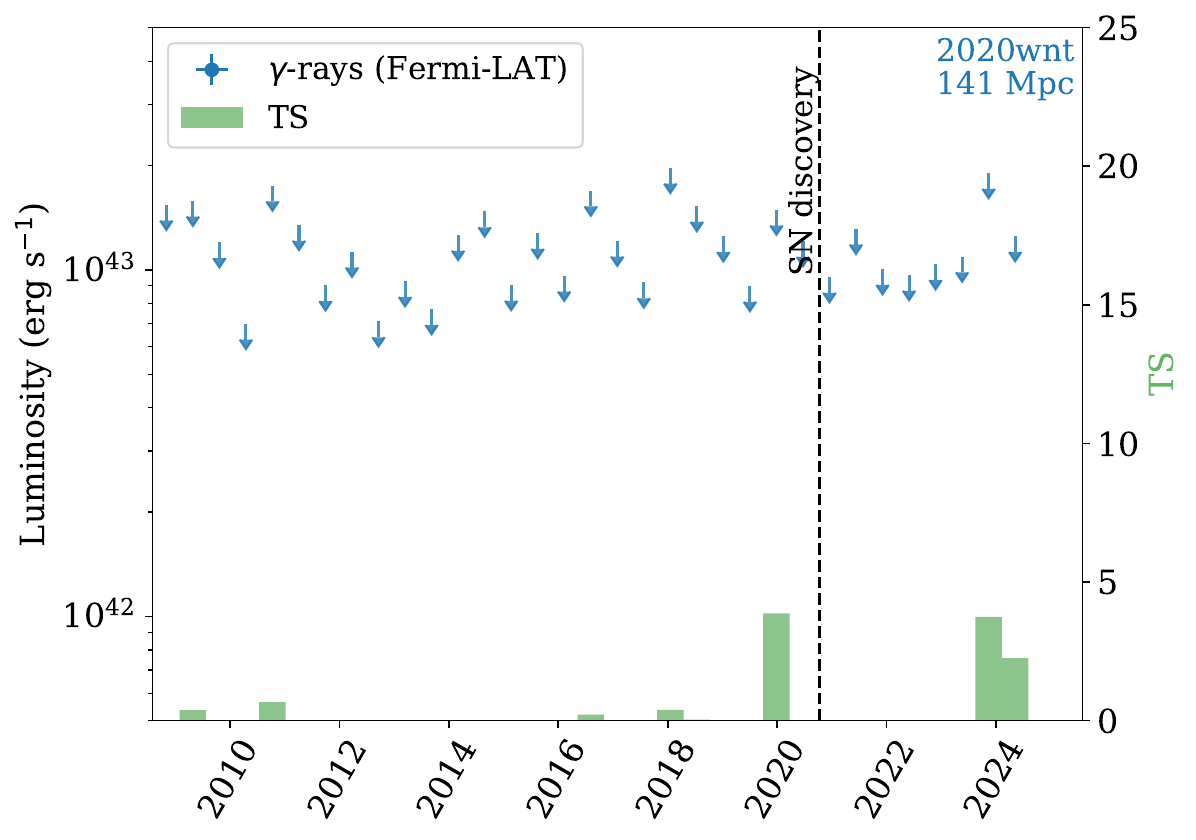} 
   \includegraphics[width=0.33\textwidth]{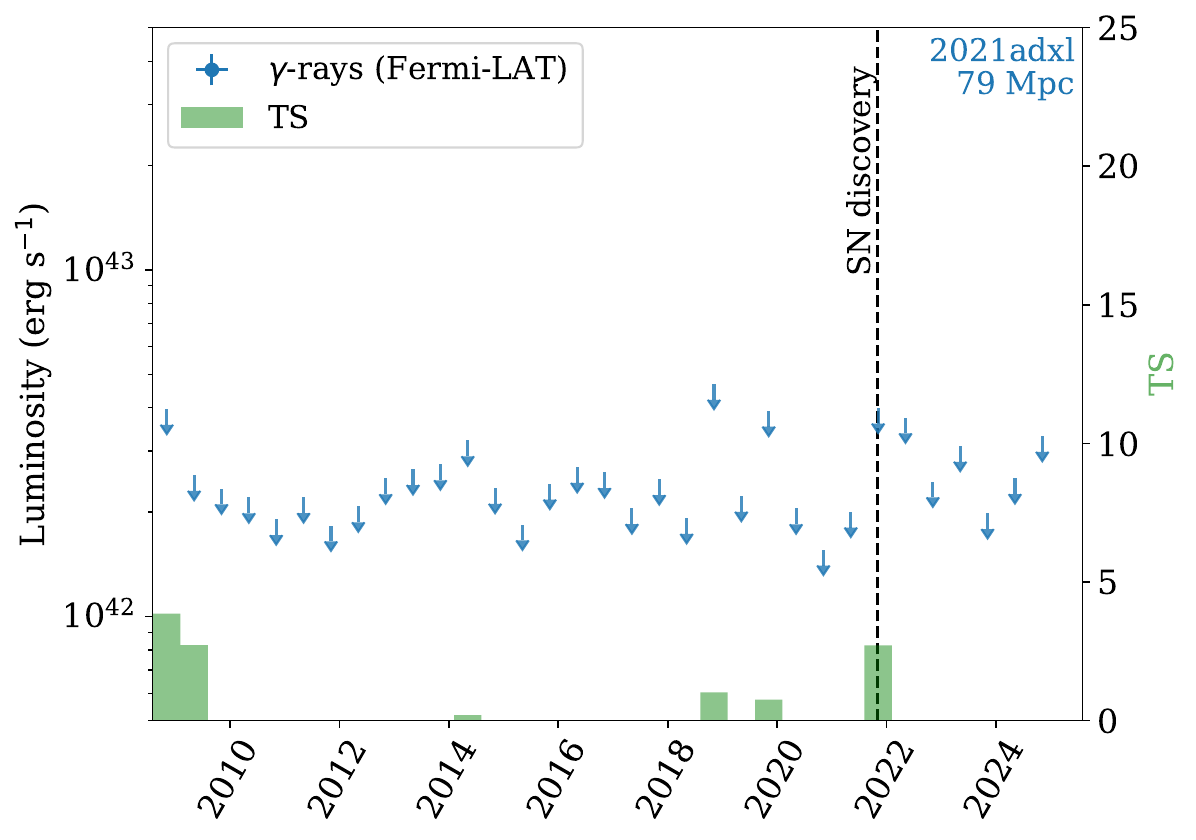}
   \includegraphics[width=0.33\textwidth]{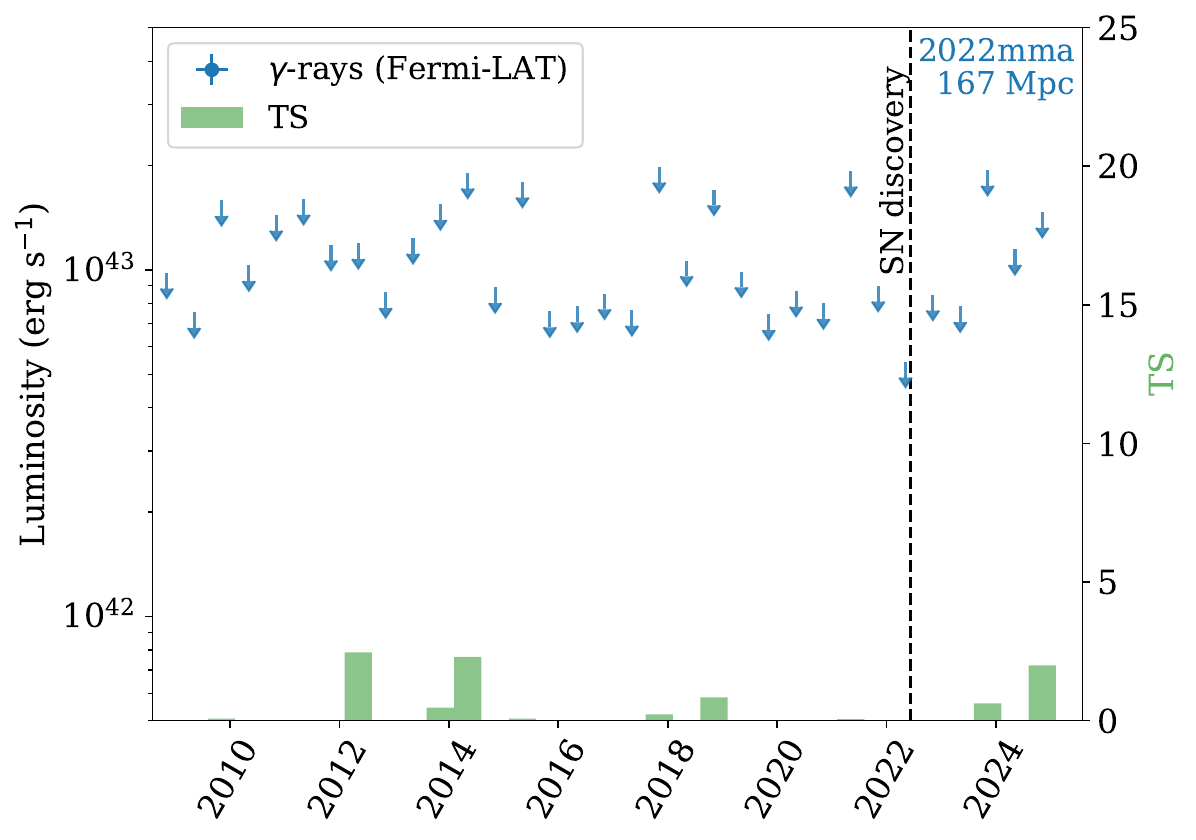}
   
   \caption{Luminosity light curves in the 100 MeV - 100 GeV energy range over 16 yrs for each SN of our sample from the \textit{Fermi} launch to August 2024 with a time bin of 6 months. A flux point is shown when the TS > 4, otherwise upper limits at the 95\% confidence level are reported. In all time bins, the spectral index of the tested source is fixed to 2. The SN discovery date is also indicated. For a comparison across the sample, the derived \textit{Fermi}-LAT flux is transformed to luminosity using the distance indicated in Table \ref{tab:sample}. }
              \label{fig:longtermLC}%
    \end{figure*}

The \textit{Fermi}-LAT is a pair conversion $\gamma$-ray instrument operating in the 30 MeV to higher than 500 GeV energy range surveying the full sky since 2008 \citep{2009ApJ...697.1071A}, which makes it the ideal high-energy SN follow-up telescope.
To study our sample of SLSNe we performed an analysis on a fixed window of one year after each SN discovery to minimize the number of trials.  Then to characterize the emission we performed two different types of \textit{Fermi}-LAT light curve analyses in the 100 MeV -- 100 GeV energy range. A first one spanning the entire dataset available (16.5 years; 2008-08-05 to 2025-02-05) with a time bin of 6 months using \textit{Pass8 R3 SOURCE} events \citep{Atwood13, Bruel18}.
Then a second light curve with finer 15-day time bins over 20 months around the SN (2 months before and 18 months after) is produced for the interesting candidates.
As the LAT point-spread function (PSF) is highly energy-dependent, varying from several degrees width in the 100 MeV to 1 GeV band down to $\sim 0.1^{\circ}$ for E $>$ 10 GeV \citep{4FGL}, this tends to dilute faint signals and reduce the source significance in the low-energy band.
To mitigate this issue, we used the summed likelihood method and jointly fit events with different angular reconstruction quality (the so called PSF0 to PSF3 event types\footnote{\url{https://Fermi.gsfc.nasa.gov/ssc/data/analysis/documentation/Cicerone/Cicerone_Data/LAT_DP.html}}).
We performed a summed likelihood with 4 components: three components (PSF1, PSF2, and PSF3) in the 100 MeV -- 1 GeV energy range, with a zenith angle cut of $< 90^{\circ}$ and a single component above 1 GeV with all event types and a broader maximum zenith angle cut of 105$^{\circ}$.
We performed a  binned analysis within a region of interest (ROI) of $10^{\circ} \times 10^{\circ}$ centered at the SLSN optical position with 8 energy bins per decade, and spatial bins of $0.05^{\circ}$.
The data reduction was performed using the LAT \textit{Fermitools} version 2.4.0\footnote{\url{https://Fermi.gsfc.nasa.gov/ssc/data/analysis/documentation/}} and \textit{fermipy} \citep{2017ICRC...35..824W} version 1.4.
To take into account the energy dispersion in the likelihood analysis\footnote{\url{https://Fermi.gsfc.nasa.gov/ssc/data/analysis/documentation/Pass8_edisp_usage.html}}, we used the science tools parameter \textit{edisp\_bins = $-2$}. This means that two energy bins above and below the analysis energy range will be added when evaluating the  model.
The Galactic diffuse emission was modeled by the standard file \texttt{gll\_iem\_v07.fits} \citep{Acero16} and the isotropic diffuse emission were described by the tabulated model in \texttt{iso\_P8R3\_SOURCE\_V3\_v1.txt}. The models are available from the \textit{Fermi} Science Support Center (FSSC)\footnote{\url{https://Fermi.gsfc.nasa.gov/ssc/data/access/lat/BackgroundModels.html}}.
This setup is common for all all three analysis presented in this study.

To evaluate the significance of a putative $\gamma$-ray counterpart to the SLSN, we added a point source to the sky model at the optical SN position with a simple power-law spectrum $E^{-\Gamma}$  with a fixed spectral index of $\Gamma=2$.
The statistical significance of the signal is estimated using 
${\rm TS}=2(\ln \mathcal{L}_1 - \ln \mathcal{L}_0)$, where $\mathcal{L}_0$ and $\mathcal{L}_1$ are the likelihoods of the background (including catalog sources, null hypothesis) and the hypothesis being tested \citep[source plus background; see][]{Mattox96}.

To account for background sources within the ROI our starting point is
the latest release of the \textit{Fermi}-LAT catalog (4FGL-DR4), which is based on 14 years of data \citep{4FGL-DR3,4FGL-DR4}. We added all sources up to a distance of $15^{\circ}$ from the center of the ROI.
We then used the $optimize$ function of \textit{fermipy}, which iteratively optimizes the parameters of the ROI in a multi-step
process\footnote{\url{https://fermipy.readthedocs.io/en/latest/fermipy.html\#fermipy.gtanalysis.GTAnalysis.optimize}}.
Finally we performed another fit leaving only the normalization of the nearby sources ($< 3^{\circ}$) free to vary, together with the isotropic and diffuse components.
To check for potential new sources, we computed a TS map over the entire dataset period (16 years).
We found no significant excess (TS$>$25) in these residual TS maps.

Once our best ROI model was obtained, we performed our light curve analysis with the \textit{lightcurve}\footnote{\url{https://fermipy.readthedocs.io/en/latest/advanced/lightcurve.html}} function of \textit{fermipy}   with 6-month and 15-day time bins where the Galactic diffuse background normalization is frozen to the best-fit value of the full time period.
In order to accommodate variable sources in the ROI over time (other than our target), the normalization was set free for any source reaching TS$>$16 in any time bin.
Because most time bins yield TS$\sim$0, we computed upper limits using a Bayesian method following the approach of \citep{Helene83}. In particular, we employed the $calc\_int$ method of the $IntegralUpperLimit$ class provided in the \textit{Fermi} Science Tools. Hereafter, upper limits at a $95\%$ confidence level are reported for detections with TS<4 and flux points otherwise.

   \begin{figure*}
   \centering
   \includegraphics[width=6.3cm]{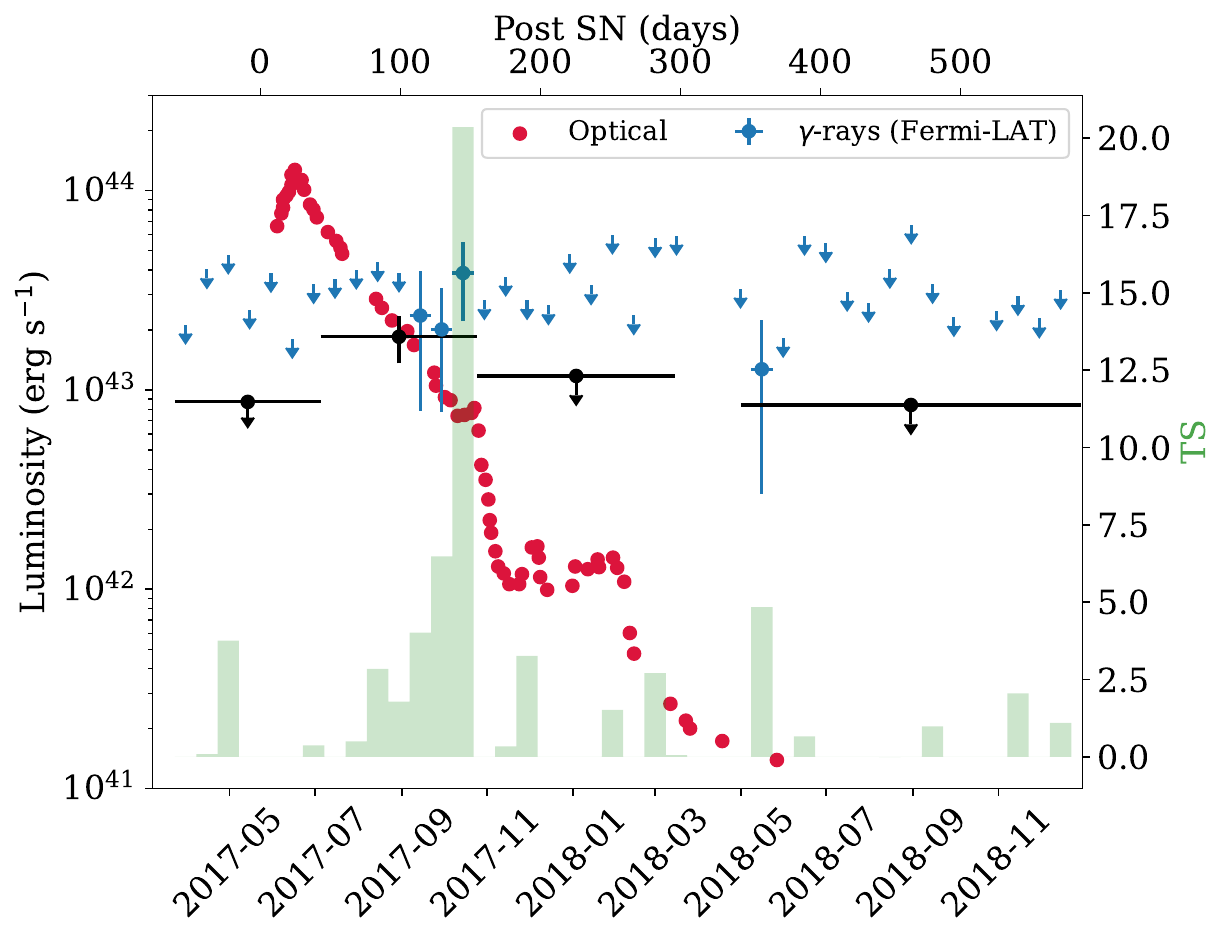} 
   \includegraphics[width=5.8cm]{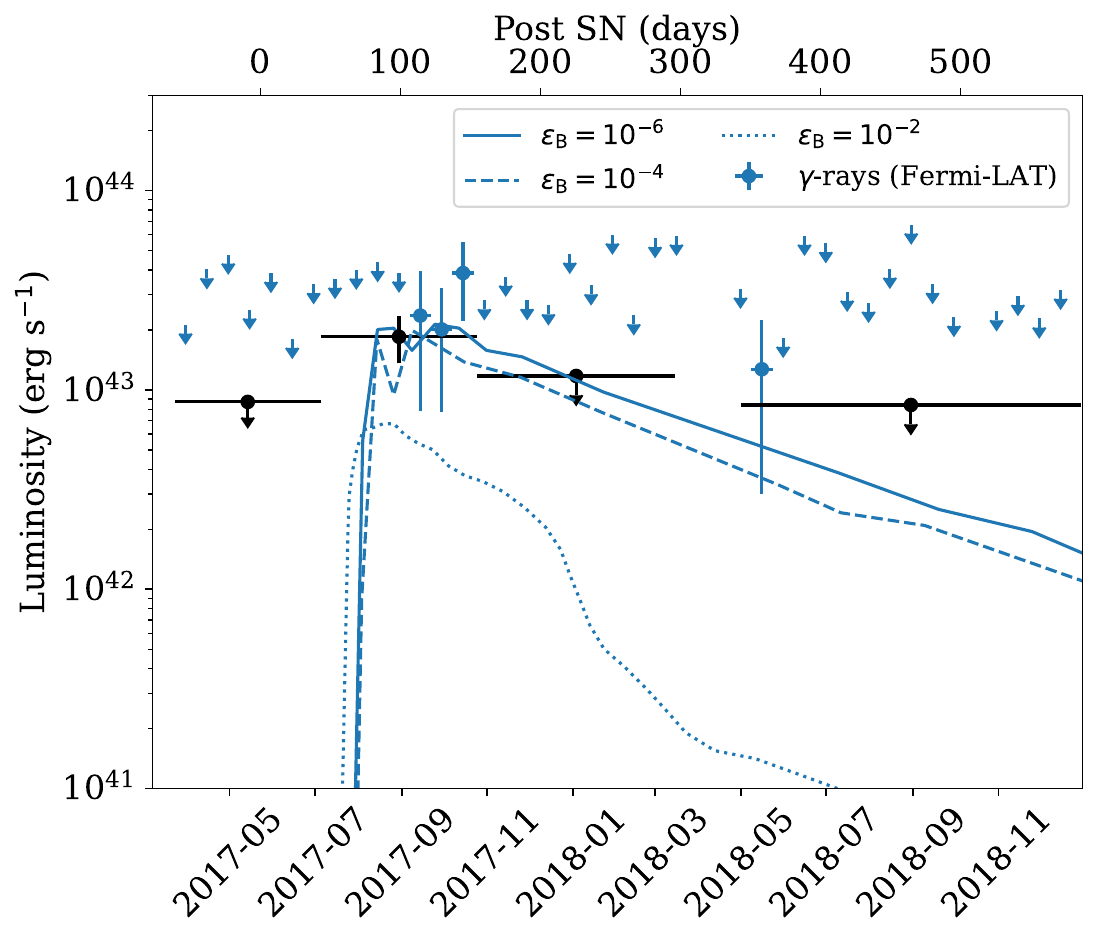}
   \includegraphics[width=5.8cm]{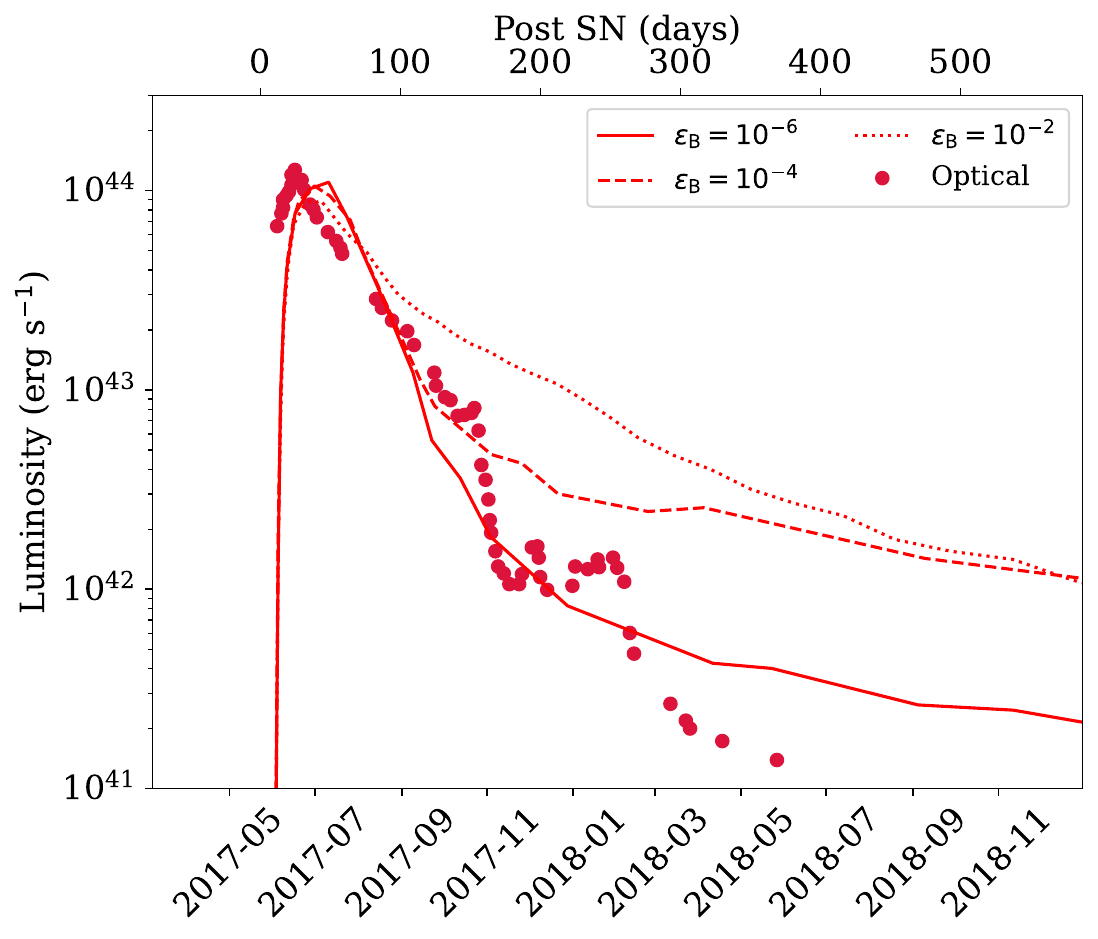} 
   
   \caption{Left: Comparison of the optical bolometric luminosity\protect\footnotemark\  taken from \citet{Lin23} and the $\gamma$-ray 100 MeV-100 GeV luminosity (this work) for SN 2017egm along with the TS in each 15-day time bin and with larger time blocks presented in Sect. \ref{sec:refinedlc} in black.
   Middle and right: $\gamma$-ray and optical luminosity observations and predictions from the magnetar model tailored to SN 2017egm by \citet{Vurm21} for different nebula magnetization parameters (see Sect. \ref{sect:magnetar_model}). Hereafter the $T_{0}$ value is taken at the discovery date 23 of March 2017. }
    \label{fig:LC-2017egm}%
    \end{figure*}

\section{Results}
\label{sect:results}

To minimize the number of trials, we first tested a simple \textit{Fermi}-LAT analysis with the aforementioned setup in which we computed the significance of a $\gamma$-ray source at the SN optical position, in a fixed time window ($T_{\rm SN} \Longrightarrow T_{\rm SN} + $ 12 months), and with a fixed spectral index of 2.0. 
The time window is motivated by the fact that the magnetar model predicts the $\gamma$-ray emission to peak in the first 12 months post-explosion \citep{Vurm21} and that dense CSM shells close to the star are needed to explain the early part of the superluminous light curve in the optical \citep[e.g.,][]{Lin23}.
The fixed spectral index is motivated by the fact that the standard diffusive shock particle acceleration (DSA) mechanism produces a particle population with an index close to 2 and that in the magnetar model the expected SED is rather flat (i.e., also with $\Gamma\sim2$) in the first months \citep{Vurm21}. The results of this first analysis are reported in Table \ref{tab:sample}.
We note that only SN 2017egm shows a significant $\gamma$-ray emission within the first year with a TS of 25.7 corresponding to $\sim 5 \sigma$ pretrial for one degree of freedom (the normalization) and 4.7$\sigma$ post-trial considering that six SLSNe were tested. When the spectral index is let free (2 degrees of freedom) for SN 2017egm, the best fitted index is $2.1\pm0.2$ for a TS of 28.1, which also corresponds to $\sim 5 \sigma$.

For SN 2019ieh, only a small hint of a signal is observed and is discussed further in the following section with a 6-month time bin light curve.  
To evaluate the impact of the fixed spectral index hypothesis, we changed it to 1.5 (2.5), which resulted in a TS  of 9.6 (4.3) for SN 2019ieh. 

\subsection{Long-term light curve of the SLSN sample}

While the $\gamma$-ray emission in the magnetar model is expected in the first year after the SN, we also considered a time range covering a larger dataset from the \textit{Fermi} launch in August 2008 to August 2024 and a time binning of 6 months.
This provides the historical light curve of the SN host galaxy. For example, faint signals over the 16 year period could indicate possible active galactic nucleus (AGN) activity in the host galaxy.
On the other hand, the detection of gamma-ray flux within the year after the SN was observed optically, combined with the lack of historical variability of the host galaxy, would strengthen the possibility that, at least temporally, the $\gamma$-ray signal could be related to the SN.

The light curves resulting from our analysis are presented in Fig. \ref{fig:longtermLC}. 
One can see that besides SN 2017egm, no time bin with TS$>$4 is observed for other targets, which include SN 2018bsz at a distance of 111 Mpc, the nearest SLSN I recorded to date \citep{Anderson2018}, SN 2020wnt, which shows the second-most luminous light curve (see Tab. \ref{tab:sample}), and SN 2021adxl, the closest event in the SLSN II sample of \citet{Pessi25}.
For SN~2019ieh, the light curve with 6 month time bins confirms that there is no significant emission.
Given the limited statistics available, the current data cannot definitively determine whether the observed signal stems from a genuine astrophysical source or is merely attributable to random statistical fluctuations.
In addition, given the fact that SN 2019ieh is the least luminous of our sample we do not consider it further in this study. 
A discussion of the possible reasons for the lack of $\gamma$-ray detection for these targets is presented in Sect. \ref{sect:discussion}.

\subsection{Refined light curve of SN 2017egm } \label{sec:refinedlc}

Given the light curves presented in the previous section, SN~2017egm is the only target for which a $\gamma$-ray signal is observed and it will be the main focus of the rest of this work.
To get a more detailed understanding of the time evolution of the $\gamma$-ray signal, we performed a refined light curve \textit{Fermi}-LAT analysis focused around the SN discovery time. The time binning is refined to 15-day time bins starting 2 months before the SN explosion and extending to 18 months after. For this finer-binned analysis, we used the same analysis configuration (summed likelihood with different PSF types) as presented in Sect. \ref{sect:Fermi-analysis}.
This more finely binned light curve confirms the detection of a delayed $\gamma$-ray emission that peaks approximately 120 days after the SN explosion, as is shown in Fig. \ref{fig:LC-2017egm}.

In the light curve shown in Fig. \ref{fig:LC-2017egm} (left panel) no signal is detected within the first $\sim$50 days after the explosion. Then the $\gamma$-ray emission rises, reaching a peak significance of TS$\sim$22 at T$\sim$130 days post-explosion. The flux is consistent with the upper limits derived by \citet{Acharyya23}, whose analysis differs in the exact time range employed and the optimization of the data selection.

\footnotetext{Note that the bolometric luminosity used here is slightly different from the one shown in Fig. \ref{fig:opticalLC} because it includes more bands. We do this because here we want to characterize its energy, while in the other case we want a fair comparison with other events that have fewer observed bands.}

To assess the duration of the peak and possible substructure, we applied the Bayesian block algorithm \citep{Scargle13}, following the likelihood formalism developed by \cite{Kerr19}. Within this framework, the ROI model was employed to compute the probability for each photon in the ROI to be associated with the new source with \textit{gtsrcprob}\footnote{\url{https://raw.githubusercontent.com/fermi-lat/fermitools-fhelp/master/fhelp_files/gtsrcprob.txt}}, and used as a weight to compute the number of blocks in our light curve ($N_b$). The addition of further blocks was penalized with a prior
$\propto N_b^{\gamma}$, where $\gamma$ parametrizes the false-positive rate. We computed this rate by redistributing photons randomly among blocks, and found that $\gamma=5$ approximately corresponds to $2\sigma$. A total of three blocks are found: two non-detections and a $\gamma$-ray signal well represented by a single block from 57939 MJD (2017-07-05) to 58051 MJD (2017-10-25) for a total duration of 112 days. No significant substructure is found for a variability threshold set at $2\sigma$. We computed upper limits for the blocks with non-detections. The first corresponds to a single bin prior to the $\gamma$-ray emission (from MJD 57835 to 57939), while the last block is divided in two bins (MJD 58051 to 58192 and MJD 58239 to 58482) as per the coverage limitations owing to recovery from the failure of one of the solar array rotator drives on {\it Fermi}\footnote{\url{https://fermi.gsfc.nasa.gov/ssc/observations/types/post_anomaly/}}. The resulting upper limits are shown in Fig. \ref{fig:LC-2017egm}.

\begin{figure}
   \centering
   \includegraphics[width=9cm]{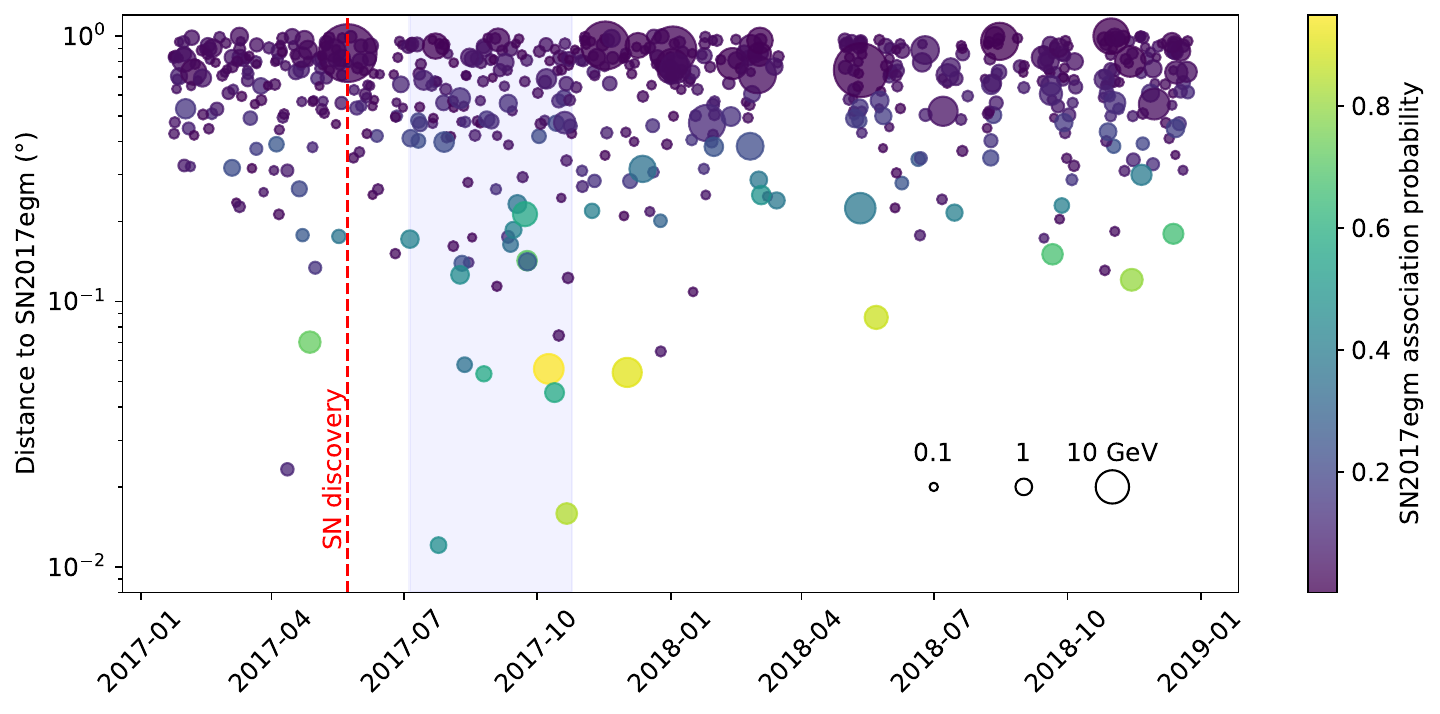} 

   \caption{Photon angular distance to SN 2017egm optical position as a function of time. Using the best-fit model of the ROI, the probability that a given photon is associated with the SN is illustrated by colors, while the size indicates the photon energy. The $\gamma$-ray time interval defined by the Bayesian block algorithm is illustrated with the shaded area. Coverage limitations due to the failure of one of the solar array rotator drives on {\it Fermi} resulted in the loss of approximately 18 days of data in March and April 2018.}
              \label{fig:photons}
    \end{figure}

For illustrative purposes, the duration of the Bayesian block is also shown in Fig. \ref{fig:photons} on top of the photon distribution as a function of time and angular distance to the SN optical position. The probability of being associated with the SN and the reconstructed energy for each photon is also added.
As was expected, the density of near-target photons with high association probability is higher within the Bayesian block relative to the full analysis period.

\subsection{Spectral analysis and source localization}
\label{sect:spectral}

Using the time block defined earlier, we carried out a spectral study using 112 days of data with the same \textit{Fermi}-LAT analysis configuration (see Sect. \ref{sect:Fermi-analysis}).
Over this short period of time, and at an extragalactic position, the level of Galactic diffuse emission cannot be well constrained, and its normalization is fixed to the best-fit value derived from the 20-month analysis ($norm=0.89$).
The isotropic diffuse emission and the normalization of the brightest source 4FGL J1015.0+4926 in the ROI was set free. The parameters of other sources were fixed to the fit obtained over the 20-month period.
Assuming a power-law spectral model, the likelihood analysis results in a significance of TS=33 for SN 2017egm when the position is fixed at the SN optical position. With two degrees of freedom for the source (normalization and spectral index), this corresponds to a $5.4 \sigma$ detection. 

We note that in the Bayesian block time interval, one photon of high energy  ($E=$ 6.8 GeV) has a high association probability (0.94) as shown in Fig. \ref{fig:photons} and Table \ref{tab:events}. To test the robustness of our detection against a single photon, we performed a likelihood analysis where this single photon is removed. This resulted in a TS=25 (instead of 33) indicating that most of the significance does not depend on a single photon.

When fitting the source localization using the \textit{fermipy} $localize$ command, the likelihood improvement is marginal ($\Delta log \mathcal{L}\ = 0.5$).
This indicates that the $\gamma$-ray signal is statistically compatible with the optical position.
The best localization is at RA$_{\rm J2000}$=154.81$^{\circ}$, Dec$_{\rm J2000}$=46.41$^{\circ}$ ($\sim0.05^{\circ}$ from the optical position; see also Sect. \ref{sect:mwl}) with a 95\% error radius of 0.12$^{\circ}$.
This localization is compatible with the position reported in \citet{Li24} with a slightly smaller 95\% error compared to their study (0.18$^{\circ}$).

\begin{figure}
   \centering
   \includegraphics[width=9.2cm]{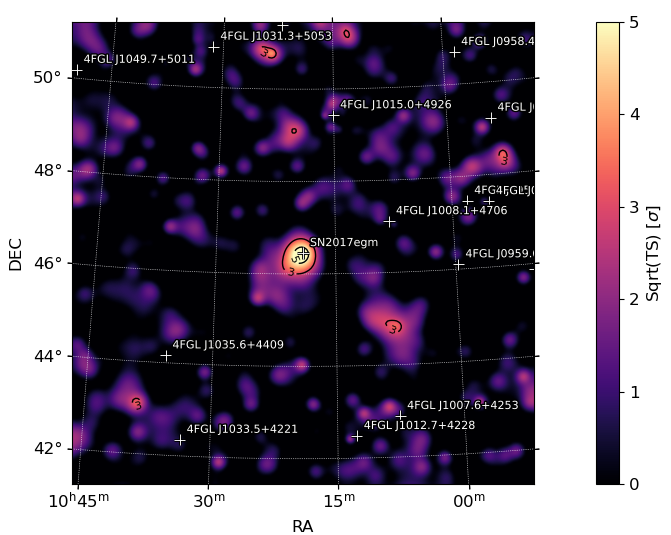} 
   \includegraphics[width=9cm]{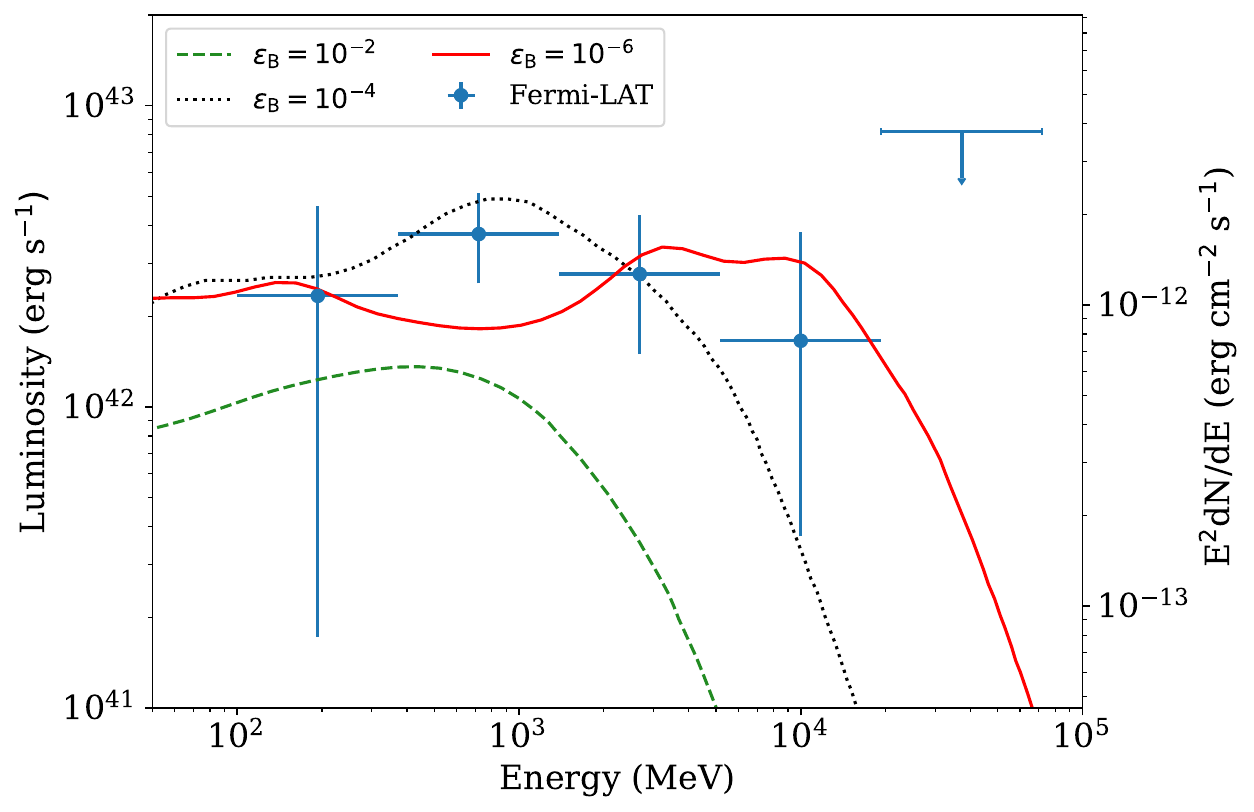}
   \vspace{-0.6cm}
   \caption{Top: Residual TS map in the 0.1--100 GeV energy range in the time interval defined by the Bayesian block algorithm (2017-07-05 to 2017-10-25). The plus symbols represent the sources from the 4FGL-DR4 catalog and the SN optical position. 
   Bottom: $\gamma$-ray luminosity SED of SN 2017egm in the same time interval. The spectral predictions from the magnetar model are shown for different nebula magnetization parameter (see Sect. \ref{sect:magnetar_model} for more details).
   }
              \label{fig:tsmap_sed}%
    \end{figure}

The residual TS map (where only SN 2017egm is excluded from the model) is presented in Fig. \ref{fig:tsmap_sed} (top panel) and shows no significant residuals in the $10^{\circ} \times 10^{\circ}$ region around the source.
The SN best-fit spectral model parameters are $\Gamma=2.17\pm0.23$ and a normalization $N_{\rm 0} = (0.82 \pm     0.22) \times 10^{-12}$ cm$^{-2}$ s$^{-1}$ MeV$^{-1}$ for a reference energy of 1 GeV.
We tested for spectral curvature with an exponential cutoff power-law model, which resulted in a marginal likelihood improvement of $\Delta log \mathcal{L}\sim 3$.
The SED shown in Fig. \ref{fig:tsmap_sed} (bottom panel) was estimated by computing the photon flux in each energy interval, assuming a power-law shape with a fixed photon index of $\Gamma=2$ for the source. A 95\% confidence level upper limit was computed when  TS $<1$ in a given energy bin. Spectral data points are given in Table \ref{tab:sed}.

\subsection{Comparison with past studies}
\label{sect:comparison}

Our detailed study confirms the evidence found by \citet{Li24}. We note that there are some variations in the analysis setup such as energy range, time range, and spatial binning size that could explain some of the differences in TS that are observed. In particular their 1-year analysis (shown in their Fig. S2) starts on  August 4, 2017, and yields a TS=23.8. As our Bayesian block analysis yields a starting date on July 5, 2017, we can assume that some signal is lost by starting in August. This difference and the fact that we carry out a joint likelihood with PSF event type  (i.e., a more sensitive analysis) might explain the slight differences observed. 
Concerning the  spectral parameters obtained in this study ($\Gamma=2.17 \pm 0.23$), they indicate a marginally harder spectrum compared to those reported by \citet{Li24} of $\Gamma=2.6 \pm 0.4$. This difference may be attributed to the differing energy ranges employed; 0.5--500 GeV in the aforementioned study, whereas our analysis was conducted within the 0.1--100 GeV range.
If we reproduce the same energy range of their analysis, we obtain similar results ($\Gamma=2.6 \pm 0.3$).
Regarding the analysis by \citet{Acharyya23}, similar variations could explain the observed differences. In particular for faint sources, the coarse spatial bin size ($0.1 \degree$)  can significantly degrade the source significance as illustrated in Table 2 of \citet{Acero22} in the case of the faint $\gamma$-ray signal of the Kepler supernova remnant. We also note that \citet{Li24} investigated the specific setup of \citet{Acharyya23} and were able to reproduce a similar TS. 

\section{Exploring other multiwavelength associations}
\label{sect:mwl}

While the giga-electronvolt association with SN 2017egm is certainly enticing, the $0.12^{\circ}$ uncertainty in the localization (95\% confidence level) requires further evaluation. Most extragalactic giga-electronvolt transients are identified as flaring blazars; therefore, AGN activity could be a suitable interpretation for the excess. \cite{Li24} searched for optical flares in the \textit{Gaia} archive \citep{GaiaDR3} and cross-matched nearby radio sources with an analysis of \textit{Chandra} data, finding three alternative counterparts \citep[see Table II in ][]{Li24}. One corresponds to a source within the host galaxy NGC~3191 \citep[Radio Source 1, coincident with WR~187; see ][]{Brinchmann2008} and another to the galaxy MCG+08-19-017 (Radio Source 2). None of them show nuclear activity, and the third source (SDSS~J101815.31+462626.6, Radio Source 3) falls outside our 95\% containment region. 

\begin{figure*}
   \centering
   \includegraphics[width=\textwidth]{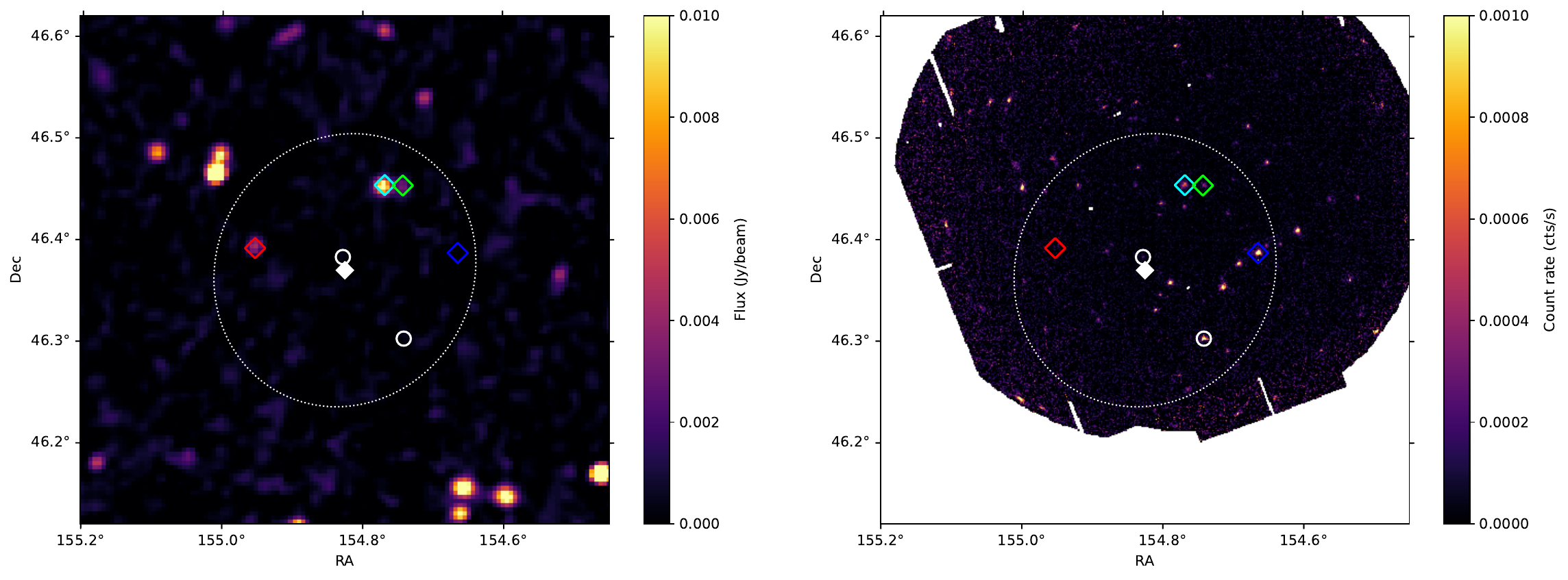}
   \vspace{-0.4cm}
   \caption{Multiwavelength maps of the region around the giga-electronvolt excess: radio (VLA; left) and X-rays (XMM; right). The white marker represents the best-fit position of the giga-electronvolt excess, and the dotted curve is its localization error ellipse at a 95\% confidence level. The cyan and green markers locate the nearby galaxies NGC~3191 and MCG+08-19-017, while the red and blue correspond to NVSS~J101948+462332 and SDSS J101839.48+462315.8, respectively. White circles mark the quasars SDSS J101858.02+461812.3 and SDSS J101918.93+462301.4 discussed by \citet[][]{Li24}.}
              \label{fig:MWL}
    \end{figure*}

However, the region contains multiple, relatively faint X-ray sources reported 
in the XMM-\textit{Newton} DR14 and \textit{Chandra} point source catalogs CSC 2.1 \citep{Webb2020,Chandra} with no known radio counterpart (see Fig. \ref{fig:MWL}). For example, \cite{Li24} discuss the optical emission of two quasars (SDSS~J101918.93+462301.4 and SDSS~J101858.02+461812.3) that are spatially consistent with detected X-ray sources (4XMM~J101918.8+462301 and 4XMM~J101858.0+461812, respectively). These sources have no reported variability in the XMM-DR14 catalog. The brightest among all X-ray sources within our localization error is the quasar 2CXO~J101839.5+462315 \citep[SDSS~J101839.48+462315.8, $z=0.941$;][]{Albareti17}, which is only reported to have mild variability in the CSC~2.1 catalog. Notably, the largest flux is found at $\sim 9.2 \times 10^{-14}$~erg/cm$^2$/s (ObsIDs 19034 and 19035; 2017-09-17 and 2017-11-09, respectively), compared with $\sim  5.6\times 10^{-14}$~erg/cm$^2$/s derived in the only other existing observation (ObsID 19036; 2018-05-21). No optical variability is found in quasi-contemporaneous ATLAS data \citep{ATLAS}, although we note that the limiting magnitude of its forced photometry is close to the brightness of the quasar. Only mild variability ($\Delta m \leq1$) is found by \textit{Gaia}, PTF, or ZTF \citep{PTF, ZTF, GaiaDR3} -- considering observations prior or after the giga-electronvolt excess peak.
Furthermore, it is worth noting that SDSS J101839.48+462315.8 lacks a known radio counterpart in any of the major radio surveys \citep[VLASS, LoTSS, RACS, FIRST, NVSS, and SUMSS; see][]{Flesch2024}. According to the \textit{Fermi}-LAT AGN catalog by \citet{Ajello2020}, the vast majority of $\gamma$-ray AGNs exhibit radio fluxes exceeding approximately 5 mJy (refer to their Fig. 14). Considering that current surveys achieve sensitivity levels in the millijansky range \citep[and down to sub-millijansky levels for VLASS,][]{Lacy2020}, the apparent radio-quiet nature of SDSS J101839.48+462315.8 suggests that it does not conform to the typical characteristics of a \textit{Fermi}-LAT blazar.

The last relevant candidate might be the unidentified VLA source NVSS~J101948+462332 \citep[4.9 mJy at 1.4 GHz; ][]{NVSS}. But the source has no X-ray or optical counterpart, and no variability was found in radio \citep{Ofek11}. Although the radio source lies outside \textit{Chandra}'s field of view, XMM observations set a limit at $F(0.2-2\;\rm{kev})< 3.9\times10^{-15}$~erg/cm$^2$/s \citep[Obs\_id: 08437401;][]{Konig22} -- for $N_H=10^{20}$~cm$^{-2}$ \citep{HI4PI} and assuming $\Gamma=2$. As in the previous case, NVSS~J101948+462332 is an unlikely association given the low X-ray flux compared with the known population of giga-electronvolt-bright AGN \citep{Ajello2020}.

Overall, although the giga-electronvolt excess is spatially consistent with the position of several quasars, the existing multiwavelength monitoring does not strongly support any association other than SN 2017egm. At the same time, the poor coverage during the Bayesian block time interval prevents an unambiguous rejection of the associations discussed.

\section{SN 2017egm $\gamma$-ray emission: CSM interaction and magnetar nebula scenarios }
\label{sect:models}

As the first SLSN to be detected at $\gamma$-ray energies, SN 2017egm represents a new  window to shed light on the mechanism powering these objects.
To understand the nature of the $\gamma$-ray emission, the temporal and spectral results from  this SN are discussed in the magnetar and CSM interaction frameworks.

\subsection{Magnetar-powered $\gamma$-ray emission}
\label{sect:magnetar_model}
In the magnetar model, the rotational energy from the millisecond magnetar is converted into a wind of electrons and positrons that inflates a magnetar wind nebula of highly energetic particles.
High-energy photons from the nebula will be injected into the ejecta via IC and synchrotron processes, where they will be partly absorbed and reprocessed into thermal optical/UV radiation.
The absorption implies mechanisms such as the Bethe-Heitler process (photon-matter interaction) and $\gamma$-$\gamma$ pair production  on thermal (optical/UV) or nonthermal (X-ray) photons that lead to an attenuation of the $\gamma$-ray flux.
These interactions govern the energy transfer and the escape of high-energy radiation.
This mechanism, where high-energy photons from the nebula convey energy to the ejecta, is a fundamental aspect of the magnetar model for maintaining the high luminosity of SLSN over an extended period.

To accurately describe the thermalization process and the optical depth as a function of both energy and time in a self-consistent way is a challenging task due to the interplay of all the interaction processes and the evolving properties of the ejecta.
To tackle this issue, \citet{Vurm21} employed a three-dimensional Monte Carlo time-dependent numerical radiative transfer model that tracks the production, transport, and interactions of photons and electron-positron pairs throughout the nebula and the ejecta.
Of particular interest for this study, the predicted observables from their model are the optical luminosity, the inferred escaping $\gamma$-ray luminosity and associated spectral distribution.

To adapt their model to the specific case of SN 2017egm, the system parameters 
 (magnetic field strength of $1\times 10^{14}$ G, pulsar period at birth of 5 ms, and ejecta mass of 3 M$_\odot$) have been fixed from the optical light curve from \citet{Nicholl17}.
Once these parameters are set, the remaining  main free parameter in their model is the nebula magnetization parameter, $\varepsilon_{\rm B}$.
This is the ratio of the magnetic energy over the injected magnetar energy \citep[see Eq. 14 in.][]{Vurm21}.
In a more magnetized nebula, a greater fraction of the energy is radiated via synchrotron emission. This emission can be more readily absorbed and thermalized by the ejecta than the higher-energy IC emission, which is more prone to escape without thermalizing \citep[see Sect 2.2 and Eq. 17 of][]{Vurm21}.
As can be seen in the models shown in Fig. \ref{fig:LC-2017egm} (middle and right panels), the direct consequence of this is that for a nebula with high  magnetization, a greater fraction of the engine power is thermalized producing greater late-time optical luminosity and lower levels of $\gamma$-rays.

From the optical and giga-electronvolt light curves, and the $\gamma$-ray spectrum, a magnetar model with low nebular magnetization is plausible with $\varepsilon_{\rm B} \sim 10^{-6}$ or smaller. This is in particular required not to largely overshoot the optical light curve at $t>$ 250 days.
As is discussed in \citet{Vurm21}, such a low magnetization value (compared to $> 10^{-2}$ for the Crab nebula but at a very different age) 
implies the existence of an extremely efficient magnetic dissipation mechanism in the nebula.
An interesting alternative scenario to reduce the late-time optical luminosity is that the spin-down luminosity decays faster than the typical magnetic dipole rate $\propto t^{-2}$; for example, due to a growing magnetic dipole moment or fall-back material accretion \citep{Metzger2018}.  

In the magnetar scenario, given a sufficiently bright source, the $\gamma$-ray observations at giga-electronvolt at late-time would allow one to trace the time evolution of the spin-down luminosity by measuring the flux decay after the light curve peak. Unfortunately this is not possible for SN 2017egm, as the upper limits derived at later times are not sufficiently constraining.
The role of tera-electronvolt observations will be discussed later in Sect. \ref{sect:cta}.

\begin{figure*}
   \centering
   \includegraphics[width=\columnwidth]{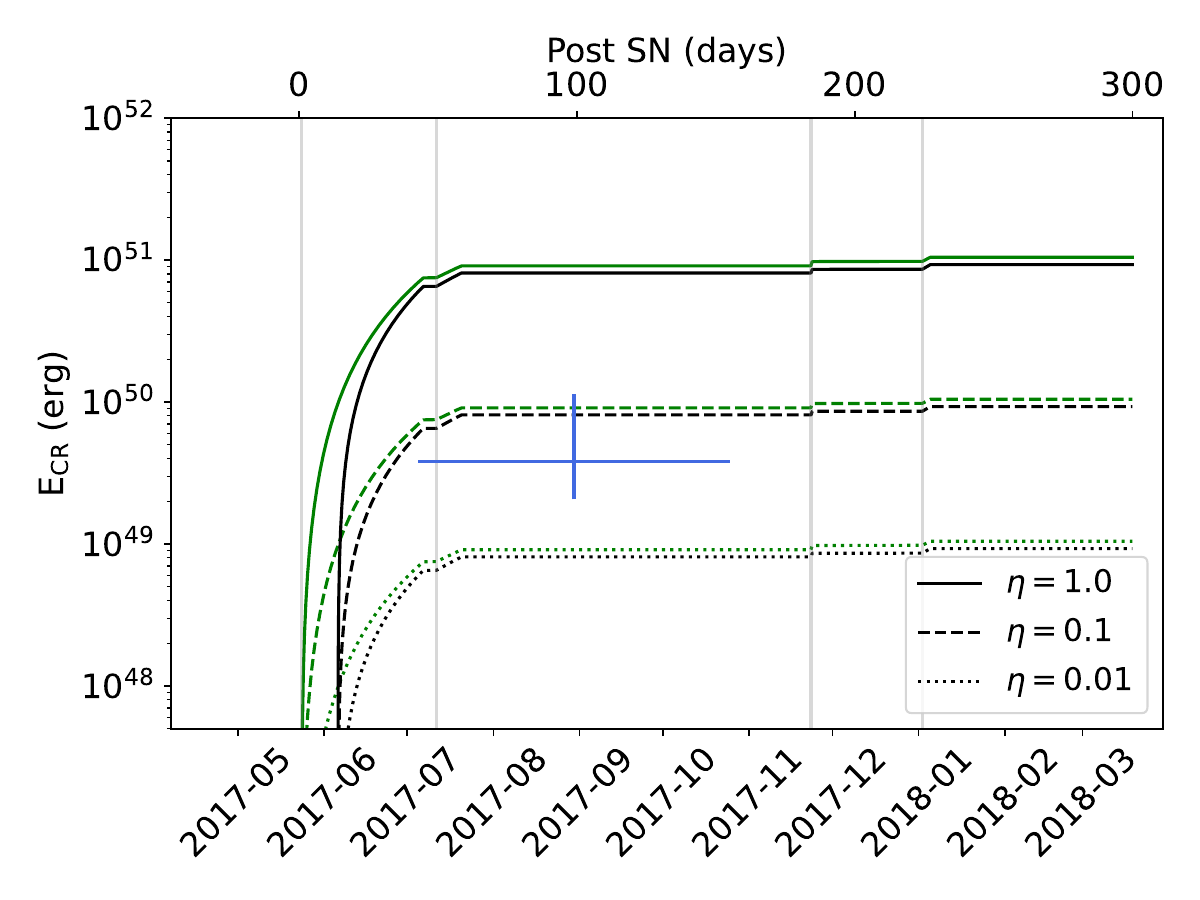}
   \includegraphics[width=\columnwidth]{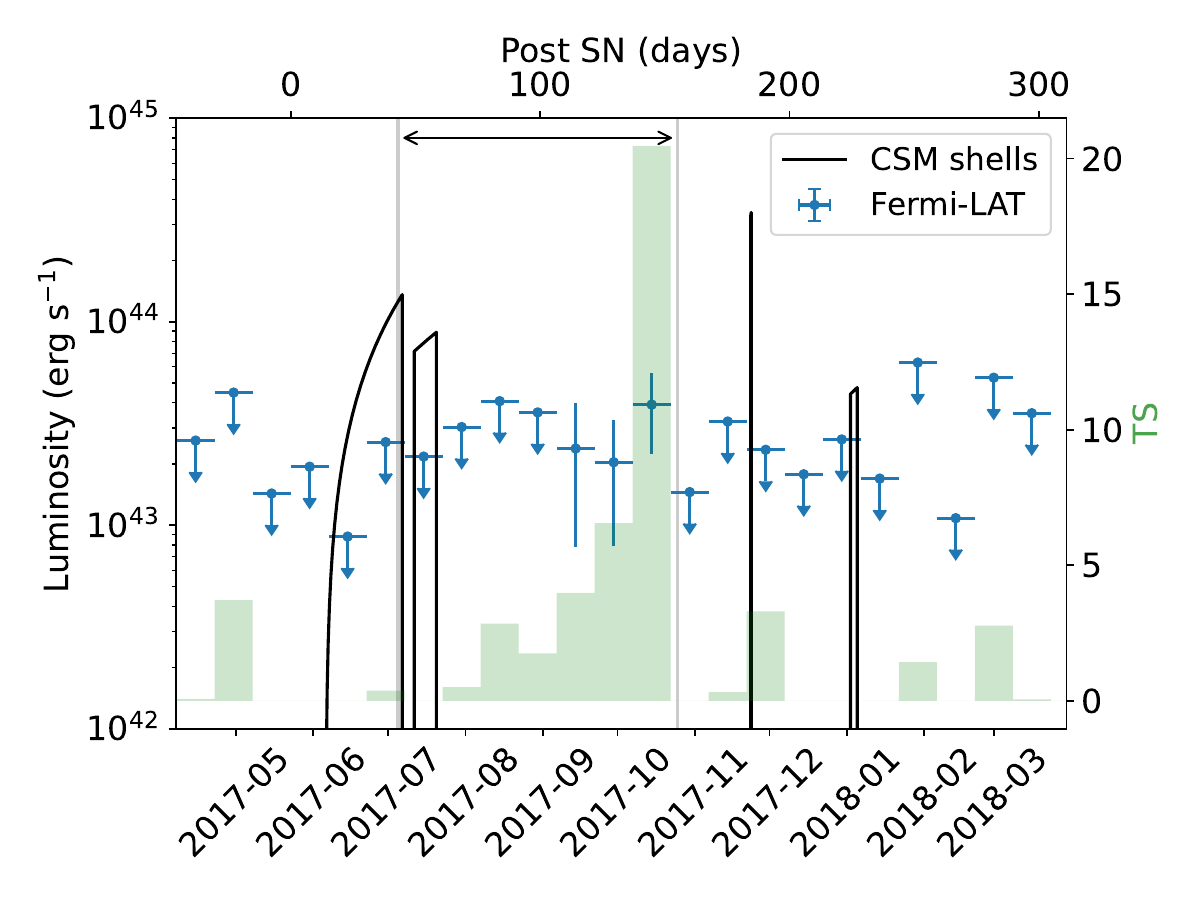} 
   
   \caption{Left: Energy conversion into cosmic rays as a function of time for different $\eta$ values (Eq. \ref{eq:ECR}), and compared with the best fit obtained from \texttt{Naima}. Vertical lines represent the crossing times of the inner radius of each shell. Scenarios for $t_i=1$~d (green) and $t_{i}=t_{br}$ (black) are displayed, both assuming $E_{\rm{cutoff}}=1$~PeV. Right: $\gamma$-ray flux expected from the interaction of each shell ($\eta=0.05$) compared with the \textit{Fermi}-LAT flux (100 MeV -- 100 GeV). The arrow represents the Bayesian block interval of the emission peak, while the TS is shown as shaded green bars. Both panels display the relevant time interval for the CSM shell model.}
              \label{fig:ECRandflux}%
    \end{figure*}

\subsection{A hadronic CSM interaction scenario}\label{sect:csmmodel}

As an alternative to the magnetar scenario, it has been proposed that the optical component and its late time bumps can be interpreted as ejecta interactions with a multi-component CSM \citep{Zhu23, Lin23}. In this scenario, hadrons from the engulfed CSM would be accelerated via diffusive shock acceleration at the forward shock, interact with the surrounding CSM and produce $\gamma$-rays through hadronic channels -- predominantly $\pi^0$-decay. In the following, we do not consider contributions from the reverse shock and will disregard leptonic contributions -- presumed to be subdominant given the high CSM densities.

Based on the CSM characterization of \cite{Lin23} in four distinct shells (Appendix \ref{sect:crossingtimes}), we constructed a density profile, $\rho(t)$, encountered by the shock assuming an expansion parameter $m=(n-3)/(n-s)=0.75$ ($n=12$, $s=0$) and an initial velocity of $V_{s,0}= 9.8\times10^3$ km/s \citep[see Eq. B2 in][]{Chatzopoulos12, Lin23}. Instead of a stellar wind profile, a constant density is assumed within each shell -- a scenario also considered by \cite{Wheeler17} for the optical peak. Diffusive shock acceleration would occur at the forward shock, although it may be suppressed if the shock is radiation dominated \citep[see, e.g.,][]{Ito2018,Levinson2020}. Here two scenarios are considered: (1) that particle acceleration starts at least one day after the explosion ($t_i=T_0+1$~d; as in \citealt{Tatischeff09}), or (2) hadrons can only gain energy beyond $R_s(t_i)\sim R_{br}$, after shock breakout \citep[a conservative assumption; see, e.g.,][]{Murase11, Katz12, Giacinti15}. The energy-conversion efficiency of accelerating cosmic rays can be parametrized as $\eta$ in the cumulative energy, $E_{\rm CR}$ (i.e., as a fraction of the shock’s kinetic energy), which is estimated as

\begin{equation}
                E_{\rm CR} = \eta \int ^{t_f} _{t_i} \frac{1}{2} \rho(t) V_s(t) ^3 4 \pi R_s(t)^2 dt \; , \label{eq:ECR}
\end{equation}

\noindent where $R_s(t)$ is the shock radius and $V_s = dR_s/dt = V_{s,0}\left[\dfrac{t}{1\;\rm{day}}\right]^{m-1}$ \citep[see, e.g.,][]{Tatischeff09, Marti-Devesa24}. Shock breakout from the first shell occurs when the optical depth is $\tau_{\rm{opt}} \sim c/V_s(t_{br})$, resulting in a breakout time of $t_{br}\sim 14$ d, assuming an opacity of $\kappa =0.2$ cm$^2$/g \citep{Lin23}.

The energy budget, $E_{\rm CR}$, should be compared with the total energy stored in protons required to explain the giga-electronvolt excess (typically expressed as $W_p$). We assume that during the time interval defined previously through Bayesian blocks (see Sect. \ref{sec:refinedlc}) the average proton distribution is well reproduced by a power law with an exponential cutoff in momentum space, i.e.,

\begin{equation}
                \dfrac{dN_{\rm{p}}}{dE} = \beta N_0\left(\dfrac{E}{E_0}\right) ^{-p} \exp{-\left(\frac{E}{E_{\rm cutoff}}\right)} \;,
  \label{eq:PLexpcut}
\end{equation}

\noindent where $N_0$ is a normalization factor, $E_0$ the scale energy, $E_{\rm cutoff}$ the cutoff energy, $p$ the proton index, and $\beta$ the particle velocity, $v/c$ -- correcting the spectral shape of a proton population below $\sim 1$~GeV when described in kinetic energy \citep{Dermer12}. At detection time, the maximum energy is not limited by proton-proton collisions. Given the lack of curvature or simultaneous VHE observations \citep{Acharyya23}, we consider again two extreme scenarios for $E_{\rm cutoff}$ -- at 1 TeV and 1 PeV. We employ the radiative process and fitting package \texttt{Naima} to perform an MCMC fit of a $\pi^0$-decay component \citep{Kafexhiu14, Zabalza15} to the \textit{Fermi}-LAT spectrum. From the previously derived $\rho(t)$ profile, we calculated the average density encountered by the shock during the Bayesian block time interval, which is $1.5\times10^{-14}$~g cm$^{-3}$. The fit results in a proton distribution with $p=2.5^{+0.4}_{-0.3}$, with a total energy content of $W_{\rm p} (>0.1\;\rm{GeV})= 3.8^{+8}_{-1.7}\times 10^{49}$~erg ($\eta \sim 0.05^{+0.09}_{-0.02}$). While energetic considerations are satisfied (Fig. \ref{fig:ECRandflux}, left panel), it is apparent that the $\gamma$-ray emission should trace the target medium in a CSM scenario.
With the cosmic-ray efficiency and spectral index derived above ($\eta=0.05$, $p=2.5$), we can estimate the $\gamma$-ray flux that would be produced by the succession of CSM shell interactions and produce the corresponding $\gamma$-ray light curve. Absorption processes (Bethe-Heitler or pair-production) are not considered here, but these do not affect our conclusions and may only impact the flux at earlier times for the LAT energy range \citep{Vurm21}.
The resulting time-evolving $\gamma$-ray emission is shown in Fig. \ref{fig:ECRandflux} (right panel) and we note that the narrow CSM shell structures provide a $\gamma$-ray emission time stamp not matched by observations (a time delay between the predicted $\gamma$-ray emission from a given shell and the observed $\gamma$-ray signal). As this CSM model was built by \citet{Lin23} to reproduce the features of the optical light curve, our observed mismatch can be interpreted in the CSM model as a temporal offset between the optical and the $\gamma$-rays.

 We note that in dense CSM shocks, the optical and $\gamma$-ray emissions originate from fundamentally different physical processes. Optical radiation is produced by the shock-powered thermal X-rays that are efficiently absorbed and reprocessed, while $\gamma$-rays arise from the nonthermal proton–proton interactions at the forward shock. In principle, this difference in emission sites creates the possibility that the optical and $\gamma$-ray  light curves may not track each other in time, because the photospheric radius and the forward-shock radius can decouple if the diffusion time becomes long enough. However, unless the diffusion time greatly exceeds the dynamical time (an optical depth of $\tau >> c/V_s$), the reprocessed luminosity still emerges broadly at the same moment with the nonthermal radiation from the accelerated particle population. Such conditions would imply a radiation-mediated shock ($\tau >> 1$), which cannot efficiently accelerate cosmic rays, as was explained earlier. Therefore, the regime that would allow long optical delays is incompatible with the one accelerating particle in a collisionless shock and producing giga-electronvolt photons.

In addition to the timing mismatch argument, one can compare the energy radiated in optical and in $\gamma$-rays at $\sim$100 days post-SN explosion, which in our case yields $L_{\gamma}/L_{\rm opt} \sim 1$ for simultaneous observations (see Fig. \ref{fig:LC-2017egm}).
As the shock converts $>90\%$ of its kinetic energy into thermal energy and $<10\%$ into cosmic rays, and then only a fraction become $\gamma$-rays, one would expect at most $L_{\gamma}/L_{\rm opt} \sim 10^{-2}-10^{-1}$, which is one or two orders of magnitude lower than the measured ratio. In other words, as the thermal channel is far more efficient than the nonthermal one, one should expect a higher optical luminosity than the $\gamma$-ray one.
This ratio can be contextualized with systems in which both the optical and $\gamma$-ray outputs are known to be powered by CSM shock interactions. For the nearby Type II SN, SN 2023ixf—one of the closest core-collapse events in the past decade—\textit{Fermi}-LAT non-detections constrain the $\gamma$-ray–to–optical luminosity ratio to ($L_{\gamma}/L_{\rm opt} < 10^{-2}$) at epochs somewhat earlier than those considered here \citep[see Fig. 9 of][]{Marti-Devesa24}. Likewise, in the recurrent nova RS~Oph, where CSM interaction is firmly established, the \textit{Fermi}-LAT detections reveal a nearly constant ratio ($L_{\gamma}/L_{\rm opt} \simeq 2-3\times10^{-3}$) over the first $\sim 50$ days \citep{Cheung2022}. These values are representative of shocks in dense environments and highlight the strong dominance of the thermal (optical) channel over the nonthermal ($\gamma$-ray) in interaction-powered transients.
As a conclusion, the CSM model faces two fundamental challenges: 1) the time delay cannot be simply explained by diffusion effects, and 2) the CSM model should produce much more optical emission than $\gamma$-rays due to the higher efficiency of the thermal channel.

  \begin{figure}
   \centering

   \includegraphics[width=9.5cm]{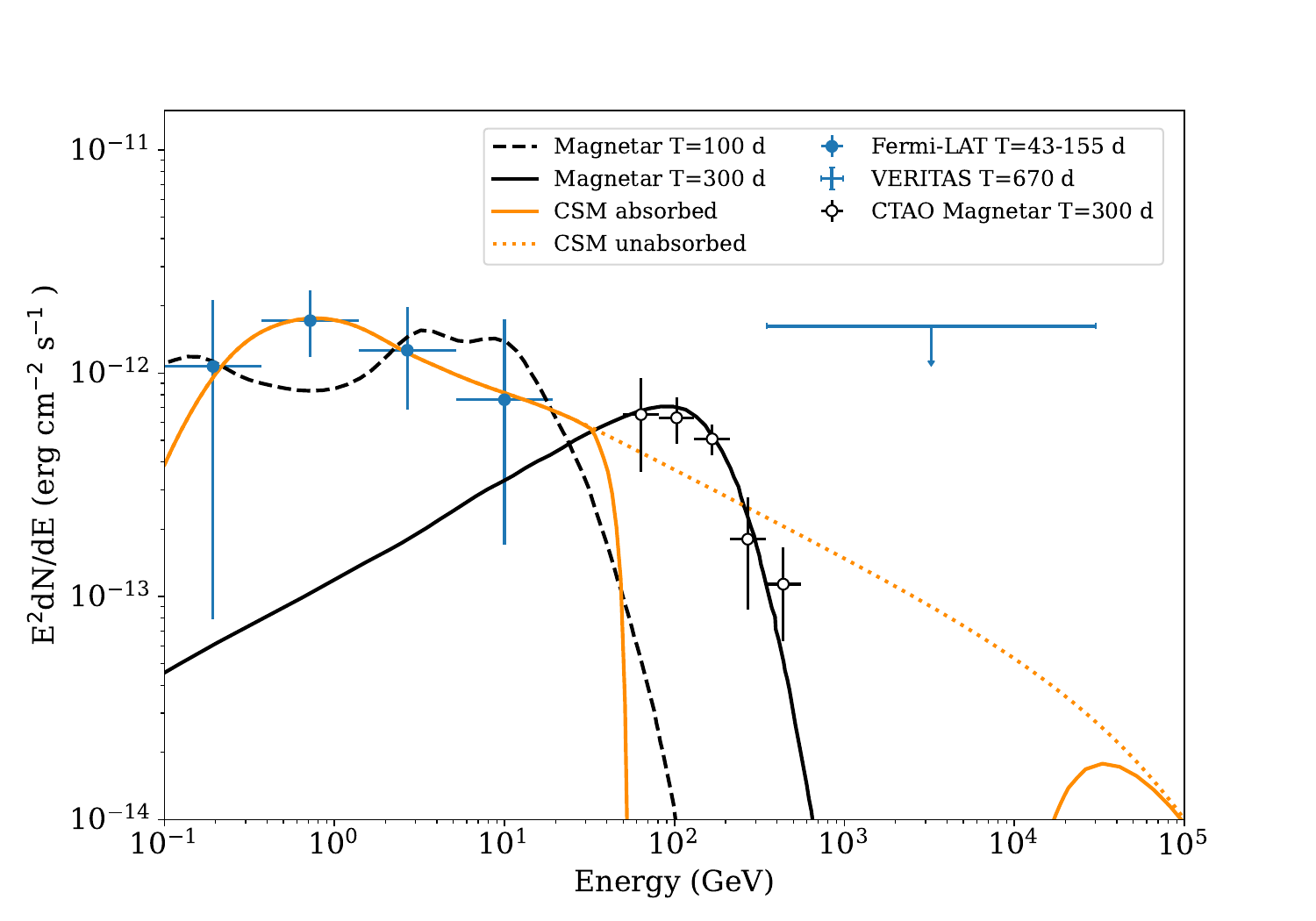} 
   \includegraphics[width=8.6cm]{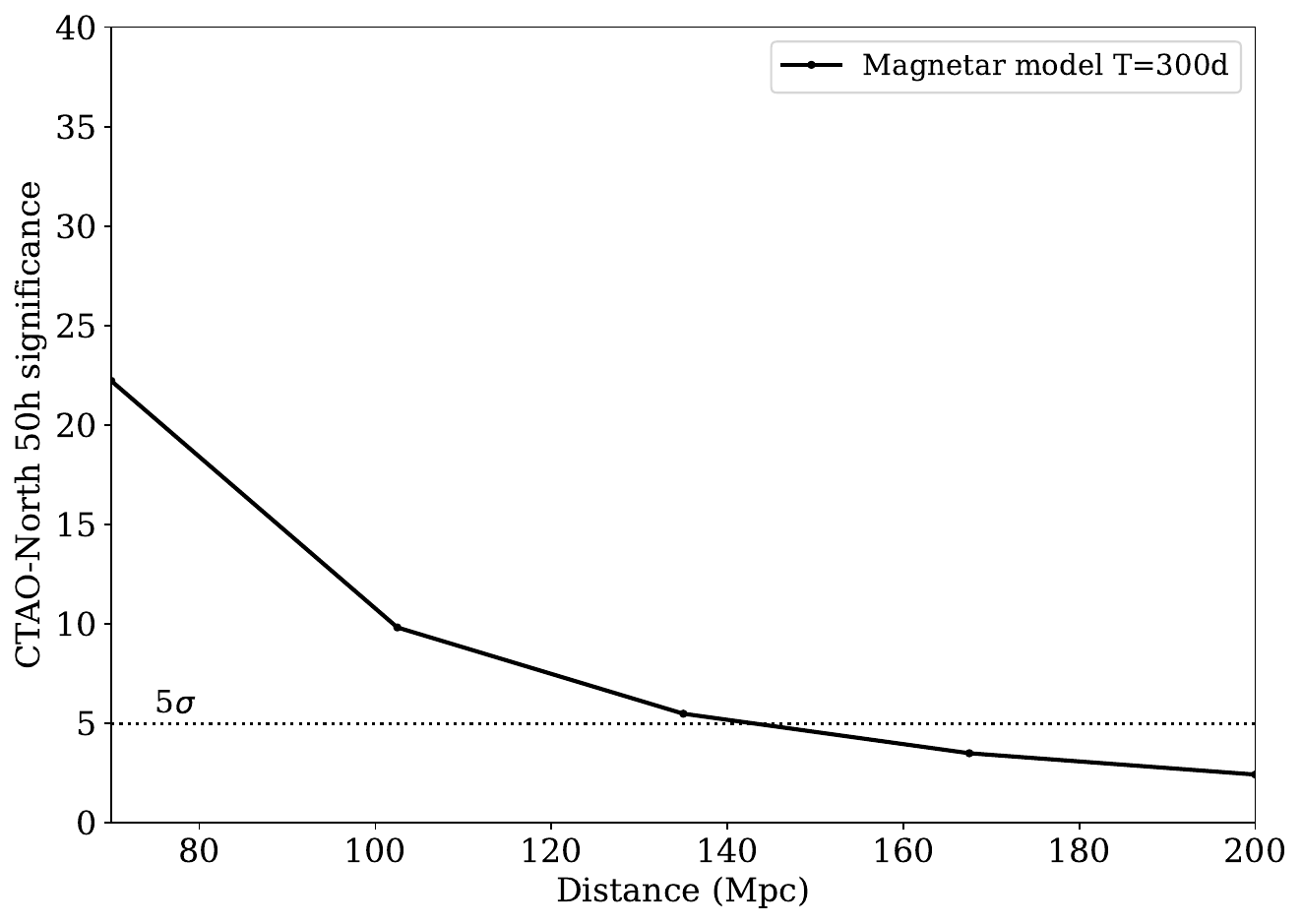} 
   
   \vspace{-0.3cm}
   \caption{Top: Comparison of CSM and magnetar (magnetization $\varepsilon_{\rm B}=10^{-6}$) spectral models of SN 2017egm at a distance of 135 Mpc together with \textit{Fermi}-LAT (this work) and  VERITAS \citep{Acharyya23} observations. In the CSM model, the predicted flux is shown with or without  $\gamma-\gamma$ absorption (see Appendix \ref{sect:gg-abs}), highlighting the fact that no emission is expected above 50 GeV.
   A simulation of 50 hours of CTAO-North observations between 50 GeV and 10 TeV is also shown for the magnetar model (see Sect. \ref{sect:cta} for more details). For clarity only detection flux points (where TS$>$4) are shown and upper limits are omitted.
   Using the CTAO simulation shown above, below we display the significance of a SN 2017egm-like source at different distances in the magnetar model only.
}
    \label{fig:cta}%

    \end{figure}

\section{Discussion}
\label{sect:discussion}

\subsection{The specific properties of SN 2017egm}

An open question remains regarding why only SN 2017egm is detected in our sample and not the other sources at similar distances or closer.
Among the SLSN I, SN 2017egm is the only source in our sample exhibiting “W-shape” O\,{\sc ii} features and  He features \citep{Zhu23},  while SN 2018bsz and SN 2020wnt stand out for their C\,{\sc ii} features (we do not discuss SN 2109ieh as it is not as luminous and this could be the sole reason why $\gamma$-rays are not detected).
The presence of this O\,{\sc ii} feature is thought to be a consequence of excitation of CNO layers by nonthermal radiation for example from the magnetar wind nebula \citep{Mazzali16}. 
These spectroscopic differences point toward different progenitor characteristics (ejecta mass, presence of a central engine, differences in the rotational velocity of the magnetar, and so on) that might impact the time evolution of the opacity and the expected $\gamma$-ray flux.
Whether SN 2017egm is representative of SLSNe in general is not yet clear and a larger SLSN population would be needed to further investigate this question.

SN 2020wnt, the second-most luminous event in our sample, has been associated both with a radioactive nickel scenario \citep{2022MNRAS.517.2056G} and a magnetar scenario \citep{Tinyanont23}, with the latter providing a better fit to the observed light curve, and thus being preferred. Given the spectral evolution, \cite{Tinyanont23}  suggests that the magnetar is
hidden inside the optically thick ejecta near peak, and only
exhibited magnetar spectroscopic signatures at later epochs ($\sim$1 year after peak). This could be due to a possibly
larger ejecta mass that might increase the $\gamma$-ray opacity
and delay the time window when $\gamma$-rays could leak from the ejecta to a moment when the spin-down luminosity had already significantly decreased and is no longer detectable.

\subsection{Limitations of the CSM and magnetar models}

All the events in our sample show some degree of interaction (light curve bumps and/or narrow spectral lines). 
While some levels of $\gamma$-ray emission can be expected in the CSM-shock interaction, we note that it is not guaranteed that the conditions for efficient particle acceleration (requiring a collisionless shock) are reached in the early times. First, the shock will be radiation-mediated, and only then, when
photons escape the system, a collisionless shock can form
and particle acceleration becomes possible \citep{Murase14}.
In addition, in a very dense CSM and photon radiation field, the $\gamma$-$\gamma$ absorption process can play an important role and drastically decrease the expected $\gamma$-ray radiation at $E>$ 10 GeV  \citep[see, e.g., ][]{cristofari2022}.

The models presented in Fig. \ref{fig:LC-2017egm} show good agreement with the initial light curve decline and the $\gamma$-ray light curve but fail to capture the bumpy late-time tail of the optical emission. Thus, a possible solution consists of a hybrid model in which the early time light curve is dominated by the magnetar model and the late time, in particular the bumpy features, by CSM interaction.
This scenario requires a very low magnetization of the nebula ($\varepsilon_{\rm B}<10^{-6}$) or a spin-down luminosity decaying faster than a magnetic dipole rate of $t^{-2}$, which would need to be explored in more depth theoretically, in a scenario where infalling accreting material could play a role \citep{Metzger2018}.

Alternatively, it has been recently proposed that the bumps in the optical light curve could be related to  modulation of the central engine luminosity in particular due to an infalling accretion disk undergoing Lense-Thirring precession as introduced in \citet{Lense18}-- see also the review by \citep{Mashhoon84}. The general concept is that around the newly formed magnetar a small, tilted accretion disk from failed ejected matter is formed. A misalignment between the accretion disk axis and the magnetar spin axis induces a Lense-Thirring torque that causes the disk to precess. This modulates the magnetar emission being injected into the ejecta, and  therefore the amount of optical emission \citep[see][]{Farah25}. Note that in this model the magnetar model explains the bulk of the luminosity and the precession of the accretion disk the small modulation.
This mechanism was successfully applied to model the light curve bump structures observed in optical in the SLSN 2024afav \citep{Farah25}.
Whether a similar scenario can quantitatively be applied to SN 2017egm remains to be seen, but such models would have the benefit of providing a self-consistent picture.

\section{Perspectives for tera-electronvolt detection}
\label{sect:cta}

As the $\gamma-\gamma$ optical depth  in the pulsar wind and ejecta decreases with time, the $\gamma$-ray window opens to $>$ 10 GeV energies within the first year after the SN and the peak energy of the escaping $\gamma$-rays shift to higher energies as can be seen in Fig. \ref{fig:cta} (top panel).
However, the optical depth still remains high above tera-electronvolt energies  \citep[see Fig. 4 of][]{Vurm21} meaning that the window for observations by atmospheric Cherenkov telescopes is mostly restricted to the $\sim$10 GeV -- 1 TeV energy band.

An upper limit on the tera-electronvolt emission from SN 2017egm was reported by the VERITAS collaboration and is presented alongside the \textit{Fermi}-LAT flux measurements in Fig. \ref{fig:cta} (top panel). The VERITAS observations correspond to a livetime of 8.7 hours and a lower-energy threshold of 350 GeV. Regardless of whether the \textit{Fermi}-LAT emission originates from a CSM shock interaction or is powered by a magnetar central engine, the predicted tera-electronvolt fluxes from both models lie well below the reported VERITAS upper limit. Notably, the magnetar model exhibits a spectral peak in the 100–200 GeV range, followed by a sharp decline at higher energies, making the VERITAS low-energy threshold of 350 GeV insufficient for detection in this scenario.
However, given the better sensitivity in the sub-tera-electronvolt energy range of the Cherenkov Telescope Array Observatory (CTAO), we explore its sensitivity to detect SN 2017egm-like objects in the future in the magnetar model at different distances.

The CSM model is not simulated as its tera-electronvolt spectrum is severely hampered by the  $\gamma-\gamma$ absorption (see Appendix \ref{sect:gg-abs}) and no emission is expected above 50 GeV.
For the magnetar case, as its model predicts a curved spectrum peaking at 100 GeV, we only simulated observations with CTAO North, which with four large-size telescopes (LSTs) is better optimized for the low-energy range. For the simulations of the expected significance, we used the package \textit{Gammapy} \citep{gammapy2023} with version 1.3 \citep{gammapy:zenodo-1.3}.
To simulate the magnetar wind-powered spectrum, we used the model of \citet{Vurm21} for the magnetar emission at T=300 days ($\varepsilon_{\rm B}=10^{-6}$) as at T=100 days no emission above 50 GeV is expected. 

The spectrum is then loaded in \textit{Gammapy} using the \texttt{TemplateSpectralModel} class.
A 3D (RA, Dec, Energy) \textit{Gammapy} dataset was constructed by assigning a point source with the aforementioned spectral model at the center of a $3^{\circ} \times 3^{\circ} $  region of interest with a bin size of 0.02$^{\circ}$ and 15 logarithmically spaced energy bins from 50 GeV to 10 TeV.
The instrumental background and responses were derived from the CTAO-North instrument response functions (IRFs\footnote{\url{https://zenodo.org/record/5499840}}) : 
\texttt{Prod5-North-20deg-AverageAz-4LSTs09MSTs.180000s}.

Assuming 50 hours of observations, the resulting simulated flux points are shown in Fig. \ref{fig:cta} (top panel) for the magnetar models, showing that a detection and a reasonable characterization of the spectral properties can be obtained with CTAO for such objects.
Interestingly, in the CSM model as the $\gamma-\gamma$ absorption is likely to kill any signal in the 50 GeV to 50 TeV energy range, a detection at tera-electronvolt energy of a SLSNe would be a strong argument in favor of the magnetar model.
However, it is important to keep in mind that these predictions are model dependent and based on the sole case of SN 2017egm, which may not be representative of the population of Type I SLSNe.

The detection significance was estimated using an Asimov dataset \citep[a data cube where each bin is set to the predicted value $N_{\rm pred}$; see][for a formal mathematical description]{Cowan2011}, which provides the expected median significance of detection without running intensive Monte-Carlo simulations.
The significance is then the likelihood ratio of the 3D \textit{Gammapy} analysis with or without the putative point source assuming a power-law model with a fixed spectral index of 2.5 and leaving only the normalization of the source free.

For the SN 2017egm magnetar model at a distance of 135 Mpc, a significance of TS=30 ($\sim 5.5\sigma$) indicates that SN 2017egm could have been detected by CTAO with 50h of observations.
To test the horizon of detectability we considered candidates in a distance range from 70 Mpc to 200 Mpc as shown in Fig. \ref{fig:cta} (bottom panel).
Our conclusion is that any SN 2017egm-like object in the magnetar model within 100 Mpc would yield a strong detection ($> 10\sigma$), while distances greater than 100 Mpc would exhibit fainter detections and be more subject to the choice of model parameters. 
Anyhow, this simulation shows that nearby SLSNe could represent a new class of source for CTAO.
Future detections in the CTAO energy range could provide constraints on the central engine powering SLSNe, the magnetar properties, and, in particular, the magnetization of the nebula.

\section{Conclusions}

Based on a curated list of nearby Type I (hydrogen poor) and Type II (hydrogen rich) SLSNe, we performed a systematic search for  $\gamma$-ray emission within the last 16 years of \textit{Fermi}-LAT observations using a summed likelihood analysis employing the LAT PSF types, optimized to search for faint emission at the optical positions of the SLSN. Then we characterized the light curve shape with a refined analysis with time bins of 15 days and a Bayesian block algorithm for the case of SN 2017egm.
Our main conclusions can be summarized as:

\begin{enumerate}

    \item From the targets in our sample, only SN 2017egm is detected, confirming the report by \citet{Li24}. We found the  likelihood TS to be TS=26 in a fixed one-year time window post-explosion or TS=33 in the Bayesian time block. 
    Both TS are reported assuming a position fixed to the SN optical one. The optimal time window starts $\sim$50 days after the explosion and lasts for 112 days up to late October of the same year. Within that time frame, the $\gamma$-ray emission can be modeled with a power-law of index $\Gamma=2.17\pm0.23$ with no evidence of spectral curvature. The localization of the $\gamma$-ray source is statistically compatible with the SN optical position.

    \item Our spectral and light curve results are visually compared against CSM and magnetar models. In the latter model of \citet{Vurm21}, the $\gamma$-ray emission from the magnetar wind nebula matches both the observed flux level and the light curve shape. In particular, the model predicts $\gamma$-rays to be visible only after $\sim$60 days, once the opacity of the ejecta has decreased, which is compatible with our findings. However, in order to explain both the optical and the $\gamma$-ray light curves, an extremely low magnetization of the nebula is required ($\varepsilon_{\rm B}<10^{-6}$) or alternatively a spin-down luminosity decaying faster than a magnetic dipole rate.

    \item Concerning the CSM model, assuming the multiple CSM shell model of \citet{Lin23}, we estimated the expected $\gamma$-ray flux stemming from each shell in a hadronic scenario. While the flux level is compatible with energetic considerations, the timing of the observed $\gamma$-ray signal and the ratio $L_{\gamma}/L_{\rm opt}\sim 1$ are inconsistent with theoretical expectations. This ratio is also not in line with measurements in other CSM-dominated objects such as novae and SNe, which yield  $L_{\gamma}/L_{\rm opt} < 10^{-2}$.

    \item Based on the previous models, we conclude that the $\gamma$-ray emission is well described by the magnetar model in order to match both the observed flux and time properties of the $\gamma$-ray signal. However, the optical light curve shows late-time bumpy structures that could either be the signature of multiple CSM shell interactions or evidence of central engine activity (perhaps magnetic reconfiguration, the Lense-Thirring effect, or other mechanisms). Plausible scenarios are therefore either a hybrid (magnetar+CSM) model or a pure magnetar model whereby both the $\gamma$-ray and optical properties are fully explained by a magnetar and an infalling accretion disk.

    \item SN 2017egm stands out both photometrically and spectroscopically from the other events in our sample. It exhibits the steepest rise and reaches the highest peak luminosity. The presence of W-shaped O\,{\sc ii} absorption lines suggests a higher radiation temperature or excitation by nonthermal radiation  from the inner magnetar wind nebula compared to the other SLSNe I in our sample, which lack these features. In contrast, the SLSNe II in the sample show narrow Balmer emission lines indicative of strong CSM interaction, suggesting a different powering mechanism. These distinctions may explain why $\gamma$-ray emission is only observed in SN 2017egm.    
    In addition, it is worth noting that there can be a large diversity of ejecta masses in SLSNe, which can range from a few solar masses up to around 40 solar masses \citep{Blanchard20,Gomez24}, leading to large differences in the $\gamma$-ray opacities and  time evolution. The $\gamma$-ray visibility is a competition between increasing transparency and decreasing pulsar spin-down energy, which implies that a low-ejecta-mass object would be favored for $\gamma$-ray detection as the needed transparency would be reached earlier.

    \item As the tera-electronvolt $\gamma$-ray emission in the CSM model is strongly suppressed due to $\gamma-\gamma$ absorption, a CTAO detection of a SLSN would strongly favor the magnetar scenario. 
    In this regard, we simulated the possible detectability of a SN 2017egm-like object 
    in the magnetar model and concluded that in 50 hours of CTAO-North observations, a maximum horizon of $\sim140$~Mpc for a 5$\sigma$ detection is obtained, which limits the SLSN samples reachable by CTAO to the nearby objects. Given the low number of known nearby SLSNe, a precise rate is hard to estimate but it is likely on the order of a few per decade within $\sim140$ Mpc.

\end{enumerate}

\medskip

\begin{acknowledgements} 
We thank the anonymous referee for the careful reading of the manuscript and for constructive and helpful criticism. 
This work made use of gammapy, numpy, scipy, astropy, and matplotlib libraries. FA wishes to thank Ismael Perez-Fournon, Matt Nicholl, and Morgan Fraser for enlightening discussions. 
The work on the Fermi-LAT analysis was supported by CNES. GMD thanks Weili Lin for generously providing further details on the CSM shells and model from \citet{Lin23}, and acknowledges support from the Beatriu de Pin\'os postdoctoral grant programme of the Departament de Recerca i Universitats de la Generalitat de Catalunya (2024 BP 00244). This work has been supported by the grant PID2024-155316NB-I00 funded by MICIU /AEI /10.13039/501100011033 / FEDER, UE and CSIC PIE 202350E189. This work was also supported by the Spanish program Unidad de Excelencia María de Maeztu CEX2020-001058-M and also supported by MCIN with funding from European Union NextGeneration EU (PRTR-C17.I1). BDM acknowledges support from the National Science Foundation (grant AST-2406637). EC would like to thank the National Science Foundation (NSF) for their support made possible by the NSF grant AST-2506735. EC would also like to thank NASA and the Smithsonian Astrophysical Observatory (SAO) for their support via the Chandra X-ray Observatory (CXO) theory grant TM4-25003X. The Flatiron Institute is supported by the Simons Foundation. IV acknowledges support by the ETAg grant PRG2159. PJP is funded by the European Union (ERC, project number 101042299, TransPIre). Views and opinions expressed are however those of the author(s) only and do not necessarily reflect those of the European Union or the European Research Council Executive Agency. Neither the European Union nor the granting authority can be held responsible for them. 

The \textit{Fermi}-LAT Collaboration acknowledges generous ongoing support from a number of agencies and institutes that have supported both the development and the operation of the LAT as well as scientific data analysis. These include the National Aeronautics and Space Administration and the Department of Energy in the United States, the Commissariat à l’Energie Atomique and the Centre National de la Recherche Scientifique / Institut National de Physique Nucléaire et de Physique des Particules in France, the Agenzia Spaziale Italiana and the Istituto Nazionale di Fisica Nucleare in Italy, the Ministry of Education, Culture, Sports, Science and Technology (MEXT), High Energy Accelerator Research Organization (KEK) and Japan Aerospace Exploration Agency (JAXA) in Japan, and the K. A. Wallenberg Foundation, the Swedish Research Council and the Swedish National Space Board in Sweden. Additional support for science analysis during the operations phase is gratefully acknowledged from the Istituto Nazionale di Astrofisica in Italy and the Centre National d’Études Spatiales in France. This work performed in part under DOE 
Contract DE-AC02-76SF00515. 

\end{acknowledgements}

\bibliographystyle{aa} 
\bibliography{reference} 

@ARTICLE{Webb2020,
       author = {{Webb}, N.~A. and {Coriat}, M. and {Traulsen}, I. and {Ballet}, J. and {Motch}, C. and {Carrera}, F.~J. and {Koliopanos}, F. and {Authier}, J. and {de la Calle}, I. and {Ceballos}, M.~T. and {Colomo}, E. and {Chuard}, D. and {Freyberg}, M. and {Garcia}, T. and {Kolehmainen}, M. and {Lamer}, G. and {Lin}, D. and {Maggi}, P. and {Michel}, L. and {Page}, C.~G. and {Page}, M.~J. and {Perea-Calderon}, J.~V. and {Pineau}, F. -X. and {Rodriguez}, P. and {Rosen}, S.~R. and {Santos Lleo}, M. and {Saxton}, R.~D. and {Schwope}, A. and {Tom{\'a}s}, L. and {Watson}, M.~G. and {Zakardjian}, A.},
        title = "{The XMM-Newton serendipitous survey. IX. The fourth XMM-Newton serendipitous source catalogue}",
      journal = {\aap},
         year = 2020,
        month = sep,
       volume = {641},
          eid = {A136},
        pages = {A136},
          doi = {10.1051/0004-6361/201937353},
archivePrefix = {arXiv},
       eprint = {2007.02899},
 primaryClass = {astro-ph.HE},
       adsurl = {https://ui.adsabs.harvard.edu/abs/2020A&A...641A.136W}
}

@ARTICLE{Anderson2018,
       author = {{Anderson}, J.~P. and {Pessi}, P.~J. and {Dessart}, L. and {Inserra}, C. and {Hiramatsu}, D. and {Taggart}, K. and {Smartt}, S.~J. and {Leloudas}, G. and {Chen}, T. -W. and {M{\"o}ller}, A. and {Roy}, R. and {Schulze}, S. and {Perley}, D. and {Selsing}, J. and {Prentice}, S.~J. and {Gal-Yam}, A. and {Angus}, C.~R. and {Arcavi}, I. and {Ashall}, C. and {Bulla}, M. and {Bray}, C. and {Burke}, J. and {Callis}, E. and {Cartier}, R. and {Chang}, S. -W. and {Chambers}, K. and {Clark}, P. and {Denneau}, L. and {Dennefeld}, M. and {Flewelling}, H. and {Fraser}, M. and {Galbany}, L. and {Gromadzki}, M. and {Guti{\'e}rrez}, C.~P. and {Heinze}, A. and {Hosseinzadeh}, G. and {Howell}, D.~A. and {Hsiao}, E.~Y. and {Kankare}, E. and {Kostrzewa-Rutkowska}, Z. and {Magnier}, E. and {Maguire}, K. and {Mazzali}, P. and {McBrien}, O. and {McCully}, C. and {Morrell}, N. and {Lowe}, T.~B. and {Onken}, C.~A. and {Onori}, F. and {Phillips}, M.~M. and {Rest}, A. and {Ridden-Harper}, R. and {Ruiter}, A.~J. and {Sand}, D.~J. and {Smith}, K.~W. and {Smith}, M. and {Stalder}, B. and {Stritzinger}, M.~D. and {Sullivan}, M. and {Tonry}, J.~L. and {Tucker}, B.~E. and {Valenti}, S. and {Wainscoat}, R. and {Waters}, C.~Z. and {Wolf}, C. and {Young}, D.},
        title = "{A nearby super-luminous supernova with a long pre-maximum \& ``plateau'' and strong C II features}",
      journal = {\aap},
         year = 2018,
        month = nov,
       volume = {620},
          eid = {A67},
        pages = {A67},
          doi = {10.1051/0004-6361/201833725},
archivePrefix = {arXiv},
       eprint = {1806.10609},
 primaryClass = {astro-ph.HE},
       adsurl = {https://ui.adsabs.harvard.edu/abs/2018A&A...620A..67A}
}

@article{gammapy2023,
    author = {{Donath}, Axel and {Terrier}, R\'egis and {Remy}, Quentin and {Sinha}, Atreyee and {Nigro}, Cosimo and
        {Pintore}, Fabio and {Kh\'elifi}, Bruno and {Olivera-Nieto}, Laura and {Ruiz}, Jose Enrique and
        {Br\"ugge}, Kai and {Linhoff}, Maximilian and {Contreras}, Jose Luis and {Acero}, Fabio and
        {Aguasca-Cabot}, Arnau and {Berge}, David and {Bhattacharjee}, Pooja and {Buchner}, Johannes and
        {Boisson}, Catherine and {Carreto Fidalgo}, David and {Chen}, Andrew and {de Bony de Lavergne}, Mathieu and
        {de Miranda Cardoso}, Jos\'e Vinicius and {Deil}, Christoph and {F\"u\ss{}ling}, Matthias and
        {Funk}, Stefan and {Giunti}, Luca and {Hinton}, Jim and {Jouvin}, L\'ea and {King}, Johannes and
        {Lefaucheur}, Julien and {Lemoine-Goumard}, Marianne and {Lenain}, Jean-Philippe and {L\'opez-Coto}, Rub\'en
        and {Mohrmann}, Lars and {Morcuende}, Daniel and {Panny}, Sebastian and {Regeard}, Maxime and {Saha}, Lab
        and {Siejkowski}, Hubert and {Siemiginowska}, Aneta and {Sip"ocz}, Brigitta M. and {Unbehaun}, Tim
        and {van Eldik}, Christopher and {Vuillaume}, Thomas and {Zanin}, Roberta},
    title = {Gammapy: A Python package for gamma-ray astronomy},
    DOI= "10.1051/0004-6361/202346488",
    url= "https://doi.org/10.1051/0004-6361/202346488",
    journal = {\aap},
    year = 2023,
    volume = 678,
    pages = "A157",
}

@misc{gammapy:zenodo-1.3,
  author       = {Acero, Fabio and
                  Aguasca-Cabot, Arnau and
                  Bernete, Juan and
                  Biederbeck, Noah and
                  Djuvsland, Julia and
                  Donath, Axel and
                  Feijen, Kirsty and
                  Fröse, Stefan and
                  Galelli, Claudio and
                  Khélifi, Bruno and
                  Konrad, Jana and
                  Kornecki, Paula and
                  Linhoff, Maximilian and
                  McKee, Kurt and
                  Mender, Simone and
                  Mohrmann, Lars and
                  Morcuende, Daniel and
                  Olivera-Nieto, Laura and
                  Peresano, Michele and
                  Pintore, Fabio and
                  Punch, Michael and
                  Regeard, Maxime and
                  Remy, Quentin and
                  Roellinghoff, Gerrit and
                  Sinha, Atreyee and
                  Sipőcz, Brigitta M and
                  Stapel, Hanna and
                  Streil, Katrin and
                  Terrier, Régis and
                  Unbehaun, Tim and
                  Wong, Samantha and
                  Yu, Pei},
  title        = {gammapy: v1.3, https://doi.org/10.5281/zenodo.14760974},
  title2        = {Gammapy: Python toolbox for gamma-ray astronomy},
  month        = jan,
  year         = 2025,
  publisher    = {Zenodo},
  version      = {v1.3},
  doi          = {10.5281/zenodo.14760974},
  url          = {https://doi.org/10.5281/zenodo.14760974},
  swhid        = {swh:1:dir:9405c2435b92c1267179543aaf913e9a3fda6ced
                   ;origin=https://doi.org/10.5281/zenodo.4701488;vis
                   it=swh:1:snp:ee94f46fa10032e542469d3614f2699ea257e
                   827;anchor=swh:1:rel:707c1b3d2919e9f78dbb649ad584c
                   30eec8359d2;path=gammapy-gammapy-c8b5337
                  },
}

@INPROCEEDINGS{Moriya24,
       author = {{Moriya}, Takashi J.},
        title = "{Superluminous supernovae}",
    booktitle = {Encyclopedia of Astrophysics, Volume 2},
         year = 2026,
       volume = {2},
        month = jan,
        pages = {720-743},
          doi = {10.1016/B978-0-443-21439-4.00017-1},
archivePrefix = {arXiv},
       eprint = {2407.12302},
 primaryClass = {astro-ph.HE},
       adsurl = {https://ui.adsabs.harvard.edu/abs/2026enap....2..720M}
}

@ARTICLE{Metzger2018,
       author = {{Metzger}, Brian D. and {Beniamini}, Paz and {Giannios}, Dimitrios},
        title = "{Effects of Fallback Accretion on Protomagnetar Outflows in Gamma-Ray Bursts and Superluminous Supernovae}",
      journal = {\apj},
         year = 2018,
        month = apr,
       volume = {857},
       number = {2},
          eid = {95},
        pages = {95},
          doi = {10.3847/1538-4357/aab70c},
archivePrefix = {arXiv},
       eprint = {1802.07750},
 primaryClass = {astro-ph.HE},
       adsurl = {https://ui.adsabs.harvard.edu/abs/2018ApJ...857...95M}
}

@ARTICLE{Lacy2020,
       author = {{Lacy}, M. and {Baum}, S.~A. and {Chandler}, C.~J. and {Chatterjee}, S. and {Clarke}, T.~E. and {Deustua}, S. and {English}, J. and {Farnes}, J. and {Gaensler}, B.~M. and {Gugliucci}, N. and {Hallinan}, G. and {Kent}, B.~R. and {Kimball}, A. and {Law}, C.~J. and {Lazio}, T.~J.~W. and {Marvil}, J. and {Mao}, S.~A. and {Medlin}, D. and {Mooley}, K. and {Murphy}, E.~J. and {Myers}, S. and {Osten}, R. and {Richards}, G.~T. and {Rosolowsky}, E. and {Rudnick}, L. and {Schinzel}, F. and {Sivakoff}, G.~R. and {Sjouwerman}, L.~O. and {Taylor}, R. and {White}, R.~L. and {Wrobel}, J. and {Andernach}, H. and {Beasley}, A.~J. and {Berger}, E. and {Bhatnager}, S. and {Birkinshaw}, M. and {Bower}, G.~C. and {Brandt}, W.~N. and {Brown}, S. and {Burke-Spolaor}, S. and {Butler}, B.~J. and {Comerford}, J. and {Demorest}, P.~B. and {Fu}, H. and {Giacintucci}, S. and {Golap}, K. and {G{\"u}th}, T. and {Hales}, C.~A. and {Hiriart}, R. and {Hodge}, J. and {Horesh}, A. and {Ivezi{\'c}}, {\v{Z}}. and {Jarvis}, M.~J. and {Kamble}, A. and {Kassim}, N. and {Liu}, X. and {Loinard}, L. and {Lyons}, D.~K. and {Masters}, J. and {Mezcua}, M. and {Moellenbrock}, G.~A. and {Mroczkowski}, T. and {Nyland}, K. and {O'Dea}, C.~P. and {O'Sullivan}, S.~P. and {Peters}, W.~M. and {Radford}, K. and {Rao}, U. and {Robnett}, J. and {Salcido}, J. and {Shen}, Y. and {Sobotka}, A. and {Witz}, S. and {Vaccari}, M. and {van Weeren}, R.~J. and {Vargas}, A. and {Williams}, P.~K.~G. and {Yoon}, I.},
        title = "{The Karl G. Jansky Very Large Array Sky Survey (VLASS). Science Case and Survey Design}",
      journal = {\pasp},
         year = 2020,
        month = mar,
       volume = {132},
       number = {1009},
          eid = {035001},
        pages = {035001},
          doi = {10.1088/1538-3873/ab63eb},
archivePrefix = {arXiv},
       eprint = {1907.01981},
 primaryClass = {astro-ph.IM},
       adsurl = {https://ui.adsabs.harvard.edu/abs/2020PASP..132c5001L}
}

@ARTICLE{Ajello2020,
       author = {{Ajello}, M. and {Angioni}, R. and {Axelsson}, M. and {Ballet}, J. and {Barbiellini}, G. and {Bastieri}, D. and {Becerra Gonzalez}, J. and {Bellazzini}, R. and {Bissaldi}, E. and {Bloom}, E.~D. and {Bonino}, R. and {Bottacini}, E. and {Bruel}, P. and {Buson}, S. and {Cafardo}, F. and {Cameron}, R.~A. and {Cavazzuti}, E. and {Chen}, S. and {Cheung}, C.~C. and {Ciprini}, S. and {Costantin}, D. and {Cutini}, S. and {D'Ammando}, F. and {de la Torre Luque}, P. and {de Menezes}, R. and {de Palma}, F. and {Desai}, A. and {Di Lalla}, N. and {Di Venere}, L. and {Dom{\'\i}nguez}, A. and {Dirirsa}, F. Fana and {Ferrara}, E.~C. and {Finke}, J. and {Franckowiak}, A. and {Fukazawa}, Y. and {Funk}, S. and {Fusco}, P. and {Gargano}, F. and {Garrappa}, S. and {Gasparrini}, D. and {Giglietto}, N. and {Giordano}, F. and {Giroletti}, M. and {Green}, D. and {Grenier}, I.~A. and {Guiriec}, S. and {Harita}, S. and {Hays}, E. and {Horan}, D. and {Itoh}, R. and {J{\'o}hannesson}, G. and {Kovac'evic'}, M. and {Krauss}, F. and {Kreter}, M. and {Kuss}, M. and {Larsson}, S. and {Leto}, C. and {Li}, J. and {Liodakis}, I. and {Longo}, F. and {Loparco}, F. and {Lott}, B. and {Lovellette}, M.~N. and {Lubrano}, P. and {Madejski}, G.~M. and {Maldera}, S. and {Manfreda}, A. and {Mart{\'\i}-Devesa}, G. and {Massaro}, F. and {Mazziotta}, M.~N. and {Mereu}, I. and {Meyer}, M. and {Migliori}, G. and {Mirabal}, N. and {Mizuno}, T. and {Monzani}, M.~E. and {Morselli}, A. and {Moskalenko}, I.~V. and {Negro}, M. and {Nemmen}, R. and {Nuss}, E. and {Ojha}, L.~S. and {Ojha}, R. and {Omodei}, N. and {Orienti}, M. and {Orlando}, E. and {Ormes}, J.~F. and {Paliya}, V.~S. and {Pei}, Z. and {Pe{\~n}a-Herazo}, H. and {Persic}, M. and {Pesce-Rollins}, M. and {Petrov}, L. and {Piron}, F. and {Poon}, H. and {Principe}, G. and {Rain{\`o}}, S. and {Rando}, R. and {Rani}, B. and {Razzano}, M. and {Razzaque}, S. and {Reimer}, A. and {Reimer}, O. and {Schinzel}, F.~K. and {Serini}, D. and {Sgr{\`o}}, C. and {Siskind}, E.~J. and {Spandre}, G. and {Spinelli}, P. and {Suson}, D.~J. and {Tachibana}, Y. and {Thompson}, D.~J. and {Torres}, D.~F. and {Torresi}, E. and {Troja}, E. and {Valverde}, J. and {van Zyl}, P. and {Yassine}, M.},
        title = "{The Fourth Catalog of Active Galactic Nuclei Detected by the Fermi Large Area Telescope}",
      journal = {\apj},
         year = 2020,
        month = apr,
       volume = {892},
       number = {2},
          eid = {105},
        pages = {105},
          doi = {10.3847/1538-4357/ab791e},
archivePrefix = {arXiv},
       eprint = {1905.10771},
 primaryClass = {astro-ph.HE},
       adsurl = {https://ui.adsabs.harvard.edu/abs/2020ApJ...892..105A}
}

@article{Flesch2024,
   title={The Millions of Optical-Radio/X-ray Associations (MORX) Catalogue, v2},
   volume={7},
   ISSN={2565-6120},
   url={http://dx.doi.org/10.21105/astro.2308.01507},
   DOI={10.21105/astro.2308.01507},
   journal={The Open Journal of Astrophysics},
   publisher={Maynooth University},
   author={Flesch, Eric Wim},
   year={2024},
   month=jan }

@ARTICLE{Brinchmann2008,
       author = {{Brinchmann}, J. and {Kunth}, D. and {Durret}, F.},
        title = "{Galaxies with Wolf-Rayet signatures in the low-redshift Universe. A survey using the Sloan Digital Sky Survey}",
      journal = {\aap},
         year = 2008,
        month = jul,
       volume = {485},
       number = {3},
        pages = {657-677},
          doi = {10.1051/0004-6361:200809783},
archivePrefix = {arXiv},
       eprint = {0805.1073},
 primaryClass = {astro-ph},
       adsurl = {https://ui.adsabs.harvard.edu/abs/2008A&A...485..657B}
}

@ARTICLE{Ackermann2015,
       author = {{Ackermann}, M. and {Arcavi}, I. and {Baldini}, L. and {Ballet}, J. and {Barbiellini}, G. and {Bastieri}, D. and {Bellazzini}, R. and {Bissaldi}, E. and {Blandford}, R.~D. and {Bonino}, R. and {Bottacini}, E. and {Brandt}, T.~J. and {Bregeon}, J. and {Bruel}, P. and {Buehler}, R. and {Buson}, S. and {Caliandro}, G.~A. and {Cameron}, R.~A. and {Caragiulo}, M. and {Caraveo}, P.~A. and {Cavazzuti}, E. and {Cecchi}, C. and {Charles}, E. and {Chekhtman}, A. and {Chiang}, J. and {Chiaro}, G. and {Ciprini}, S. and {Claus}, R. and {Cohen-Tanugi}, J. and {Cutini}, S. and {D'Ammando}, F. and {de Angelis}, A. and {de Palma}, F. and {Desiante}, R. and {Di Venere}, L. and {Drell}, P.~S. and {Favuzzi}, C. and {Fegan}, S.~J. and {Franckowiak}, A. and {Funk}, S. and {Fusco}, P. and {Gal-Yam}, A. and {Gargano}, F. and {Gasparrini}, D. and {Giglietto}, N. and {Giordano}, F. and {Giroletti}, M. and {Glanzman}, T. and {Godfrey}, G. and {Grenier}, I.~A. and {Grove}, J.~E. and {Guiriec}, S. and {Harding}, A.~K. and {Hayashi}, K. and {Hewitt}, J.~W. and {Hill}, A.~B. and {Horan}, D. and {Jogler}, T. and {J{\'o}hannesson}, G. and {Kocevski}, D. and {Kuss}, M. and {Larsson}, S. and {Lashner}, J. and {Latronico}, L. and {Li}, J. and {Li}, L. and {Longo}, F. and {Loparco}, F. and {Lovellette}, M.~N. and {Lubrano}, P. and {Malyshev}, D. and {Mayer}, M. and {Mazziotta}, M.~N. and {McEnery}, J.~E. and {Michelson}, P.~F. and {Mizuno}, T. and {Monzani}, M.~E. and {Morselli}, A. and {Murase}, K. and {Nugent}, P. and {Nuss}, E. and {Ofek}, E. and {Ohsugi}, T. and {Orienti}, M. and {Orlando}, E. and {Ormes}, J.~F. and {Paneque}, D. and {Pesce-Rollins}, M. and {Piron}, F. and {Pivato}, G. and {Rain{\`o}}, S. and {Rando}, R. and {Razzano}, M. and {Reimer}, A. and {Reimer}, O. and {Schulz}, A. and {Sgr{\`o}}, C. and {Siskind}, E.~J. and {Spada}, F. and {Spandre}, G. and {Spinelli}, P. and {Suson}, D.~J. and {Takahashi}, H. and {Thayer}, J.~B. and {Tibaldo}, L. and {Torres}, D.~F. and {Troja}, E. and {Vianello}, G. and {Werner}, M. and {Wood}, K.~S. and {Wood}, M.},
        title = "{Search for Early Gamma-ray Production in Supernovae Located in a Dense Circumstellar Medium with the Fermi LAT}",
      journal = {\apj},
         year = 2015,
        month = jul,
       volume = {807},
       number = {2},
          eid = {169},
        pages = {169},
          doi = {10.1088/0004-637X/807/2/169},
archivePrefix = {arXiv},
       eprint = {1506.01647},
 primaryClass = {astro-ph.HE},
       adsurl = {https://ui.adsabs.harvard.edu/abs/2015ApJ...807..169A}
}

@ARTICLE{2009ApJ...697.1071A,
   author = {{Atwood}, W.~B. and {Abdo}, A.~A. and {Ackermann}, M. and {Althouse}, W. and 
	{Anderson}, B. and {Axelsson}, M. and {Baldini}, L. and {Ballet}, J. and 
	{Band}, D.~L. and {Barbiellini}, G. and et al.},
    title = "{The Large Area Telescope on the Fermi Gamma-Ray Space Telescope Mission}",
  journal = {\apj},
archivePrefix = "arXiv",
   eprint = {0902.1089},
 primaryClass = "astro-ph.IM",
     year = 2009,
    month = jun,
   volume = 697,
    pages = {1071-1102},
      doi = {10.1088/0004-637X/697/2/1071},
   adsurl = {http://adsabs.harvard.edu/abs/2009ApJ...697.1071A}
}

@INPROCEEDINGS{2017ICRC...35..824W,
       author = {{Wood}, M. and {Caputo}, R. and {Charles}, E. and {Di Mauro}, M. and
         {Magill}, J. and {Perkins}, J.~S. and {Fermi-LAT Collaboration}},
        title = "{Fermipy: An open-source Python package for analysis of Fermi-LAT Data}",
    booktitle = {35th International Cosmic Ray Conference (ICRC2017)},
         year = 2017,
       series = {International Cosmic Ray Conference},
       volume = {301},
        month = jan,
          eid = {824},
        pages = {824},
archivePrefix = {arXiv},
       eprint = {1707.09551},
 primaryClass = {astro-ph.IM},
       adsurl = {https://ui.adsabs.harvard.edu/abs/2017ICRC...35..824W}
}

@ARTICLE{4FGL-DR4,
       author = {{Ballet}, J. and {Bruel}, P. and {Burnett}, T.~H. and {Lott}, B. and {The Fermi-LAT collaboration}},
        title = "{Fermi Large Area Telescope Fourth Source Catalog Data Release 4 (4FGL-DR4)}",
      journal = {arXiv e-prints},
         year = 2023,
        month = jul,
          eid = {arXiv:2307.12546},
        pages = {arXiv:2307.12546},
          doi = {10.48550/arXiv.2307.12546},
archivePrefix = {arXiv},
       eprint = {2307.12546},
 primaryClass = {astro-ph.HE},
       adsurl = {https://ui.adsabs.harvard.edu/abs/2023arXiv230712546B}
}

@ARTICLE{4FGL-DR3,
       author = {{Abdollahi}, S. and {Acero}, F. and {Baldini}, L. and {Ballet}, J. and {Bastieri}, D. and {Bellazzini}, R. and {Berenji}, B. and {Berretta}, A. and {Bissaldi}, E. and {Blandford}, R.~D. and {Bloom}, E. and {Bonino}, R. and {Brill}, A. and {Britto}, R.~J. and {Bruel}, P. and {Burnett}, T.~H. and {Buson}, S. and {Cameron}, R.~A. and {Caputo}, R. and {Caraveo}, P.~A. and {Castro}, D. and {Chaty}, S. and {Cheung}, C.~C. and {Chiaro}, G. and {Cibrario}, N. and {Ciprini}, S. and {Coronado-Bl{\'a}zquez}, J. and {Crnogorcevic}, M. and {Cutini}, S. and {D'Ammando}, F. and {De Gaetano}, S. and {Digel}, S.~W. and {Di Lalla}, N. and {Dirirsa}, F. and {Di Venere}, L. and {Dom{\'\i}nguez}, A. and {Fallah Ramazani}, V. and {Fegan}, S.~J. and {Ferrara}, E.~C. and {Fiori}, A. and {Fleischhack}, H. and {Franckowiak}, A. and {Fukazawa}, Y. and {Funk}, S. and {Fusco}, P. and {Galanti}, G. and {Gammaldi}, V. and {Gargano}, F. and {Garrappa}, S. and {Gasparrini}, D. and {Giacchino}, F. and {Giglietto}, N. and {Giordano}, F. and {Giroletti}, M. and {Glanzman}, T. and {Green}, D. and {Grenier}, I.~A. and {Grondin}, M. -H. and {Guillemot}, L. and {Guiriec}, S. and {Gustafsson}, M. and {Harding}, A.~K. and {Hays}, E. and {Hewitt}, J.~W. and {Horan}, D. and {Hou}, X. and {J{\'o}hannesson}, G. and {Karwin}, C. and {Kayanoki}, T. and {Kerr}, M. and {Kuss}, M. and {Landriu}, D. and {Larsson}, S. and {Latronico}, L. and {Lemoine-Goumard}, M. and {Li}, J. and {Liodakis}, I. and {Longo}, F. and {Loparco}, F. and {Lott}, B. and {Lubrano}, P. and {Maldera}, S. and {Malyshev}, D. and {Manfreda}, A. and {Mart{\'\i}-Devesa}, G. and {Mazziotta}, M.~N. and {Mereu}, I. and {Meyer}, M. and {Michelson}, P.~F. and {Mirabal}, N. and {Mitthumsiri}, W. and {Mizuno}, T. and {Moiseev}, A.~A. and {Monzani}, M.~E. and {Morselli}, A. and {Moskalenko}, I.~V. and {Negro}, M. and {Nuss}, E. and {Omodei}, N. and {Orienti}, M. and {Orlando}, E. and {Paneque}, D. and {Pei}, Z. and {Perkins}, J.~S. and {Persic}, M. and {Pesce-Rollins}, M. and {Petrosian}, V. and {Pillera}, R. and {Poon}, H. and {Porter}, T.~A. and {Principe}, G. and {Rain{\`o}}, S. and {Rando}, R. and {Rani}, B. and {Razzano}, M. and {Razzaque}, S. and {Reimer}, A. and {Reimer}, O. and {Reposeur}, T. and {S{\'a}nchez-Conde}, M. and {Saz Parkinson}, P.~M. and {Scotton}, L. and {Serini}, D. and {Sgr{\`o}}, C. and {Siskind}, E.~J. and {Smith}, D.~A. and {Spandre}, G. and {Spinelli}, P. and {Sueoka}, K. and {Suson}, D.~J. and {Tajima}, H. and {Tak}, D. and {Thayer}, J.~B. and {Thompson}, D.~J. and {Torres}, D.~F. and {Troja}, E. and {Valverde}, J. and {Wood}, K. and {Zaharijas}, G.},
        title = "{Incremental Fermi Large Area Telescope Fourth Source Catalog}",
      journal = {\apjs},
         year = 2022,
        month = jun,
       volume = {260},
       number = {2},
          eid = {53},
        pages = {53},
          doi = {10.3847/1538-4365/ac6751},
archivePrefix = {arXiv},
       eprint = {2201.11184},
 primaryClass = {astro-ph.HE},
       adsurl = {https://ui.adsabs.harvard.edu/abs/2022ApJS..260...53A}
}

@ARTICLE{4FGL,
       author = {{Abdollahi}, S. and {Acero}, F. and {Ackermann}, M. and {Ajello}, M. and {Atwood}, W.~B. and {Axelsson}, M. and {Baldini}, L. and {Ballet}, J. and {Barbiellini}, G. and {Bastieri}, D. and {Becerra Gonzalez}, J. and {Bellazzini}, R. and {Berretta}, A. and {Bissaldi}, E. and {Blandford}, R.~D. and {Bloom}, E.~D. and {Bonino}, R. and {Bottacini}, E. and {Brandt}, T.~J. and {Bregeon}, J. and {Bruel}, P. and {Buehler}, R. and {Burnett}, T.~H. and {Buson}, S. and {Cameron}, R.~A. and {Caputo}, R. and {Caraveo}, P.~A. and {Casandjian}, J.~M. and {Castro}, D. and {Cavazzuti}, E. and {Charles}, E. and {Chaty}, S. and {Chen}, S. and {Cheung}, C.~C. and {Chiaro}, G. and {Ciprini}, S. and {Cohen-Tanugi}, J. and {Cominsky}, L.~R. and {Coronado-Bl{\'a}zquez}, J. and {Costantin}, D. and {Cuoco}, A. and {Cutini}, S. and {D'Ammando}, F. and {DeKlotz}, M. and {de la Torre Luque}, P. and {de Palma}, F. and {Desai}, A. and {Digel}, S.~W. and {Di Lalla}, N. and {Di Mauro}, M. and {Di Venere}, L. and {Dom{\'\i}nguez}, A. and {Dumora}, D. and {Fana Dirirsa}, F. and {Fegan}, S.~J. and {Ferrara}, E.~C. and {Franckowiak}, A. and {Fukazawa}, Y. and {Funk}, S. and {Fusco}, P. and {Gargano}, F. and {Gasparrini}, D. and {Giglietto}, N. and {Giommi}, P. and {Giordano}, F. and {Giroletti}, M. and {Glanzman}, T. and {Green}, D. and {Grenier}, I.~A. and {Griffin}, S. and {Grondin}, M.-H. and {Grove}, J.~E. and {Guiriec}, S. and {Harding}, A.~K. and {Hayashi}, K. and {Hays}, E. and {Hewitt}, J.~W. and {Horan}, D. and {J{\'o}hannesson}, G. and {Johnson}, T.~J. and {Kamae}, T. and {Kerr}, M. and {Kocevski}, D. and {Kovac'evic'}, M. and {Kuss}, M. and {Landriu}, D. and {Larsson}, S. and {Latronico}, L. and {Lemoine-Goumard}, M. and {Li}, J. and {Liodakis}, I. and {Longo}, F. and {Loparco}, F. and {Lott}, B. and {Lovellette}, M.~N. and {Lubrano}, P. and {Madejski}, G.~M. and {Maldera}, S. and {Malyshev}, D. and {Manfreda}, A. and {Marchesini}, E.~J. and {Marcotulli}, L. and {Mart{\'\i}-Devesa}, G. and {Martin}, P. and {Massaro}, F. and {Mazziotta}, M.~N. and {McEnery}, J.~E. and {Mereu}, I. and {Meyer}, M. and {Michelson}, P.~F. and {Mirabal}, N. and {Mizuno}, T. and {Monzani}, M.~E. and {Morselli}, A. and {Moskalenko}, I.~V. and {Negro}, M. and {Nuss}, E. and {Ojha}, R. and {Omodei}, N. and {Orienti}, M. and {Orlando}, E. and {Ormes}, J.~F. and {Palatiello}, M. and {Paliya}, V.~S. and {Paneque}, D. and {Pei}, Z. and {Pe{\~n}a-Herazo}, H. and {Perkins}, J.~S. and {Persic}, M. and {Pesce-Rollins}, M. and {Petrosian}, V. and {Petrov}, L. and {Piron}, F. and {Poon}, H. and {Porter}, T.~A. and {Principe}, G. and {Rain{\`o}}, S. and {Rando}, R. and {Razzano}, M. and {Razzaque}, S. and {Reimer}, A. and {Reimer}, O. and {Remy}, Q. and {Reposeur}, T. and {Romani}, R.~W. and {Saz Parkinson}, P.~M. and {Schinzel}, F.~K. and {Serini}, D. and {Sgr{\`o}}, C. and {Siskind}, E.~J. and {Smith}, D.~A. and {Spandre}, G. and {Spinelli}, P. and {Strong}, A.~W. and {Suson}, D.~J. and {Tajima}, H. and {Takahashi}, M.~N. and {Tak}, D. and {Thayer}, J.~B. and {Thompson}, D.~J. and {Tibaldo}, L. and {Torres}, D.~F. and {Torresi}, E. and {Valverde}, J. and {Van Klaveren}, B. and {van Zyl}, P. and {Wood}, K. and {Yassine}, M. and {Zaharijas}, G.},
        title = "{Fermi Large Area Telescope Fourth Source Catalog}",
      journal = {\apjs},
         year = 2020,
        month = mar,
       volume = {247},
       number = {1},
          eid = {33},
        pages = {33},
          doi = {10.3847/1538-4365/ab6bcb},
archivePrefix = {arXiv},
       eprint = {1902.10045},
 primaryClass = {astro-ph.HE},
       adsurl = {https://ui.adsabs.harvard.edu/abs/2020ApJS..247...33A}
}

@ARTICLE{Nicholl17,
       author = {{Nicholl}, Matt and {Berger}, Edo and {Margutti}, Raffaella and {Blanchard}, Peter K. and {Guillochon}, James and {Leja}, Joel and {Chornock}, Ryan},
        title = "{The Superluminous Supernova SN 2017egm in the Nearby Galaxy NGC 3191: A Metal-rich Environment Can Support a Typical SLSN Evolution}",
      journal = {\apjl},
         year = 2017,
        month = aug,
       volume = {845},
       number = {1},
          eid = {L8},
        pages = {L8},
          doi = {10.3847/2041-8213/aa82b1},
archivePrefix = {arXiv},
       eprint = {1706.08517},
 primaryClass = {astro-ph.HE},
       adsurl = {https://ui.adsabs.harvard.edu/abs/2017ApJ...845L...8N}
}

@ARTICLE{Cheung2022,
       author = {{Cheung}, C.~C. and {Johnson}, T.~J. and {Jean}, P. and {Kerr}, M. and {Page}, K.~L. and {Osborne}, J.~P. and {Beardmore}, A.~P. and {Sokolovsky}, K.~V. and {Teyssier}, F. and {Ciprini}, S. and {Mart{\'\i}-Devesa}, G. and {Mereu}, I. and {Razzaque}, S. and {Wood}, K.~S. and {Shore}, S.~N. and {Korotkiy}, S. and {Levina}, A. and {Blumenzweig}, A.},
        title = "{Fermi LAT Gamma-ray Detection of the Recurrent Nova RS Ophiuchi during its 2021 Outburst}",
      journal = {\apj},
         year = 2022,
        month = aug,
       volume = {935},
       number = {1},
          eid = {44},
        pages = {44},
          doi = {10.3847/1538-4357/ac7eb7},
archivePrefix = {arXiv},
       eprint = {2207.02921},
 primaryClass = {astro-ph.HE},
       adsurl = {https://ui.adsabs.harvard.edu/abs/2022ApJ...935...44C}
}

@ARTICLE{Vurm21,
       author = {{Vurm}, Indrek and {Metzger}, Brian D.},
        title = "{Gamma-Ray Thermalization and Leakage from Millisecond Magnetar Nebulae: Toward a Self-consistent Model for Superluminous Supernovae}",
      journal = {\apj},
         year = 2021,
        month = aug,
       volume = {917},
       number = {2},
          eid = {77},
        pages = {77},
          doi = {10.3847/1538-4357/ac0826},
archivePrefix = {arXiv},
       eprint = {2101.05299},
 primaryClass = {astro-ph.HE},
       adsurl = {https://ui.adsabs.harvard.edu/abs/2021ApJ...917...77V}
}

@ARTICLE{Acharyya23,
       author = {{Acharyya}, A. and {Adams}, C.~B. and {Bangale}, P. and {Benbow}, W. and {Buckley}, J.~H. and {Capasso}, M. and {Dwarkadas}, V.~V. and {Errando}, M. and {Falcone}, A. and {Feng}, Q. and {Finley}, J.~P. and {Foote}, Juniper and {Fortson}, L. and {Furniss}, A. and {Gallagher}, G. and {Gent}, A. and {Hanlon}, W.~F. and {Hervet}, O. and {Holder}, J. and {Humensky}, T.~B. and {Jin}, W. and {Kaaret}, P. and {Kertzman}, M. and {Kherlakian}, M. and {Kieda}, D. and {Kleiner}, T.~K. and {Kumar}, S. and {Lang}, M.~J. and {Lundy}, M. and {Maier}, G. and {McGrath}, C.~E. and {Millis}, J. and {Moriarty}, P. and {Mukherjee}, R. and {Nievas-Rosillo}, M. and {O'Brien}, S. and {Ong}, R.~A. and {Patel}, S.~R. and {Pfrang}, K. and {Pohl}, M. and {Pueschel}, E. and {Quinn}, J. and {Ragan}, K. and {Reynolds}, P.~T. and {Ribeiro}, D. and {Roache}, E. and {Ryan}, J.~L. and {Sadeh}, I. and {Santander}, M. and {Sembroski}, G.~H. and {Shang}, R. and {Splettstoesser}, M. and {Tak}, D. and {Tucci}, J.~V. and {Weinstein}, A. and {Williams}, D.~A. and {Metzger}, B.~D. and {Nicholl}, M. and {Vurm}, I.},
        title = "{VERITAS and Fermi-LAT Constraints on the Gamma-Ray Emission from Superluminous Supernovae SN2015bn and SN2017egm}",
      journal = {\apj},
         year = 2023,
        month = mar,
       volume = {945},
       number = {1},
          eid = {30},
        pages = {30},
          doi = {10.3847/1538-4357/acb7e6},
archivePrefix = {arXiv},
       eprint = {2302.06686},
 primaryClass = {astro-ph.HE},
       adsurl = {https://ui.adsabs.harvard.edu/abs/2023ApJ...945...30A}
}

@ARTICLE{Mazzali16,
       author = {{Mazzali}, P.~A. and {Sullivan}, M. and {Pian}, E. and {Greiner}, J. and {Kann}, D.~A.},
        title = "{Spectrum formation in superluminous supernovae (Type I)}",
      journal = {\mnras},
         year = 2016,
        month = jun,
       volume = {458},
       number = {4},
        pages = {3455-3465},
          doi = {10.1093/mnras/stw512},
archivePrefix = {arXiv},
       eprint = {1603.00388},
 primaryClass = {astro-ph.HE},
       adsurl = {https://ui.adsabs.harvard.edu/abs/2016MNRAS.458.3455M}
}

@ARTICLE{cristofari2022,
       author = {{Cristofari}, P. and {Marcowith}, A. and {Renaud}, M. and {Dwarkadas}, V.~V. and {Tatischeff}, V. and {Giacinti}, G. and {Peretti}, E. and {Sol}, H.},
        title = "{The first days of Type II-P core collapse supernovae in the gamma-ray range}",
      journal = {\mnras},
         year = 2022,
        month = apr,
       volume = {511},
       number = {3},
        pages = {3321-3329},
          doi = {10.1093/mnras/stac217},
archivePrefix = {arXiv},
       eprint = {2201.09583},
 primaryClass = {astro-ph.HE},
       adsurl = {https://ui.adsabs.harvard.edu/abs/2022MNRAS.511.3321C}
}

@ARTICLE{Nicholl21,
       author = {{Nicholl}, Matt},
        title = "{Superluminous supernovae: an explosive decade}",
      journal = {Astronomy and Geophysics},
         year = 2021,
        month = oct,
       volume = {62},
       number = {5},
        pages = {5.34-5.42},
          doi = {10.1093/astrogeo/atab092},
archivePrefix = {arXiv},
       eprint = {2109.08697},
 primaryClass = {astro-ph.HE},
       adsurl = {https://ui.adsabs.harvard.edu/abs/2021A&G....62.5.34N}
}

@ARTICLE{Blanchard20,
       author = {{Blanchard}, Peter K. and {Berger}, Edo and {Nicholl}, Matt and {Villar}, V. Ashley},
        title = "{The Pre-explosion Mass Distribution of Hydrogen-poor Superluminous Supernova Progenitors and New Evidence for a Mass-Spin Correlation}",
      journal = {\apj},
         year = 2020,
        month = jul,
       volume = {897},
       number = {2},
          eid = {114},
        pages = {114},
          doi = {10.3847/1538-4357/ab9638},
archivePrefix = {arXiv},
       eprint = {2002.09508},
 primaryClass = {astro-ph.HE},
       adsurl = {https://ui.adsabs.harvard.edu/abs/2020ApJ...897..114B}
}

@ARTICLE{Lin23,
       author = {{Lin}, Weili and {Wang}, Xiaofeng and {Yan}, Lin and {Gal-Yam}, Avishay and {Mo}, Jun and {Brink}, Thomas G. and {Filippenko}, Alexei V. and {Xiang}, Danfeng and {Lunnan}, Ragnhild and {Zheng}, Weikang and {Brown}, Peter and {Kasliwal}, Mansi and {Fremling}, Christoffer and {Blagorodnova}, Nadejda and {Mirzaqulov}, Davron and {Ehgamberdiev}, Shuhrat A. and {Lin}, Han and {Zhang}, Kaicheng and {Zhang}, Jicheng and {Yan}, Shengyu and {Zhang}, Jujia and {Chen}, Zhihao and {Deng}, Licai and {Wang}, Kun and {Xiao}, Lin and {Wang}, Lingjun},
        title = "{A superluminous supernova lightened by collisions with pulsational pair-instability shells}",
      journal = {Nature Astronomy},
         year = 2023,
        month = jul,
       volume = {7},
        pages = {779-789},
          doi = {10.1038/s41550-023-01957-3},
archivePrefix = {arXiv},
       eprint = {2304.10416},
 primaryClass = {astro-ph.HE},
       adsurl = {https://ui.adsabs.harvard.edu/abs/2023NatAs...7..779L}
}

@ARTICLE{Brennan24,
       author = {{Brennan}, S.~J. and {Schulze}, S. and {Lunnan}, R. and {Sollerman}, J. and {Yan}, L. and {Fransson}, C. and {Irani}, I. and {Melinder}, J. and {Chen}, T. -W. and {De}, K. and {Fremling}, C. and {Kim}, Y. -L. and {Perley}, D. and {Pessi}, P.~J. and {Drake}, A.~J. and {Graham}, M.~J. and {Laher}, R.~R. and {Masci}, F.~J. and {Purdum}, J. and {Rodriguez}, H.},
        title = "{SN 2021adxl: A luminous nearby interacting supernova in an extremely low-metallicity environment}",
      journal = {\aap},
         year = 2024,
        month = oct,
       volume = {690},
          eid = {A259},
        pages = {A259},
          doi = {10.1051/0004-6361/202349036},
archivePrefix = {arXiv},
       eprint = {2312.13280},
 primaryClass = {astro-ph.HE},
       adsurl = {https://ui.adsabs.harvard.edu/abs/2024A&A...690A.259B}
}

@ARTICLE{Helene83,
       author = {{Helene}, O.},
        title = "{Upper limit of peak area}",
      journal = {Nuclear Instruments and Methods in Physics Research},
         year = 1983,
        month = jul,
       volume = {212},
       number = {1-3},
        pages = {319-322},
          doi = {10.1016/0167-5087(83)90709-3},
       adsurl = {https://ui.adsabs.harvard.edu/abs/1983NIMPR.212..319H}
}

@ARTICLE{Ito2018,
       author = {{Ito}, Hirotaka and {Levinson}, Amir and {Stern}, Boris E. and {Nagataki}, Shigehiro},
        title = "{Monte Carlo simulations of relativistic radiation-mediated shocks - I. Photon-rich regime}",
      journal = {\mnras},
         year = 2018,
        month = feb,
       volume = {474},
       number = {2},
        pages = {2828-2851},
          doi = {10.1093/mnras/stx2722},
archivePrefix = {arXiv},
       eprint = {1709.08955},
 primaryClass = {astro-ph.HE},
       adsurl = {https://ui.adsabs.harvard.edu/abs/2018MNRAS.474.2828I}
}

@ARTICLE{Levinson2020,
       author = {{Levinson}, Amir and {Nakar}, Ehud},
        title = "{Physics of radiation mediated shocks and its applications to GRBs, supernovae, and neutron star mergers}",
      journal = {\physrep},
         year = 2020,
        month = jun,
       volume = {866},
        pages = {1-46},
          doi = {10.1016/j.physrep.2020.04.003},
archivePrefix = {arXiv},
       eprint = {1909.10288},
 primaryClass = {astro-ph.HE},
       adsurl = {https://ui.adsabs.harvard.edu/abs/2020PhR...866....1L}
}

@ARTICLE{Lense18,
       author = {{Lense}, J.},
        title = "{{\"U}ber Relativit{\"a}tseinfl{\"u}sse in den Mondsystemen}",
      journal = {Astronomische Nachrichten},
         year = 1918,
        month = mar,
       volume = {206},
       number = {14},
        pages = {117},
          doi = {10.1002/asna.19182061402},
       adsurl = {https://ui.adsabs.harvard.edu/abs/1918AN....206..117L}
}

@ARTICLE{Mashhoon84,
       author = {{Mashhoon}, B. and {Hehl}, F.~W. and {Theiss}, D.~S.},
        title = "{On the gravitational effects of rotating masses: the Thirring-Lense papers.}",
      journal = {General Relativity and Gravitation},
         year = 1984,
        month = aug,
       volume = {16},
       number = {8},
        pages = {711-750},
          doi = {10.1007/BF00762913},
       adsurl = {https://ui.adsabs.harvard.edu/abs/1984GReGr..16..711M}
}

@ARTICLE{Farah25,
       author = {{Farah}, Joseph R. and {Prust}, Logan J. and {Howell}, D. Andrew and {Ni}, Yuan Qi and {McCully}, Curtis and {Andrews}, Moira and {Kumar}, Harsh and {Hiramatsu}, Daichi and {Wynn}, Sebastian Gomez Kathryn and {Filippenko}, Alexei V. and {Bostroem}, K. Azalee and {Berger}, Edo and {Blanchard}, Peter},
        title = "{Lense-Thirring precessing magnetar engine drives a superluminous supernova}",
      journal = {Submitted to Nature},
         year = 2025,
        month = sep,
          eid = {arXiv:2509.08051},
        pages = {arXiv:2509.08051},
          doi = {10.48550/arXiv.2509.08051},
archivePrefix = {arXiv},
       eprint = {2509.08051},
 primaryClass = {astro-ph.HE},
       adsurl = {https://ui.adsabs.harvard.edu/abs/2025arXiv250908051F}
}

@ARTICLE{Tinyanont23,
       author = {{Tinyanont}, Samaporn and {Woosley}, Stan E. and {Taggart}, Kirsty and {Foley}, Ryan J. and {Yan}, Lin and {Lunnan}, Ragnhild and {Davis}, Kyle W. and {Kilpatrick}, Charles D. and {Siebert}, Matthew R. and {Schulze}, Steve and {Ashall}, Chris and {Chen}, Ting-Wan and {De}, Kishalay and {Dimitriadis}, Georgios and {Dong}, Dillon Z. and {Fremling}, Christoffer and {Gagliano}, Alexander and {Jha}, Saurabh W. and {Jones}, David O. and {Kasliwal}, Mansi M. and {Miao}, Hao-Yu and {Pan}, Yen-Chen and {Perley}, Daniel A. and {Ravi}, Vikram and {Rojas-Bravo}, C{\'e}sar and {Sfaradi}, Itai and {Sollerman}, Jesper and {Alarcon}, Vanessa and {Angulo}, Rodrigo and {Clever}, Karoli E. and {Crawford}, Payton and {Couch}, Cirilla and {Dandu}, Srujan and {Dhara}, Atirath and {Johnson}, Jessica and {Lai}, Zhisen and {Smith}, Carli},
        title = "{Supernova 2020wnt: An Atypical Superluminous Supernova with a Hidden Central Engine}",
      journal = {\apj},
         year = 2023,
        month = jul,
       volume = {951},
       number = {1},
          eid = {34},
        pages = {34},
          doi = {10.3847/1538-4357/acc6c3},
archivePrefix = {arXiv},
       eprint = {2212.00177},
 primaryClass = {astro-ph.HE},
       adsurl = {https://ui.adsabs.harvard.edu/abs/2023ApJ...951...34T}
}

@ARTICLE{2022MNRAS.517.2056G,
       author = {{Guti{\'e}rrez}, C.~P. and {Pastorello}, A. and {Bersten}, M. and {Benetti}, S. and {Orellana}, M. and {Fiore}, A. and {Karamehmetoglu}, E. and {Kravtsov}, T. and {Reguitti}, A. and {Reynolds}, T.~M. and {Valerin}, G. and {Mazzali}, P. and {Sullivan}, M. and {Cai}, Y. -Z. and {Elias-Rosa}, N. and {Fraser}, M. and {Hsiao}, E.~Y. and {Kankare}, E. and {Kotak}, R. and {Kuncarayakti}, H. and {Li}, Z. and {Mattila}, S. and {Mo}, J. and {Moran}, S. and {Ochner}, P. and {Shahbandeh}, M. and {Tomasella}, L. and {Wang}, X. and {Yan}, S. and {Zhang}, J. and {Zhang}, T. and {Stritzinger}, M.~D.},
        title = "{SN 2020wnt: a slow-evolving carbon-rich superluminous supernova with no O II lines and a bumpy light curve}",
      journal = {\mnras},
         year = 2022,
        month = dec,
       volume = {517},
       number = {2},
        pages = {2056-2075},
          doi = {10.1093/mnras/stac2747},
archivePrefix = {arXiv},
       eprint = {2206.01662},
 primaryClass = {astro-ph.HE},
       adsurl = {https://ui.adsabs.harvard.edu/abs/2022MNRAS.517.2056G}
}

@ARTICLE{Chen21,
       author = {{Chen}, T. -W. and {Brennan}, S.~J. and {Wesson}, R. and {Fraser}, M. and {Schweyer}, T. and {Inserra}, C. and {Schulze}, S. and {Nicholl}, M. and {Anderson}, J.~P. and {Hsiao}, E.~Y. and {Jerkstrand}, A. and {Kankare}, E. and {Kool}, E.~C. and {Kravtsov}, T. and {Kuncarayakti}, H. and {Leloudas}, G. and {Li}, C. -J. and {Matsuura}, M. and {Pursiainen}, M. and {Roy}, R. and {Ruiter}, A.~J. and {Schady}, P. and {Seitenzahl}, I. and {Sollerman}, J. and {Tartaglia}, L. and {Wang}, L. and {Yates}, R.~M. and {Yang}, S. and {Baade}, D. and {Carini}, R. and {Gal-Yam}, A. and {Galbany}, L. and {Gonzalez-Gaitan}, S. and {Gromadzki}, M. and {Gutierrez}, C.~P. and {Kotak}, R. and {Maguire}, K. and {Mazzali}, P.~A. and {Mueller-Bravo}, T.~E. and {Paraskeva}, E. and {Pessi}, P.~J. and {Pignata}, G. and {Rau}, A. and {Young}, D.~R.},
        title = "{SN 2018bsz: significant dust formation in a nearby superluminous supernova}",
      journal = {arXiv e-prints},
         year = 2021,
        month = sep,
          eid = {arXiv:2109.07942},
        pages = {arXiv:2109.07942},
          doi = {10.48550/arXiv.2109.07942},
archivePrefix = {arXiv},
       eprint = {2109.07942},
 primaryClass = {astro-ph.HE},
       adsurl = {https://ui.adsabs.harvard.edu/abs/2021arXiv210907942C}
}

@ARTICLE{Murase14,
       author = {{Murase}, Kohta and {Thompson}, Todd A. and {Ofek}, Eran O.},
        title = "{Probing cosmic ray ion acceleration with radio-submm and gamma-ray emission from interaction-powered supernovae}",
      journal = {\mnras},
         year = 2014,
        month = may,
       volume = {440},
       number = {3},
        pages = {2528-2543},
          doi = {10.1093/mnras/stu384},
archivePrefix = {arXiv},
       eprint = {1311.6778},
 primaryClass = {astro-ph.HE},
       adsurl = {https://ui.adsabs.harvard.edu/abs/2014MNRAS.440.2528M}
}

@ARTICLE{Coppejans18,
       author = {{Coppejans}, D.~L. and {Margutti}, R. and {Guidorzi}, C. and {Chomiuk}, L. and {Alexander}, K.~D. and {Berger}, E. and {Bietenholz}, M.~F. and {Blanchard}, P.~K. and {Challis}, P. and {Chornock}, R. and {Drout}, M. and {Fong}, W. and {MacFadyen}, A. and {Migliori}, G. and {Milisavljevic}, D. and {Nicholl}, M. and {Parrent}, J.~T. and {Terreran}, G. and {Zauderer}, B.~A.},
        title = "{Jets in Hydrogen-poor Superluminous Supernovae: Constraints from a Comprehensive Analysis of Radio Observations}",
      journal = {\apj},
         year = 2018,
        month = mar,
       volume = {856},
       number = {1},
          eid = {56},
        pages = {56},
          doi = {10.3847/1538-4357/aab36e},
archivePrefix = {arXiv},
       eprint = {1711.03428},
 primaryClass = {astro-ph.HE},
       adsurl = {https://ui.adsabs.harvard.edu/abs/2018ApJ...856...56C}
}

@ARTICLE{Zhu23,
       author = {{Zhu}, Jiazheng and {Jiang}, Ning and {Dong}, Subo and {Filippenko}, Alexei V. and {Rudy}, Richard J. and {Pastorello}, A. and {Ashall}, Christopher and {Bose}, Subhash and {Post}, R.~S. and {Bersier}, D. and {Benetti}, Stefano and {Brink}, Thomas G. and {Chen}, Ping and {Dou}, Liming and {Elias-Rosa}, N. and {Lundqvist}, Peter and {Mattila}, Seppo and {Russell}, Ray W. and {Sitko}, Michael L. and {Somero}, Auni and {Stritzinger}, M.~D. and {Wang}, Tinggui and {Brown}, Peter J. and {Cappellaro}, E. and {Fraser}, Morgan and {Kankare}, Erkki and {Moran}, S. and {Prentice}, Simon and {Pursimo}, Tapio and {Reynolds}, T.~M. and {Zheng}, WeiKang},
        title = "{SN 2017egm: A Helium-rich Superluminous Supernova with Multiple Bumps in the Light Curves}",
      journal = {\apj},
         year = 2023,
        month = may,
       volume = {949},
       number = {1},
          eid = {23},
        pages = {23},
          doi = {10.3847/1538-4357/acc2c3},
archivePrefix = {arXiv},
       eprint = {2303.03424},
 primaryClass = {astro-ph.HE},
       adsurl = {https://ui.adsabs.harvard.edu/abs/2023ApJ...949...23Z}
}

@ARTICLE{Pessi25,
       author = {{Pessi}, P.~J. and {Lunnan}, R. and {Sollerman}, J. and {Schulze}, S. and {Gkini}, A. and {Gangopadhyay}, A. and {Yan}, L. and {Gal-Yam}, A. and {Perley}, D.~A. and {Chen}, T. -W. and {Hinds}, K.~R. and {Brennan}, S.~J. and {Hu}, Y. and {Singh}, A. and {Andreoni}, I. and {Cook}, D.~O. and {Fremling}, C. and {Ho}, A.~Y.~Q. and {Sharma}, Y. and {van Velzen}, S. and {Kangas}, T. and {Wold}, A. and {Bellm}, E.~C. and {Bloom}, J.~S. and {Graham}, M.~J. and {Kasliwal}, M.~M. and {Kulkarni}, S.~R. and {Riddle}, R. and {Rusholme}, B.},
        title = "{Sample of hydrogen-rich superluminous supernovae from the Zwicky Transient Facility}",
      journal = {\aap},
         year = 2025,
        month = mar,
       volume = {695},
          eid = {A142},
        pages = {A142},
          doi = {10.1051/0004-6361/202452014},
archivePrefix = {arXiv},
       eprint = {2408.15086},
 primaryClass = {astro-ph.HE},
       adsurl = {https://ui.adsabs.harvard.edu/abs/2025A&A...695A.142P}
}

@ARTICLE{Gomez22,
       author = {{Gomez}, Sebastian and {Berger}, Edo and {Nicholl}, Matt and {Blanchard}, Peter K. and {Hosseinzadeh}, Griffin},
        title = "{Luminous Supernovae: Unveiling a Population between Superluminous and Normal Core-collapse Supernovae}",
      journal = {\apj},
         year = 2022,
        month = dec,
       volume = {941},
       number = {2},
          eid = {107},
        pages = {107},
          doi = {10.3847/1538-4357/ac9842},
archivePrefix = {arXiv},
       eprint = {2204.08486},
 primaryClass = {astro-ph.HE},
       adsurl = {https://ui.adsabs.harvard.edu/abs/2022ApJ...941..107G}
}

@ARTICLE{Gomez24,
       author = {{Gomez}, Sebastian and {Nicholl}, Matt and {Berger}, Edo and {Blanchard}, Peter K. and {Villar}, V. Ashley and {Rest}, Sofia and {Hosseinzadeh}, Griffin and {Aamer}, Aysha and {Ajay}, Yukta and {Athukoralalage}, Wasundara and {Coulter}, David C. and {Eftekhari}, Tarraneh and {Fiore}, Achille and {Franz}, Noah and {Fox}, Ori and {Gagliano}, Alexander and {Hiramatsu}, Daichi and {Howell}, D. Andrew and {Hsu}, Brian and {Karmen}, Mitchell and {Siebert}, Matthew R. and {K{\"o}nyves-T{\'o}th}, R{\'e}ka and {Kumar}, Harsh and {McCully}, Curtis and {Pellegrino}, Craig and {Pierel}, Justin and {Rest}, Armin and {Wang}, Qinan},
        title = "{The Type I superluminous supernova catalogue I: light-curve properties, models, and catalogue description}",
      journal = {\mnras},
         year = 2024,
        month = nov,
       volume = {535},
       number = {1},
        pages = {471-515},
          doi = {10.1093/mnras/stae2270},
archivePrefix = {arXiv},
       eprint = {2407.07946},
 primaryClass = {astro-ph.HE},
       adsurl = {https://ui.adsabs.harvard.edu/abs/2024MNRAS.535..471G}
}

@ARTICLE{Li24,
       author = {{Li}, Shang and {Liang}, Yun-Feng and {Liao}, Neng-Hui and {Lei}, Lei and {Fan}, Yi-Zhong},
        title = "{Evidence for GeV Emission from the Superluminous Supernova SN 2017egm}",
      journal = {\prl},
         year = 2026,
        month = mar,
       volume = {136},
       number = {11},
          eid = {111402},
        pages = {111402},
          doi = {10.1103/jhhk-ywtl},
       adsurl = {https://ui.adsabs.harvard.edu/abs/2026PhRvL.136k1402L}
}

@ARTICLE{Gal-Yam19,
       author = {{Gal-Yam}, Avishay},
        title = "{The Most Luminous Supernovae}",
      journal = {\araa},
         year = 2019,
        month = aug,
       volume = {57},
        pages = {305-333},
          doi = {10.1146/annurev-astro-081817-051819},
archivePrefix = {arXiv},
       eprint = {1812.01428},
 primaryClass = {astro-ph.HE},
       adsurl = {https://ui.adsabs.harvard.edu/abs/2019ARA&A..57..305G}
}

@ARTICLE{Inserra19,
       author = {{Inserra}, C.},
        title = "{Observational properties of extreme supernovae}",
      journal = {Nature Astronomy},
         year = 2019,
        month = aug,
       volume = {3},
        pages = {697-705},
          doi = {10.1038/s41550-019-0854-4},
archivePrefix = {arXiv},
       eprint = {1908.02314},
 primaryClass = {astro-ph.HE},
       adsurl = {https://ui.adsabs.harvard.edu/abs/2019NatAs...3..697I}
}

@ARTICLE{Cowan2011,
       author = {{Cowan}, Glen and {Cranmer}, Kyle and {Gross}, Eilam and {Vitells}, Ofer},
        title = "{Asymptotic formulae for likelihood-based tests of new physics}",
      journal = {European Physical Journal C},
         year = 2011,
        month = feb,
       volume = {71},
       number = {2},
          eid = {1554},
        pages = {1554},
          doi = {10.1140/epjc/s10052-011-1554-0},
archivePrefix = {arXiv},
       eprint = {1007.1727},
 primaryClass = {physics.data-an},
       adsurl = {https://ui.adsabs.harvard.edu/abs/2011EPJC...71.1554C}
}

@ARTICLE{Chatzopoulos13,
       author = {{Chatzopoulos}, E. and {Wheeler}, J. Craig and {Vinko}, J. and {Horvath}, Z.~L. and {Nagy}, A.},
        title = "{Analytical Light Curve Models of Superluminous Supernovae: {\ensuremath{\chi}}$^{2}$-minimization of Parameter Fits}",
      journal = {\apj},
         year = 2013,
        month = aug,
       volume = {773},
       number = {1},
          eid = {76},
        pages = {76},
          doi = {10.1088/0004-637X/773/1/76},
archivePrefix = {arXiv},
       eprint = {1306.3447},
 primaryClass = {astro-ph.HE},
       adsurl = {https://ui.adsabs.harvard.edu/abs/2013ApJ...773...76C}
}

@ARTICLE{Dessart12,
       author = {{Dessart}, Luc and {Hillier}, D. John and {Waldman}, Roni and {Livne}, Eli and {Blondin}, St{\'e}phane},
        title = "{Superluminous supernovae: $^{56}$Ni power versus magnetar radiation}",
      journal = {\mnras},
         year = 2012,
        month = oct,
       volume = {426},
       number = {1},
        pages = {L76-L80},
          doi = {10.1111/j.1745-3933.2012.01329.x},
archivePrefix = {arXiv},
       eprint = {1208.1214},
 primaryClass = {astro-ph.SR},
       adsurl = {https://ui.adsabs.harvard.edu/abs/2012MNRAS.426L..76D}
}

@ARTICLE{Metzger15,
       author = {{Metzger}, Brian D. and {Margalit}, Ben and {Kasen}, Daniel and {Quataert}, Eliot},
        title = "{The diversity of transients from magnetar birth in core collapse supernovae}",
      journal = {\mnras},
         year = 2015,
        month = dec,
       volume = {454},
       number = {3},
        pages = {3311-3316},
          doi = {10.1093/mnras/stv2224},
archivePrefix = {arXiv},
       eprint = {1508.02712},
 primaryClass = {astro-ph.HE},
       adsurl = {https://ui.adsabs.harvard.edu/abs/2015MNRAS.454.3311M}
}

@ARTICLE{Moriya18-review,
       author = {{Moriya}, Takashi J. and {Sorokina}, Elena I. and {Chevalier}, Roger A.},
        title = "{Superluminous Supernovae}",
      journal = {\ssr},
         year = 2018,
        month = mar,
       volume = {214},
       number = {2},
          eid = {59},
        pages = {59},
          doi = {10.1007/s11214-018-0493-6},
archivePrefix = {arXiv},
       eprint = {1803.01875},
 primaryClass = {astro-ph.HE},
       adsurl = {https://ui.adsabs.harvard.edu/abs/2018SSRv..214...59M}
}

@ARTICLE{Moriya18,
       author = {{Moriya}, Takashi J. and {Nicholl}, Matt and {Guillochon}, James},
        title = "{Systematic Investigation of the Fallback Accretion-powered Model for Hydrogen-poor Superluminous Supernovae}",
      journal = {\apj},
         year = 2018,
        month = nov,
       volume = {867},
       number = {2},
          eid = {113},
        pages = {113},
          doi = {10.3847/1538-4357/aae53d},
archivePrefix = {arXiv},
       eprint = {1806.00090},
 primaryClass = {astro-ph.HE},
       adsurl = {https://ui.adsabs.harvard.edu/abs/2018ApJ...867..113M}
}

@ARTICLE{Yuan18,
       author = {{Yuan}, Qiang and {Liao}, Neng-Hui and {Xin}, Yu-Liang and {Li}, Ye and {Fan}, Yi-Zhong and {Zhang}, Bing and {Hu}, Hong-Bo and {Bi}, Xiao-Jun},
        title = "{Fermi Large Area Telescope Detection of Gamma-Ray Emission from the Direction of Supernova iPTF14hls}",
      journal = {\apjl},
         year = 2018,
        month = feb,
       volume = {854},
       number = {2},
          eid = {L18},
        pages = {L18},
          doi = {10.3847/2041-8213/aaacc9},
archivePrefix = {arXiv},
       eprint = {1712.01043},
 primaryClass = {astro-ph.HE},
       adsurl = {https://ui.adsabs.harvard.edu/abs/2018ApJ...854L..18Y}
}

@ARTICLE{Prokhorov21,
       author = {{Prokhorov}, D.~A. and {Moraghan}, A. and {Vink}, J.},
        title = "{Search for gamma rays from SNe with a variable-size sliding-time-window analysis of the Fermi-LAT data}",
      journal = {\mnras},
         year = 2021,
        month = jul,
       volume = {505},
       number = {1},
        pages = {1413-1421},
          doi = {10.1093/mnras/stab1313},
archivePrefix = {arXiv},
       eprint = {2105.04007},
 primaryClass = {astro-ph.HE},
       adsurl = {https://ui.adsabs.harvard.edu/abs/2021MNRAS.505.1413P}
}

@ARTICLE{Bose2018,
       author = {{Bose}, Subhash and {Dong}, Subo and {Pastorello}, A. and {Filippenko}, Alexei V. and {Kochanek}, C.~S. and {Mauerhan}, Jon and {Romero-Ca{\~n}izales}, C. and {Brink}, Thomas G. and {Chen}, Ping and {Prieto}, J.~L. and {Post}, R. and {Ashall}, Christopher and {Grupe}, Dirk and {Tomasella}, L. and {Benetti}, Stefano and {Shappee}, B.~J. and {Stanek}, K.~Z. and {Cai}, Zheng and {Falco}, E. and {Lundqvist}, Peter and {Mattila}, Seppo and {Mutel}, Robert and {Ochner}, Paolo and {Pooley}, David and {Stritzinger}, M.~D. and {Villanueva}, Jr., S. and {Zheng}, WeiKang and {Beswick}, R.~J. and {Brown}, Peter J. and {Cappellaro}, E. and {Davis}, Scott and {Fraser}, Morgan and {de Jaeger}, Thomas and {Elias-Rosa}, N. and {Gall}, C. and {Gaudi}, B. Scott and {Herczeg}, Gregory J. and {Hestenes}, Julia and {Holoien}, T.~W. -S. and {Hosseinzadeh}, Griffin and {Hsiao}, E.~Y. and {Hu}, Shaoming and {Jaejin}, Shin and {Jeffers}, Ben and {Koff}, R.~A. and {Kumar}, Sahana and {Kurtenkov}, Alexander and {Lau}, Marie Wingyee and {Prentice}, Simon and {Reynolds}, T. and {Rudy}, Richard J. and {Shahbandeh}, Melissa and {Somero}, Auni and {Stassun}, Keivan G. and {Thompson}, Todd A. and {Valenti}, Stefano and {Woo}, Jong-Hak and {Yunus}, Sameen},
        title = "{Gaia17biu/SN 2017egm in NGC 3191: The Closest Hydrogen-poor Superluminous Supernova to Date Is in a {\textquotedblleft}Normal,{\textquotedblright} Massive, Metal-rich Spiral Galaxy}",
      journal = {\apj},
         year = 2018,
        month = jan,
       volume = {853},
       number = {1},
          eid = {57},
        pages = {57},
          doi = {10.3847/1538-4357/aaa298},
archivePrefix = {arXiv},
       eprint = {1708.00864},
 primaryClass = {astro-ph.HE},
       adsurl = {https://ui.adsabs.harvard.edu/abs/2018ApJ...853...57B}
}

@ARTICLE{Renault18,
       author = {{Renault-Tinacci}, N. and {Kotera}, K. and {Neronov}, A. and {Ando}, S.},
        title = "{Search for {\ensuremath{\gamma}}-ray emission from superluminous supernovae with the Fermi-LAT}",
      journal = {\aap},
         year = 2018,
        month = mar,
       volume = {611},
          eid = {A45},
        pages = {A45},
          doi = {10.1051/0004-6361/201730741},
archivePrefix = {arXiv},
       eprint = {1708.08971},
 primaryClass = {astro-ph.HE},
       adsurl = {https://ui.adsabs.harvard.edu/abs/2018A&A...611A..45R}
}

@ARTICLE{Quimby11,
       author = {{Quimby}, R.~M. and {Kulkarni}, S.~R. and {Kasliwal}, M.~M. and {Gal-Yam}, A. and {Arcavi}, I. and {Sullivan}, M. and {Nugent}, P. and {Thomas}, R. and {Howell}, D.~A. and {Nakar}, E. and {Bildsten}, L. and {Theissen}, C. and {Law}, N.~M. and {Dekany}, R. and {Rahmer}, G. and {Hale}, D. and {Smith}, R. and {Ofek}, E.~O. and {Zolkower}, J. and {Velur}, V. and {Walters}, R. and {Henning}, J. and {Bui}, K. and {McKenna}, D. and {Poznanski}, D. and {Cenko}, S.~B. and {Levitan}, D.},
        title = "{Hydrogen-poor superluminous stellar explosions}",
      journal = {\nat},
         year = 2011,
        month = jun,
       volume = {474},
       number = {7352},
        pages = {487-489},
          doi = {10.1038/nature10095},
archivePrefix = {arXiv},
       eprint = {0910.0059},
 primaryClass = {astro-ph.CO},
       adsurl = {https://ui.adsabs.harvard.edu/abs/2011Natur.474..487Q}
}

@ARTICLE{Hosseinzadeh22,
       author = {{Hosseinzadeh}, Griffin and {Berger}, Edo and {Metzger}, Brian D. and {Gomez}, Sebastian and {Nicholl}, Matt and {Blanchard}, Peter},
        title = "{Bumpy Declining Light Curves Are Common in Hydrogen-poor Superluminous Supernovae}",
      journal = {\apj},
         year = 2022,
        month = jul,
       volume = {933},
       number = {1},
          eid = {14},
        pages = {14},
          doi = {10.3847/1538-4357/ac67dd},
archivePrefix = {arXiv},
       eprint = {2109.09743},
 primaryClass = {astro-ph.HE},
       adsurl = {https://ui.adsabs.harvard.edu/abs/2022ApJ...933...14H}
}

@ARTICLE{Woosley07,
       author = {{Woosley}, S.~E. and {Blinnikov}, S. and {Heger}, Alexander},
        title = "{Pulsational pair instability as an explanation for the most luminous supernovae}",
      journal = {\nat},
         year = 2007,
        month = nov,
       volume = {450},
       number = {7168},
        pages = {390-392},
          doi = {10.1038/nature06333},
archivePrefix = {arXiv},
       eprint = {0710.3314},
 primaryClass = {astro-ph},
       adsurl = {https://ui.adsabs.harvard.edu/abs/2007Natur.450..390W}
}

@ARTICLE{Woosley17,
       author = {{Woosley}, S.~E.},
        title = "{Pulsational Pair-instability Supernovae}",
      journal = {\apj},
         year = 2017,
        month = feb,
       volume = {836},
       number = {2},
          eid = {244},
        pages = {244},
          doi = {10.3847/1538-4357/836/2/244},
archivePrefix = {arXiv},
       eprint = {1608.08939},
 primaryClass = {astro-ph.HE},
       adsurl = {https://ui.adsabs.harvard.edu/abs/2017ApJ...836..244W}
}

@ARTICLE{Chatzopoulos12,
       author = {{Chatzopoulos}, E. and {Wheeler}, J. Craig},
        title = "{Hydrogen-poor Circumstellar Shells from Pulsational Pair-instability Supernovae with Rapidly Rotating Progenitors}",
      journal = {\apj},
         year = 2012,
        month = dec,
       volume = {760},
       number = {2},
          eid = {154},
        pages = {154},
          doi = {10.1088/0004-637X/760/2/154},
archivePrefix = {arXiv},
       eprint = {1210.1617},
 primaryClass = {astro-ph.HE},
       adsurl = {https://ui.adsabs.harvard.edu/abs/2012ApJ...760..154C}
}

@ARTICLE{Renzo20,
       author = {{Renzo}, M. and {Farmer}, R. and {Justham}, S. and {G{\"o}tberg}, Y. and {de Mink}, S.~E. and {Zapartas}, E. and {Marchant}, P. and {Smith}, N.},
        title = "{Predictions for the hydrogen-free ejecta of pulsational pair-instability supernovae}",
      journal = {\aap},
         year = 2020,
        month = aug,
       volume = {640},
          eid = {A56},
        pages = {A56},
          doi = {10.1051/0004-6361/202037710},
archivePrefix = {arXiv},
       eprint = {2002.05077},
 primaryClass = {astro-ph.SR},
       adsurl = {https://ui.adsabs.harvard.edu/abs/2020A&A...640A..56R}
}

@ARTICLE{Scargle13,
       author = {{Scargle}, Jeffrey D. and {Norris}, Jay P. and {Jackson}, Brad and {Chiang}, James},
        title = "{Studies in Astronomical Time Series Analysis. VI. Bayesian Block Representations}",
      journal = {\apj},
         year = 2013,
        month = feb,
       volume = {764},
       number = {2},
          eid = {167},
        pages = {167},
          doi = {10.1088/0004-637X/764/2/167},
archivePrefix = {arXiv},
       eprint = {1207.5578},
 primaryClass = {astro-ph.IM},
       adsurl = {https://ui.adsabs.harvard.edu/abs/2013ApJ...764..167S}
}

@ARTICLE{Kerr19,
       author = {{Kerr}, M.},
        title = "{Multiscale Time- and Frequency-domain Likelihood Analysis with Photon Weights}",
      journal = {\apj},
         year = 2019,
        month = nov,
       volume = {885},
       number = {1},
          eid = {92},
        pages = {92},
          doi = {10.3847/1538-4357/ab459f},
archivePrefix = {arXiv},
       eprint = {1910.00140},
 primaryClass = {astro-ph.HE},
       adsurl = {https://ui.adsabs.harvard.edu/abs/2019ApJ...885...92K}
}

@ARTICLE{Dermer12,
       author = {{Dermer}, Charles D.},
        title = "{Diffuse Galactic Gamma Rays from Shock-Accelerated Cosmic Rays}",
      journal = {\prl},
         year = 2012,
        month = aug,
       volume = {109},
       number = {9},
          eid = {091101},
        pages = {091101},
          doi = {10.1103/PhysRevLett.109.091101},
archivePrefix = {arXiv},
       eprint = {1206.2899},
 primaryClass = {astro-ph.HE},
       adsurl = {https://ui.adsabs.harvard.edu/abs/2012PhRvL.109i1101D}
}

@INPROCEEDINGS{Zabalza15,
       author = {{Zabalza}, V.},
        title = "{Naima: a Python package for inference of particle distribution properties from nonthermal spectra}",
    booktitle = {34th International Cosmic Ray Conference (ICRC2015)},
         year = 2015,
       series = {International Cosmic Ray Conference},
       volume = {34},
        month = jul,
          eid = {922},
        pages = {922},
          doi = {10.22323/1.236.0922},
archivePrefix = {arXiv},
       eprint = {1509.03319},
 primaryClass = {astro-ph.HE},
       adsurl = {https://ui.adsabs.harvard.edu/abs/2015ICRC...34..922Z}
}

@INPROCEEDINGS{Katz12,
       author = {{Katz}, Boaz and {Sapir}, Nir and {Waxman}, Eli},
        title = "{X-rays, {\ensuremath{\gamma}}-rays and neutrinos from collisionless shocks in supernova wind breakouts}",
    booktitle = {Death of Massive Stars: Supernovae and Gamma-Ray Bursts},
         year = 2012,
       editor = {{Roming}, P. and {Kawai}, N. and {Pian}, E.},
       series = {IAU Symposium},
       volume = {279},
        month = sep,
        pages = {274-281},
          doi = {10.1017/S174392131201304X},
archivePrefix = {arXiv},
       eprint = {1106.1898},
 primaryClass = {astro-ph.HE},
       adsurl = {https://ui.adsabs.harvard.edu/abs/2012IAUS..279..274K}
}

@ARTICLE{Tatischeff09,
       author = {{Tatischeff}, V.},
        title = "{Radio emission and nonlinear diffusive shock acceleration of cosmic rays in the supernova SN 1993J}",
      journal = {\aap},
         year = 2009,
        month = may,
       volume = {499},
       number = {1},
        pages = {191-213},
          doi = {10.1051/0004-6361/200811511},
archivePrefix = {arXiv},
       eprint = {0903.2944},
 primaryClass = {astro-ph.HE},
       adsurl = {https://ui.adsabs.harvard.edu/abs/2009A&A...499..191T}
}

@ARTICLE{Giacinti15,
       author = {{Giacinti}, G. and {Bell}, A.~R.},
        title = "{Collisionless shocks and TeV neutrinos before Supernova shock breakout from an optically thick wind}",
      journal = {\mnras},
         year = 2015,
        month = jun,
       volume = {449},
       number = {4},
        pages = {3693-3699},
          doi = {10.1093/mnras/stv561},
archivePrefix = {arXiv},
       eprint = {1503.04170},
 primaryClass = {astro-ph.HE},
       adsurl = {https://ui.adsabs.harvard.edu/abs/2015MNRAS.449.3693G}
}

@ARTICLE{Kafexhiu14,
       author = {{Kafexhiu}, Ervin and {Aharonian}, Felix and {Taylor}, Andrew M. and {Vila}, Gabriela S.},
        title = "{Parametrization of gamma-ray production cross sections for p p interactions in a broad proton energy range from the kinematic threshold to PeV energies}",
      journal = {\prd},
         year = 2014,
        month = dec,
       volume = {90},
       number = {12},
          eid = {123014},
        pages = {123014},
          doi = {10.1103/PhysRevD.90.123014},
archivePrefix = {arXiv},
       eprint = {1406.7369},
 primaryClass = {astro-ph.HE},
       adsurl = {https://ui.adsabs.harvard.edu/abs/2014PhRvD..90l3014K}
}

@ARTICLE{Wheeler17,
       author = {{Wheeler}, J. Craig and {Chatzopoulos}, Emmanouil and {Vink{\'o}}, Jozsef and {Tuminello}, Richard},
        title = "{Circumstellar Interaction Models for the Bolometric Light Curve of Type I Superluminous SN 2017egm}",
      journal = {\apjl},
         year = 2017,
        month = dec,
       volume = {851},
       number = {1},
          eid = {L14},
        pages = {L14},
          doi = {10.3847/2041-8213/aa9d84},
archivePrefix = {arXiv},
       eprint = {1710.04994},
 primaryClass = {astro-ph.HE},
       adsurl = {https://ui.adsabs.harvard.edu/abs/2017ApJ...851L..14W}
}

@ARTICLE{Acero22,
       author = {{Acero}, F. and {Lemoine-Goumard}, M. and {Ballet}, J.},
        title = "{Characterization of the Gamma-ray Emission from the Kepler Supernova Remnant with Fermi-LAT}",
      journal = {\aap},
         year = 2022,
        month = apr,
       volume = {660},
          eid = {A129},
        pages = {A129},
          doi = {10.1051/0004-6361/202142200},
archivePrefix = {arXiv},
       eprint = {2201.05567},
 primaryClass = {astro-ph.HE},
       adsurl = {https://ui.adsabs.harvard.edu/abs/2022A&A...660A.129A}
}

@ARTICLE{Murase11,
       author = {{Murase}, Kohta and {Thompson}, Todd A. and {Lacki}, Brian C. and {Beacom}, John F.},
        title = "{New class of high-energy transients from crashes of supernova ejecta with massive circumstellar material shells}",
      journal = {\prd},
         year = 2011,
        month = aug,
       volume = {84},
       number = {4},
          eid = {043003},
        pages = {043003},
          doi = {10.1103/PhysRevD.84.043003},
archivePrefix = {arXiv},
       eprint = {1012.2834},
 primaryClass = {astro-ph.HE},
       adsurl = {https://ui.adsabs.harvard.edu/abs/2011PhRvD..84d3003M}
}

@ARTICLE{Margutti2014,
       author = {{Margutti}, R. and {Milisavljevic}, D. and {Soderberg}, A.~M. and {Chornock}, R. and {Zauderer}, B.~A. and {Murase}, K. and {Guidorzi}, C. and {Sanders}, N.~E. and {Kuin}, P. and {Fransson}, C. and {Levesque}, E.~M. and {Chandra}, P. and {Berger}, E. and {Bianco}, F.~B. and {Brown}, P.~J. and {Challis}, P. and {Chatzopoulos}, E. and {Cheung}, C.~C. and {Choi}, C. and {Chomiuk}, L. and {Chugai}, N. and {Contreras}, C. and {Drout}, M.~R. and {Fesen}, R. and {Foley}, R.~J. and {Fong}, W. and {Friedman}, A.~S. and {Gall}, C. and {Gehrels}, N. and {Hjorth}, J. and {Hsiao}, E. and {Kirshner}, R. and {Im}, M. and {Leloudas}, G. and {Lunnan}, R. and {Marion}, G.~H. and {Martin}, J. and {Morrell}, N. and {Neugent}, K.~F. and {Omodei}, N. and {Phillips}, M.~M. and {Rest}, A. and {Silverman}, J.~M. and {Strader}, J. and {Stritzinger}, M.~D. and {Szalai}, T. and {Utterback}, N.~B. and {Vinko}, J. and {Wheeler}, J.~C. and {Arnett}, D. and {Campana}, S. and {Chevalier}, R. and {Ginsburg}, A. and {Kamble}, A. and {Roming}, P.~W.~A. and {Pritchard}, T. and {Stringfellow}, G.},
        title = "{A Panchromatic View of the Restless SN 2009ip Reveals the Explosive Ejection of a Massive Star Envelope}",
      journal = {\apj},
         year = 2014,
        month = jan,
       volume = {780},
       number = {1},
          eid = {21},
        pages = {21},
          doi = {10.1088/0004-637X/780/1/21},
archivePrefix = {arXiv},
       eprint = {1306.0038},
 primaryClass = {astro-ph.HE},
       adsurl = {https://ui.adsabs.harvard.edu/abs/2014ApJ...780...21M}
}

@ARTICLE{Marti-Devesa24,
       author = {{Mart{\'\i}-Devesa}, G. and {Cheung}, C.~C. and {Di Lalla}, N. and {Renaud}, M. and {Principe}, G. and {Omodei}, N. and {Acero}, F.},
        title = "{Early-time {\ensuremath{\gamma}}-ray constraints on cosmic-ray acceleration in the core-collapse SN 2023ixf with the Fermi Large Area Telescope}",
      journal = {\aap},
         year = 2024,
        month = jun,
       volume = {686},
          eid = {A254},
        pages = {A254},
          doi = {10.1051/0004-6361/202349061},
archivePrefix = {arXiv},
       eprint = {2404.10487},
 primaryClass = {astro-ph.HE},
       adsurl = {https://ui.adsabs.harvard.edu/abs/2024A&A...686A.254M}
}

@ARTICLE{Albareti17,
       author = {{Albareti}, Franco D. and {Allende Prieto}, Carlos and {Almeida}, Andres and {Anders}, Friedrich and {Anderson}, Scott and {Andrews}, Brett H. and {Arag{\'o}n-Salamanca}, Alfonso and {Argudo-Fern{\'a}ndez}, Maria and {Armengaud}, Eric and {Aubourg}, Eric and {Avila-Reese}, Vladimir and {Badenes}, Carles and {Bailey}, Stephen and {Barbuy}, Beatriz and {Barger}, Kat and {Barrera-Ballesteros}, Jorge and {Bartosz}, Curtis and {Basu}, Sarbani and {Bates}, Dominic and {Battaglia}, Giuseppina and {Baumgarten}, Falk and {Baur}, Julien and {Bautista}, Julian and {Beers}, Timothy C. and {Belfiore}, Francesco and {Bershady}, Matthew and {Bertran de Lis}, Sara and {Bird}, Jonathan C. and {Bizyaev}, Dmitry and {Blanc}, Guillermo A. and {Blanton}, Michael and {Blomqvist}, Michael and {Bolton}, Adam S. and {Borissova}, J. and {Bovy}, Jo and {Brandt}, William Nielsen and {Brinkmann}, Jonathan and {Brownstein}, Joel R. and {Bundy}, Kevin and {Burtin}, Etienne and {Busca}, Nicol{\'a}s G. and {Camacho Chavez}, Hugo Orlando and {Cano D{\'\i}az}, M. and {Cappellari}, Michele and {Carrera}, Ricardo and {Chen}, Yanping and {Cherinka}, Brian and {Cheung}, Edmond and {Chiappini}, Cristina and {Chojnowski}, Drew and {Chuang}, Chia-Hsun and {Chung}, Haeun and {Cirolini}, Rafael Fernando and {Clerc}, Nicolas and {Cohen}, Roger E. and {Comerford}, Julia M. and {Comparat}, Johan and {Correa do Nascimento}, Janaina and {Cousinou}, Marie-Claude and {Covey}, Kevin and {Crane}, Jeffrey D. and {Croft}, Rupert and {Cunha}, Katia and {Darling}, Jeremy and {Davidson}, Jr., James W. and {Dawson}, Kyle and {Da Costa}, Luiz and {Da Silva Ilha}, Gabriele and {Deconto Machado}, Alice and {Delubac}, Timoth{\'e}e and {De Lee}, Nathan and {De la Macorra}, Axel and {De la Torre}, Sylvain and {Diamond-Stanic}, Aleksandar M. and {Donor}, John and {Downes}, Juan Jose and {Drory}, Niv and {Du}, Cheng and {Du Mas des Bourboux}, H{\'e}lion and {Dwelly}, Tom and {Ebelke}, Garrett and {Eigenbrot}, Arthur and {Eisenstein}, Daniel J. and {Elsworth}, Yvonne P. and {Emsellem}, Eric and {Eracleous}, Michael and {Escoffier}, Stephanie and {Evans}, Michael L. and {Falc{\'o}n-Barroso}, Jes{\'u}s and {Fan}, Xiaohui and {Favole}, Ginevra and {Fernandez-Alvar}, Emma and {Fernandez-Trincado}, J.~G. and {Feuillet}, Diane and {Fleming}, Scott W. and {Font-Ribera}, Andreu and {Freischlad}, Gordon and {Frinchaboy}, Peter and {Fu}, Hai and {Gao}, Yang and {Garcia}, Rafael A. and {Garcia-Dias}, R. and {Garcia-Hern{\'a}ndez}, D.~A. and {Garcia P{\'e}rez}, Ana E. and {Gaulme}, Patrick and {Ge}, Junqiang and {Geisler}, Douglas and {Gillespie}, Bruce and {Gil Marin}, Hector and {Girardi}, L{\'e}o and {Goddard}, Daniel and {Gomez Maqueo Chew}, Yilen and {Gonzalez-Perez}, Violeta and {Grabowski}, Kathleen and {Green}, Paul and {Grier}, Catherine J. and {Grier}, Thomas and {Guo}, Hong and {Guy}, Julien and {Hagen}, Alex and {Hall}, Matt and {Harding}, Paul and {Harley}, R.~E. and {Hasselquist}, Sten and {Hawley}, Suzanne and {Hayes}, Christian R. and {Hearty}, Fred and {Hekker}, Saskia and {Hernandez Toledo}, Hector and {Ho}, Shirley and {Hogg}, David W. and {Holley-Bockelmann}, Kelly and {Holtzman}, Jon A. and {Holzer}, Parker H. and {Hu}, Jian and {Huber}, Daniel and {Hutchinson}, Timothy Alan and {Hwang}, Ho Seong and {Ibarra-Medel}, H{\'e}ctor J. and {Ivans}, Inese I. and {Ivory}, KeShawn and {Jaehnig}, Kurt and {Jensen}, Trey W. and {Johnson}, Jennifer A. and {Jones}, Amy and {Jullo}, Eric and {Kallinger}, T. and {Kinemuchi}, Karen and {Kirkby}, David and {Klaene}, Mark and {Kneib}, Jean-Paul and {Kollmeier}, Juna A. and {Lacerna}, Ivan and {Lane}, Richard R. and {Lang}, Dustin and {Laurent}, Pierre and {Law}, David R. and {Leauthaud}, Alexie and {Le Goff}, Jean-Marc and {Li}, Chen and {Li}, Cheng and {Li}, Niu and {Li}, Ran and {Liang}, Fu-Heng and {Liang}, Yu and {Lima}, Marcos and {Lin}, Lihwai and {Lin}, Lin and {Lin}, Yen-Ting and {Liu}, Chao and {Long}, Dan and {Lucatello}, Sara and {MacDonald}, Nicholas and {MacLeod}, Chelsea L. and {Mackereth}, J. Ted and {Mahadevan}, Suvrath and {Maia}, Marcio Antonio Geimba and {Maiolino}, Roberto and {Majewski}, Steven R. and {Malanushenko}, Olena and {Malanushenko}, Viktor and {Mallmann}, N{\'\i}colas Dullius and {Manchado}, Arturo and {Maraston}, Claudia and {Marques-Chaves}, Rui and {Martinez Valpuesta}, Inma and {Masters}, Karen L. and {Mathur}, Savita and {McGreer}, Ian D. and {Merloni}, Andrea and {Merrifield}, Michael R. and {M{\'e}sz{\'a}ros}, Szabolcs and {Meza}, Andres and {Miglio}, Andrea and {Minchev}, Ivan and {Molaverdikhani}, Karan and {Montero-Dorta}, Antonio D. and {Mosser}, Benoit and {Muna}, Demitri and {Myers}, Adam},
        title = "{The 13th Data Release of the Sloan Digital Sky Survey: First Spectroscopic Data from the SDSS-IV Survey Mapping Nearby Galaxies at Apache Point Observatory}",
      journal = {\apjs},
         year = 2017,
        month = dec,
       volume = {233},
       number = {2},
          eid = {25},
        pages = {25},
          doi = {10.3847/1538-4365/aa8992},
archivePrefix = {arXiv},
       eprint = {1608.02013},
 primaryClass = {astro-ph.GA},
       adsurl = {https://ui.adsabs.harvard.edu/abs/2017ApJS..233...25A}
}

@ARTICLE{GaiaDR3,
       author = {{Gaia Collaboration} and {Vallenari}, A. and {Brown}, A.~G.~A. and {Prusti}, T. and {de Bruijne}, J.~H.~J. and {Arenou}, F. and {Babusiaux}, C. and {Biermann}, M. and {Creevey}, O.~L. and {Ducourant}, C. and {Evans}, D.~W. and {Eyer}, L. and {Guerra}, R. and {Hutton}, A. and {Jordi}, C. and {Klioner}, S.~A. and {Lammers}, U.~L. and {Lindegren}, L. and {Luri}, X. and {Mignard}, F. and {Panem}, C. and {Pourbaix}, D. and {Randich}, S. and {Sartoretti}, P. and {Soubiran}, C. and {Tanga}, P. and {Walton}, N.~A. and {Bailer-Jones}, C.~A.~L. and {Bastian}, U. and {Drimmel}, R. and {Jansen}, F. and {Katz}, D. and {Lattanzi}, M.~G. and {van Leeuwen}, F. and {Bakker}, J. and {Cacciari}, C. and {Casta{\~n}eda}, J. and {De Angeli}, F. and {Fabricius}, C. and {Fouesneau}, M. and {Fr{\'e}mat}, Y. and {Galluccio}, L. and {Guerrier}, A. and {Heiter}, U. and {Masana}, E. and {Messineo}, R. and {Mowlavi}, N. and {Nicolas}, C. and {Nienartowicz}, K. and {Pailler}, F. and {Panuzzo}, P. and {Riclet}, F. and {Roux}, W. and {Seabroke}, G.~M. and {Sordo}, R. and {Th{\'e}venin}, F. and {Gracia-Abril}, G. and {Portell}, J. and {Teyssier}, D. and {Altmann}, M. and {Andrae}, R. and {Audard}, M. and {Bellas-Velidis}, I. and {Benson}, K. and {Berthier}, J. and {Blomme}, R. and {Burgess}, P.~W. and {Busonero}, D. and {Busso}, G. and {C{\'a}novas}, H. and {Carry}, B. and {Cellino}, A. and {Cheek}, N. and {Clementini}, G. and {Damerdji}, Y. and {Davidson}, M. and {de Teodoro}, P. and {Nu{\~n}ez Campos}, M. and {Delchambre}, L. and {Dell'Oro}, A. and {Esquej}, P. and {Fern{\'a}ndez-Hern{\'a}ndez}, J. and {Fraile}, E. and {Garabato}, D. and {Garc{\'\i}a-Lario}, P. and {Gosset}, E. and {Haigron}, R. and {Halbwachs}, J. -L. and {Hambly}, N.~C. and {Harrison}, D.~L. and {Hern{\'a}ndez}, J. and {Hestroffer}, D. and {Hodgkin}, S.~T. and {Holl}, B. and {Jan{\ss}en}, K. and {Jevardat de Fombelle}, G. and {Jordan}, S. and {Krone-Martins}, A. and {Lanzafame}, A.~C. and {L{\"o}ffler}, W. and {Marchal}, O. and {Marrese}, P.~M. and {Moitinho}, A. and {Muinonen}, K. and {Osborne}, P. and {Pancino}, E. and {Pauwels}, T. and {Recio-Blanco}, A. and {Reyl{\'e}}, C. and {Riello}, M. and {Rimoldini}, L. and {Roegiers}, T. and {Rybizki}, J. and {Sarro}, L.~M. and {Siopis}, C. and {Smith}, M. and {Sozzetti}, A. and {Utrilla}, E. and {van Leeuwen}, M. and {Abbas}, U. and {{\'A}brah{\'a}m}, P. and {Abreu Aramburu}, A. and {Aerts}, C. and {Aguado}, J.~J. and {Ajaj}, M. and {Aldea-Montero}, F. and {Altavilla}, G. and {{\'A}lvarez}, M.~A. and {Alves}, J. and {Anders}, F. and {Anderson}, R.~I. and {Anglada Varela}, E. and {Antoja}, T. and {Baines}, D. and {Baker}, S.~G. and {Balaguer-N{\'u}{\~n}ez}, L. and {Balbinot}, E. and {Balog}, Z. and {Barache}, C. and {Barbato}, D. and {Barros}, M. and {Barstow}, M.~A. and {Bartolom{\'e}}, S. and {Bassilana}, J. -L. and {Bauchet}, N. and {Becciani}, U. and {Bellazzini}, M. and {Berihuete}, A. and {Bernet}, M. and {Bertone}, S. and {Bianchi}, L. and {Binnenfeld}, A. and {Blanco-Cuaresma}, S. and {Blazere}, A. and {Boch}, T. and {Bombrun}, A. and {Bossini}, D. and {Bouquillon}, S. and {Bragaglia}, A. and {Bramante}, L. and {Breedt}, E. and {Bressan}, A. and {Brouillet}, N. and {Brugaletta}, E. and {Bucciarelli}, B. and {Burlacu}, A. and {Butkevich}, A.~G. and {Buzzi}, R. and {Caffau}, E. and {Cancelliere}, R. and {Cantat-Gaudin}, T. and {Carballo}, R. and {Carlucci}, T. and {Carnerero}, M.~I. and {Carrasco}, J.~M. and {Casamiquela}, L. and {Castellani}, M. and {Castro-Ginard}, A. and {Chaoul}, L. and {Charlot}, P. and {Chemin}, L. and {Chiaramida}, V. and {Chiavassa}, A. and {Chornay}, N. and {Comoretto}, G. and {Contursi}, G. and {Cooper}, W.~J. and {Cornez}, T. and {Cowell}, S. and {Crifo}, F. and {Cropper}, M. and {Crosta}, M. and {Crowley}, C. and {Dafonte}, C. and {Dapergolas}, A. and {David}, M. and {David}, P. and {de Laverny}, P. and {De Luise}, F. and {De March}, R.},
        title = "{Gaia Data Release 3. Summary of the content and survey properties}",
      journal = {\aap},
         year = 2023,
        month = jun,
       volume = {674},
          eid = {A1},
        pages = {A1},
          doi = {10.1051/0004-6361/202243940},
archivePrefix = {arXiv},
       eprint = {2208.00211},
 primaryClass = {astro-ph.GA},
       adsurl = {https://ui.adsabs.harvard.edu/abs/2023A&A...674A...1G}
}

@ARTICLE{Chandra,
       author = {{Evans}, Ian N. and {Evans}, Janet D. and {Mart{\'\i}nez-Galarza}, J. Rafael and {Miller}, Joseph B. and {Primini}, Francis A. and {Azadi}, Mojegan and {Burke}, Douglas J. and {Civano}, Francesca M. and {D'Abrusco}, Raffaele and {Fabbiano}, Giuseppina and {Graessle}, Dale E. and {Grier}, John D. and {Houck}, John C. and {Lauer}, Jennifer and {McCollough}, Michael L. and {Nowak}, Michael A. and {Plummer}, David A. and {Rots}, Arnold H. and {Siemiginowska}, Aneta and {Tibbetts}, Michael S.},
        title = "{The Chandra Source Catalog Release 2 Series}",
      journal = {\apjs},
         year = 2024,
        month = oct,
       volume = {274},
       number = {2},
          eid = {22},
        pages = {22},
          doi = {10.3847/1538-4365/ad6319},
archivePrefix = {arXiv},
       eprint = {2407.10799},
 primaryClass = {astro-ph.HE},
       adsurl = {https://ui.adsabs.harvard.edu/abs/2024ApJS..274...22E}
}

@ARTICLE{ATLAS,
       author = {{Tonry}, J.~L. and {Denneau}, L. and {Heinze}, A.~N. and {Stalder}, B. and {Smith}, K.~W. and {Smartt}, S.~J. and {Stubbs}, C.~W. and {Weiland}, H.~J. and {Rest}, A.},
        title = "{ATLAS: A High-cadence All-sky Survey System}",
      journal = {\pasp},
         year = 2018,
        month = jun,
       volume = {130},
       number = {988},
        pages = {064505},
          doi = {10.1088/1538-3873/aabadf},
archivePrefix = {arXiv},
       eprint = {1802.00879},
 primaryClass = {astro-ph.IM},
       adsurl = {https://ui.adsabs.harvard.edu/abs/2018PASP..130f4505T}
}

@ARTICLE{ZTF,
       author = {{Masci}, Frank J. and {Laher}, Russ R. and {Rusholme}, Ben and {Shupe}, David L. and {Groom}, Steven and {Surace}, Jason and {Jackson}, Edward and {Monkewitz}, Serge and {Beck}, Ron and {Flynn}, David and {Terek}, Scott and {Landry}, Walter and {Hacopians}, Eugean and {Desai}, Vandana and {Howell}, Justin and {Brooke}, Tim and {Imel}, David and {Wachter}, Stefanie and {Ye}, Quan-Zhi and {Lin}, Hsing-Wen and {Cenko}, S. Bradley and {Cunningham}, Virginia and {Rebbapragada}, Umaa and {Bue}, Brian and {Miller}, Adam A. and {Mahabal}, Ashish and {Bellm}, Eric C. and {Patterson}, Maria T. and {Juri{\'c}}, Mario and {Golkhou}, V. Zach and {Ofek}, Eran O. and {Walters}, Richard and {Graham}, Matthew and {Kasliwal}, Mansi M. and {Dekany}, Richard G. and {Kupfer}, Thomas and {Burdge}, Kevin and {Cannella}, Christopher B. and {Barlow}, Tom and {Van Sistine}, Angela and {Giomi}, Matteo and {Fremling}, Christoffer and {Blagorodnova}, Nadejda and {Levitan}, David and {Riddle}, Reed and {Smith}, Roger M. and {Helou}, George and {Prince}, Thomas A. and {Kulkarni}, Shrinivas R.},
        title = "{The Zwicky Transient Facility: Data Processing, Products, and Archive}",
      journal = {\pasp},
         year = 2019,
        month = jan,
       volume = {131},
       number = {995},
        pages = {018003},
          doi = {10.1088/1538-3873/aae8ac},
archivePrefix = {arXiv},
       eprint = {1902.01872},
 primaryClass = {astro-ph.IM},
       adsurl = {https://ui.adsabs.harvard.edu/abs/2019PASP..131a8003M}
}

@ARTICLE{NVSS,
       author = {{Condon}, J.~J. and {Cotton}, W.~D. and {Greisen}, E.~W. and {Yin}, Q.~F. and {Perley}, R.~A. and {Taylor}, G.~B. and {Broderick}, J.~J.},
        title = "{The NRAO VLA Sky Survey}",
      journal = {\aj},
         year = 1998,
        month = may,
       volume = {115},
       number = {5},
        pages = {1693-1716},
          doi = {10.1086/300337},
       adsurl = {https://ui.adsabs.harvard.edu/abs/1998AJ....115.1693C}
}

@ARTICLE{Ofek11,
       author = {{Ofek}, Eran O. and {Frail}, Dale A.},
        title = "{The Structure Function of Variable 1.4 GHz Radio Sources Based on NVSS and FIRST Observations}",
      journal = {\apj},
         year = 2011,
        month = aug,
       volume = {737},
       number = {1},
          eid = {45},
        pages = {45},
          doi = {10.1088/0004-637X/737/1/45},
archivePrefix = {arXiv},
       eprint = {1105.3479},
 primaryClass = {astro-ph.GA},
       adsurl = {https://ui.adsabs.harvard.edu/abs/2011ApJ...737...45O}
}

@ARTICLE{PTF,
       author = {{Rau}, Arne and {Kulkarni}, Shrinivas R. and {Law}, Nicholas M. and {Bloom}, Joshua S. and {Ciardi}, David and {Djorgovski}, George S. and {Fox}, Derek B. and {Gal-Yam}, Avishay and {Grillmair}, Carl C. and {Kasliwal}, Mansi M. and {Nugent}, Peter E. and {Ofek}, Eran O. and {Quimby}, Robert M. and {Reach}, William T. and {Shara}, Michael and {Bildsten}, Lars and {Cenko}, S. Bradley and {Drake}, Andrew J. and {Filippenko}, Alexei V. and {Helfand}, David J. and {Helou}, George and {Howell}, D. Andrew and {Poznanski}, Dovi and {Sullivan}, Mark},
        title = "{Exploring the Optical Transient Sky with the Palomar Transient Factory}",
      journal = {\pasp},
         year = 2009,
        month = dec,
       volume = {121},
       number = {886},
        pages = {1334},
          doi = {10.1086/605911},
archivePrefix = {arXiv},
       eprint = {0906.5355},
 primaryClass = {astro-ph.CO},
       adsurl = {https://ui.adsabs.harvard.edu/abs/2009PASP..121.1334R}
}

@ARTICLE{Konig22,
       author = {{K{\"o}nig}, O. and {Saxton}, R.~D. and {Kretschmar}, P. and {Angelini}, L. and {Belanger}, G. and {Evans}, P.~A. and {Freyberg}, M.~J. and {Savchenko}, V. and {Traulsen}, I. and {Wilms}, J.},
        title = "{HILIGT, Upper Limit Servers II - Implementing the data servers}",
      journal = {Astronomy and Computing},
         year = 2022,
        month = jan,
       volume = {38},
          eid = {100529},
        pages = {100529},
          doi = {10.1016/j.ascom.2021.100529},
archivePrefix = {arXiv},
       eprint = {2111.13563},
 primaryClass = {astro-ph.HE},
       adsurl = {https://ui.adsabs.harvard.edu/abs/2022A&C....3800529K}
}

@ARTICLE{HI4PI,
       author = {{HI4PI Collaboration} and {Ben Bekhti}, N. and {Fl{\"o}er}, L. and {Keller}, R. and {Kerp}, J. and {Lenz}, D. and {Winkel}, B. and {Bailin}, J. and {Calabretta}, M.~R. and {Dedes}, L. and {Ford}, H.~A. and {Gibson}, B.~K. and {Haud}, U. and {Janowiecki}, S. and {Kalberla}, P.~M.~W. and {Lockman}, F.~J. and {McClure-Griffiths}, N.~M. and {Murphy}, T. and {Nakanishi}, H. and {Pisano}, D.~J. and {Staveley-Smith}, L.},
        title = "{HI4PI: A full-sky H I survey based on EBHIS and GASS}",
      journal = {\aap},
         year = 2016,
        month = oct,
       volume = {594},
          eid = {A116},
        pages = {A116},
          doi = {10.1051/0004-6361/201629178},
archivePrefix = {arXiv},
       eprint = {1610.06175},
 primaryClass = {astro-ph.GA},
       adsurl = {https://ui.adsabs.harvard.edu/abs/2016A&A...594A.116H}
}

@ARTICLE{Fang20,
       author = {{Fang}, Ke and {Metzger}, Brian D. and {Vurm}, Indrek and {Aydi}, Elias and {Chomiuk}, Laura},
        title = "{High-energy Neutrinos and Gamma Rays from Nonrelativistic Shock-powered Transients}",
      journal = {\apj},
         year = 2020,
        month = nov,
       volume = {904},
       number = {1},
          eid = {4},
        pages = {4},
          doi = {10.3847/1538-4357/abbc6e},
archivePrefix = {arXiv},
       eprint = {2007.15742},
 primaryClass = {astro-ph.HE},
       adsurl = {https://ui.adsabs.harvard.edu/abs/2020ApJ...904....4F}
}

@ARTICLE{Atwood13,
       author = {{Atwood}, W. and {Albert}, A. and {Baldini}, L. and {Tinivella}, M. and {Bregeon}, J. and {Pesce-Rollins}, M. and {Sgr{\`o}}, C. and {Bruel}, P. and {Charles}, E. and {Drlica-Wagner}, A. and {Franckowiak}, A. and {Jogler}, T. and {Rochester}, L. and {Usher}, T. and {Wood}, M. and {Cohen-Tanugi}, J. and {Zimmer}, S.},
        title = "{Pass 8: Toward the Full Realization of the Fermi-LAT Scientific Potential}",
      series = {2012 Fermi Symposium proceedings},
         year = 2013,
        month = mar,
          eid = {arXiv:1303.3514},
        pages = {arXiv:1303.3514},
          doi = {10.48550/arXiv.1303.3514},
archivePrefix = {arXiv},
       eprint = {1303.3514},
 primaryClass = {astro-ph.IM},
       adsurl = {https://ui.adsabs.harvard.edu/abs/2013arXiv1303.3514A}
}

@ARTICLE{Bruel18,
       author = {{Bruel}, P. and {Burnett}, T.~H. and {Digel}, S.~W. and {Johannesson}, G. and {Omodei}, N. and {Wood}, M.},
        title = "{Fermi-LAT improved Pass\raisebox{-0.5ex}\textasciitilde8 event selection}",
      journal = {arXiv e-prints},
         year = 2018,
        month = oct,
          eid = {arXiv:1810.11394},
        pages = {arXiv:1810.11394},
          doi = {10.48550/arXiv.1810.11394},
archivePrefix = {arXiv},
       eprint = {1810.11394},
 primaryClass = {astro-ph.IM},
       adsurl = {https://ui.adsabs.harvard.edu/abs/2018arXiv181011394B}
}

@ARTICLE{Acero16,
       author = {{Acero}, F. and {Ackermann}, M. and {Ajello}, M. and {Albert}, A. and {Baldini}, L. and {Ballet}, J. and {Barbiellini}, G. and {Bastieri}, D. and {Bellazzini}, R. and {Bissaldi}, E. and {Bloom}, E.~D. and {Bonino}, R. and {Bottacini}, E. and {Brandt}, T.~J. and {Bregeon}, J. and {Bruel}, P. and {Buehler}, R. and {Buson}, S. and {Caliandro}, G.~A. and {Cameron}, R.~A. and {Caragiulo}, M. and {Caraveo}, P.~A. and {Casandjian}, J.~M. and {Cavazzuti}, E. and {Cecchi}, C. and {Charles}, E. and {Chekhtman}, A. and {Chiang}, J. and {Chiaro}, G. and {Ciprini}, S. and {Claus}, R. and {Cohen-Tanugi}, J. and {Conrad}, J. and {Cuoco}, A. and {Cutini}, S. and {D'Ammando}, F. and {de Angelis}, A. and {de Palma}, F. and {Desiante}, R. and {Digel}, S.~W. and {Di Venere}, L. and {Drell}, P.~S. and {Favuzzi}, C. and {Fegan}, S.~J. and {Ferrara}, E.~C. and {Focke}, W.~B. and {Franckowiak}, A. and {Funk}, S. and {Fusco}, P. and {Gargano}, F. and {Gasparrini}, D. and {Giglietto}, N. and {Giordano}, F. and {Giroletti}, M. and {Glanzman}, T. and {Godfrey}, G. and {Grenier}, I.~A. and {Guiriec}, S. and {Hadasch}, D. and {Harding}, A.~K. and {Hayashi}, K. and {Hays}, E. and {Hewitt}, J.~W. and {Hill}, A.~B. and {Horan}, D. and {Hou}, X. and {Jogler}, T. and {J{\'o}hannesson}, G. and {Kamae}, T. and {Kuss}, M. and {Landriu}, D. and {Larsson}, S. and {Latronico}, L. and {Li}, J. and {Li}, L. and {Longo}, F. and {Loparco}, F. and {Lovellette}, M.~N. and {Lubrano}, P. and {Maldera}, S. and {Malyshev}, D. and {Manfreda}, A. and {Martin}, P. and {Mayer}, M. and {Mazziotta}, M.~N. and {McEnery}, J.~E. and {Michelson}, P.~F. and {Mirabal}, N. and {Mizuno}, T. and {Monzani}, M.~E. and {Morselli}, A. and {Nuss}, E. and {Ohsugi}, T. and {Omodei}, N. and {Orienti}, M. and {Orlando}, E. and {Ormes}, J.~F. and {Paneque}, D. and {Pesce-Rollins}, M. and {Piron}, F. and {Pivato}, G. and {Rain{\`o}}, S. and {Rando}, R. and {Razzano}, M. and {Razzaque}, S. and {Reimer}, A. and {Reimer}, O. and {Remy}, Q. and {Renault}, N. and {S{\'a}nchez-Conde}, M. and {Schaal}, M. and {Schulz}, A. and {Sgr{\`o}}, C. and {Siskind}, E.~J. and {Spada}, F. and {Spandre}, G. and {Spinelli}, P. and {Strong}, A.~W. and {Suson}, D.~J. and {Tajima}, H. and {Takahashi}, H. and {Thayer}, J.~B. and {Thompson}, D.~J. and {Tibaldo}, L. and {Tinivella}, M. and {Torres}, D.~F. and {Tosti}, G. and {Troja}, E. and {Vianello}, G. and {Werner}, M. and {Wood}, K.~S. and {Wood}, M. and {Zaharijas}, G. and {Zimmer}, S.},
        title = "{Development of the Model of Galactic Interstellar Emission for Standard Point-source Analysis of Fermi Large Area Telescope Data}",
      journal = {\apjs},
         year = 2016,
        month = apr,
       volume = {223},
       number = {2},
          eid = {26},
        pages = {26},
          doi = {10.3847/0067-0049/223/2/26},
archivePrefix = {arXiv},
       eprint = {1602.07246},
 primaryClass = {astro-ph.HE},
       adsurl = {https://ui.adsabs.harvard.edu/abs/2016ApJS..223...26A}
}

@ARTICLE{Mattox96,
       author = {{Mattox}, J.~R. and {Bertsch}, D.~L. and {Chiang}, J. and {Dingus}, B.~L. and {Digel}, S.~W. and {Esposito}, J.~A. and {Fierro}, J.~M. and {Hartman}, R.~C. and {Hunter}, S.~D. and {Kanbach}, G. and {Kniffen}, D.~A. and {Lin}, Y.~C. and {Macomb}, D.~J. and {Mayer-Hasselwander}, H.~A. and {Michelson}, P.~F. and {von Montigny}, C. and {Mukherjee}, R. and {Nolan}, P.~L. and {Ramanamurthy}, P.~V. and {Schneid}, E. and {Sreekumar}, P. and {Thompson}, D.~J. and {Willis}, T.~D.},
        title = "{The Likelihood Analysis of EGRET Data}",
      journal = {\apj},
         year = 1996,
        month = apr,
       volume = {461},
        pages = {396},
          doi = {10.1086/177068},
       adsurl = {https://ui.adsabs.harvard.edu/abs/1996ApJ...461..396M}
}

@ARTICLE{Murase19,
       author = {{Murase}, Kohta and {Franckowiak}, Anna and {Maeda}, Keiichi and {Margutti}, Raffaella and {Beacom}, John F.},
        title = "{High-energy Emission from Interacting Supernovae: New Constraints on Cosmic-Ray Acceleration in Dense Circumstellar Environments}",
      journal = {\apj},
         year = 2019,
        month = mar,
       volume = {874},
       number = {1},
          eid = {80},
        pages = {80},
          doi = {10.3847/1538-4357/ab0422},
archivePrefix = {arXiv},
       eprint = {1807.01460},
 primaryClass = {astro-ph.HE},
       adsurl = {https://ui.adsabs.harvard.edu/abs/2019ApJ...874...80M}
}

@ARTICLE{Wiserep,
       author = {{Yaron}, Ofer and {Gal-Yam}, Avishay},
        title = "{WISeREP{\textemdash}An Interactive Supernova Data Repository}",
      journal = {\pasp},
         year = 2012,
        month = jul,
       volume = {124},
       number = {917},
        pages = {668},
          doi = {10.1086/666656},
archivePrefix = {arXiv},
       eprint = {1204.1891},
 primaryClass = {astro-ph.IM},
       adsurl = {https://ui.adsabs.harvard.edu/abs/2012PASP..124..668Y}
}

@ARTICLE{Planck18,
       author = {{Planck Collaboration} and {Aghanim}, N. and {Akrami}, Y. and {Arroja}, F. and {Ashdown}, M. and {Aumont}, J. and {Baccigalupi}, C. and {Ballardini}, M. and {Banday}, A.~J. and {Barreiro}, R.~B. and {Bartolo}, N. and {Basak}, S. and {Battye}, R. and {Benabed}, K. and {Bernard}, J.-P. and {Bersanelli}, M. and {Bielewicz}, P. and {Bock}, J.~J. and {Bond}, J.~R. and {Borrill}, J. and {Bouchet}, F.~R. and {Boulanger}, F. and {Bucher}, M. and {Burigana}, C. and {Butler}, R.~C. and {Calabrese}, E. and {Cardoso}, J.-F. and {Carron}, J. and {Casaponsa}, B. and {Challinor}, A. and {Chiang}, H.~C. and {Colombo}, L.~P.~L. and {Combet}, C. and {Contreras}, D. and {Crill}, B.~P. and {Cuttaia}, F. and {de Bernardis}, P. and {de Zotti}, G. and {Delabrouille}, J. and {Delouis}, J.-M. and {D{\'e}sert}, F.-X. and {Di Valentino}, E. and {Dickinson}, C. and {Diego}, J.~M. and {Donzelli}, S. and {Dor{\'e}}, O. and {Douspis}, M. and {Ducout}, A. and {Dupac}, X. and {Efstathiou}, G. and {Elsner}, F. and {En{\ss}lin}, T.~A. and {Eriksen}, H.~K. and {Falgarone}, E. and {Fantaye}, Y. and {Fergusson}, J. and {Fernandez-Cobos}, R. and {Finelli}, F. and {Forastieri}, F. and {Frailis}, M. and {Franceschi}, E. and {Frolov}, A. and {Galeotta}, S. and {Galli}, S. and {Ganga}, K. and {G{\'e}nova-Santos}, R.~T. and {Gerbino}, M. and {Ghosh}, T. and {Gonz{\'a}lez-Nuevo}, J. and {G{\'o}rski}, K.~M. and {Gratton}, S. and {Gruppuso}, A. and {Gudmundsson}, J.~E. and {Hamann}, J. and {Handley}, W. and {Hansen}, F.~K. and {Helou}, G. and {Herranz}, D. and {Hildebrandt}, S.~R. and {Hivon}, E. and {Huang}, Z. and {Jaffe}, A.~H. and {Jones}, W.~C. and {Karakci}, A. and {Keih{\"a}nen}, E. and {Keskitalo}, R. and {Kiiveri}, K. and {Kim}, J. and {Kisner}, T.~S. and {Knox}, L. and {Krachmalnicoff}, N. and {Kunz}, M. and {Kurki-Suonio}, H. and {Lagache}, G. and {Lamarre}, J.-M. and {Langer}, M. and {Lasenby}, A. and {Lattanzi}, M. and {Lawrence}, C.~R. and {Le Jeune}, M. and {Leahy}, J.~P. and {Lesgourgues}, J. and {Levrier}, F. and {Lewis}, A. and {Liguori}, M. and {Lilje}, P.~B. and {Lilley}, M. and {Lindholm}, V. and {L{\'o}pez-Caniego}, M. and {Lubin}, P.~M. and {Ma}, Y.-Z. and {Mac{\'\i}as-P{\'e}rez}, J.~F. and {Maggio}, G. and {Maino}, D. and {Mandolesi}, N. and {Mangilli}, A. and {Marcos-Caballero}, A. and {Maris}, M. and {Martin}, P.~G. and {Martinelli}, M. and {Mart{\'\i}nez-Gonz{\'a}lez}, E. and {Matarrese}, S. and {Mauri}, N. and {McEwen}, J.~D. and {Meerburg}, P.~D. and {Meinhold}, P.~R. and {Melchiorri}, A. and {Mennella}, A. and {Migliaccio}, M. and {Millea}, M. and {Mitra}, S. and {Miville-Desch{\^e}nes}, M.-A. and {Molinari}, D. and {Moneti}, A. and {Montier}, L. and {Morgante}, G. and {Moss}, A. and {Mottet}, S. and {M{\"u}nchmeyer}, M. and {Natoli}, P. and {N{\o}rgaard-Nielsen}, H.~U. and {Oxborrow}, C.~A. and {Pagano}, L. and {Paoletti}, D. and {Partridge}, B. and {Patanchon}, G. and {Pearson}, T.~J. and {Peel}, M. and {Peiris}, H.~V. and {Perrotta}, F. and {Pettorino}, V. and {Piacentini}, F. and {Polastri}, L. and {Polenta}, G. and {Puget}, J.-L. and {Rachen}, J.~P. and {Reinecke}, M. and {Remazeilles}, M. and {Renault}, C. and {Renzi}, A. and {Rocha}, G. and {Rosset}, C. and {Roudier}, G. and {Rubi{\~n}o-Mart{\'\i}n}, J.~A. and {Ruiz-Granados}, B. and {Salvati}, L. and {Sandri}, M. and {Savelainen}, M. and {Scott}, D. and {Shellard}, E.~P.~S. and {Shiraishi}, M. and {Sirignano}, C. and {Sirri}, G. and {Spencer}, L.~D. and {Sunyaev}, R. and {Suur-Uski}, A.-S. and {Tauber}, J.~A. and {Tavagnacco}, D. and {Tenti}, M. and {Terenzi}, L. and {Toffolatti}, L. and {Tomasi}, M. and {Trombetti}, T. and {Valiviita}, J. and {Van Tent}, B. and {Vibert}, L. and {Vielva}, P. and {Villa}, F. and {Vittorio}, N. and {Wandelt}, B.~D. and {Wehus}, I.~K. and {White}, M. and {White}, S.~D.~M. and {Zacchei}, A. and {Zonca}, A.},
        title = "{Planck 2018 results. I. Overview and the cosmological legacy of Planck}",
      journal = {\aap},
         year = 2020,
        month = sep,
       volume = {641},
          eid = {A1},
        pages = {A1},
          doi = {10.1051/0004-6361/201833880},
archivePrefix = {arXiv},
       eprint = {1807.06205},
 primaryClass = {astro-ph.CO},
       adsurl = {https://ui.adsabs.harvard.edu/abs/2020A&A...641A...1P}
}

\begin{appendix}

\section{\textit{Fermi}-LAT extended data}

\begin{table}[t]
\caption{Arrival time, energy, and distance from the nominal optical position of SN 2017egm for the list of events shown in Fig. \ref{fig:photons} with an association probability larger than 50\%. }
\label{tab:events}
\centering    
\begin{tabular}{lccc}
\hline \hline
Time & Probability & Energy & Distance  \\
MET &  & GeV & deg  \\
\hline
514977340.8 & 0.72  & 2.34 &  0.071\\
525330848.1 & 0.57  & 0.76 &  0.053\\
527775943.4 & 0.57  & 3.55 &  0.213\\
527900219.8 & 0.71  & 1.74 &  0.142\\
529193069.7 & 0.95  & 6.87 &  0.056\\
529535373.0 & 0.58  & 1.56 &  0.045\\
530258550.6 & 0.84  & 2.07 &  0.016\\
533858313.2 & 0.91  & 6.41 &  0.054\\
548673784.0 & 0.87  & 3.05 &  0.087\\
559162934.5 & 0.67  & 2.03 &  0.151\\
563875293.0 & 0.80  & 2.52 &  0.121\\
566353118.7 & 0.66  & 1.89 &  0.180\\

\end{tabular}
\tablefoot{Time is provided in mission elapsed time (MET).}
\end{table}

The list of events with a likelihood probability of association with SN 2017egm corresponding to Fig. \ref{fig:photons} is provided in Table \ref{tab:events}. Only events with probabilities larger than 50\% according to the procedure described in Sect. \ref{sec:refinedlc} are included.

The \textit{Fermi}-LAT data points of the SED shown in Fig. \ref{fig:tsmap_sed} (bottom panel) are presented in Table \ref{tab:sed}.
These were estimated with the analysis configuration described in Sect. \ref{sect:Fermi-analysis} and in the Bayesian block time interval.

\begin{table}
\caption{\textit{Fermi}-LAT energy flux data points in the time interval defined by the Bayesian block algorithm (2017-07-05 to 2017-10-25; see Fig. \ref{fig:tsmap_sed}). }
\label{tab:sed}
\centering    
\begin{tabular}{lcc}
\hline \hline
Energy band & $E^{2}dN/dE$ & TS \\
GeV & $10^{-12}$ erg cm$^{-2}$ s$^{-1}$ &  \\
\hline
0.19 (0.10,0.37) & 1.07 (-0.99, 1.06)  & 1.2  \\
0.72 (0.37,1.39) & 1.72 (-0.54, 0.63)  & 18.2 \\
2.68 (1.39,5.18) & 1.27 (-0.58, 0.72)  & 11.6 \\
10.00 (5.18,19.31) & 0.76 (-0.59, 0.98)  & 5.5 \\
37.28 (19.31,71.97) & $<3.8$  & 0.0 \\

\end{tabular}
\tablefoot{The energy flux errors in parenthesis are statistical and at the 1$\sigma$ level, while upper limits are provided at $2\sigma$ level. }
\end{table}

\section{CSM model: Shock crossing times}
\label{sect:crossingtimes}

The CSM model assumes a shock propagating through a series of shells with $V_s = dR_s/dt = V_{s,0}\left[\dfrac{t}{1\;\rm{day}}\right]^{m-1}$, where the expansion parameter is $m=0.75$ and the initial shock velocity  $V_{s,0}= 9.8\times10^3$ km/s \citep{Lin23}. Table \ref{tab:times} provides average densities as well as forward shock crossing times for each shell, derived from Extended Data Table 1 \cite{Lin23}. As in Sects. \ref{sect:results} and \ref{sect:csmmodel}, $T_0$ is set to the discovery time (57896 MJD).

\begin{table}
\caption{Densities and shock crossing times (in and out) for each shell. }
\label{tab:times}
\centering    
\begin{tabular}{lccc}
\hline \hline
Shell & $\rho_i$ ($10^{-14}$ g cm$^{-3}$) & $T_{\rm in}-T_0$ (d) & $T_{\rm out}-T_0$ (d) \\
\hline
1 & 26.1 & 1.0  & 44.8  \\
2 & 13.7 & 49.4  & 58.5 \\
3 & 50.5 & 184.3  & 184.7 \\
4 & 6.35 & 224.3  & 227.2 \\

\end{tabular}
\tablefoot{Shells are numbered as in \cite{Lin23}.}
\end{table}

\section{Estimate of $\gamma$-$\gamma$ absorption in a CSM interaction scenario}
\label{sect:gg-abs}

\begin{figure}[h]
   \centering
   
   \includegraphics[width=0.775\columnwidth]{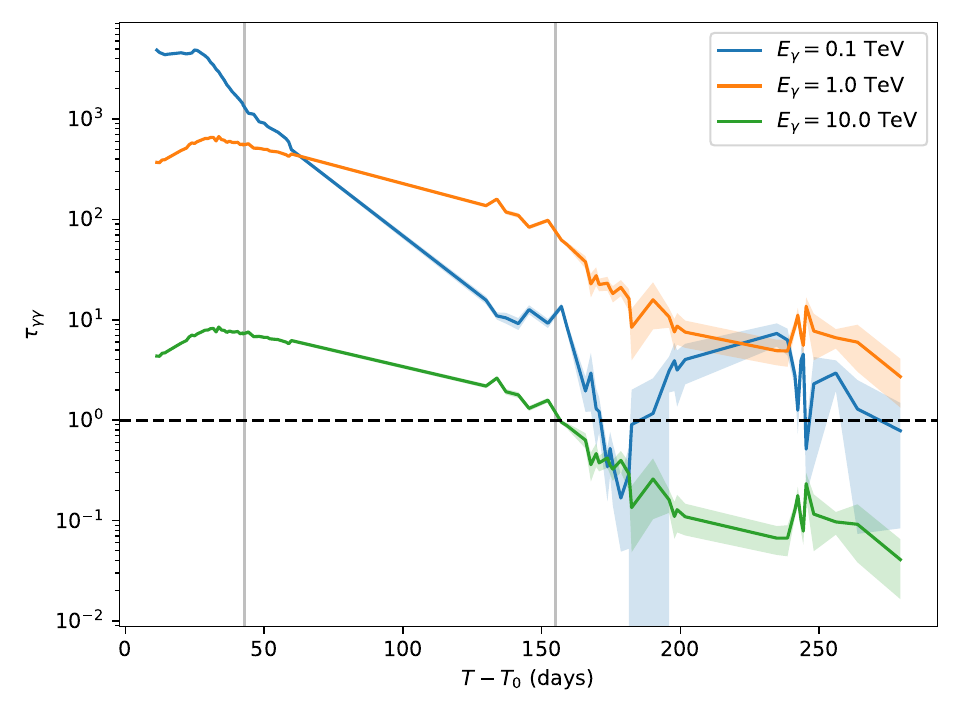} 

   \includegraphics[width=0.775\columnwidth]{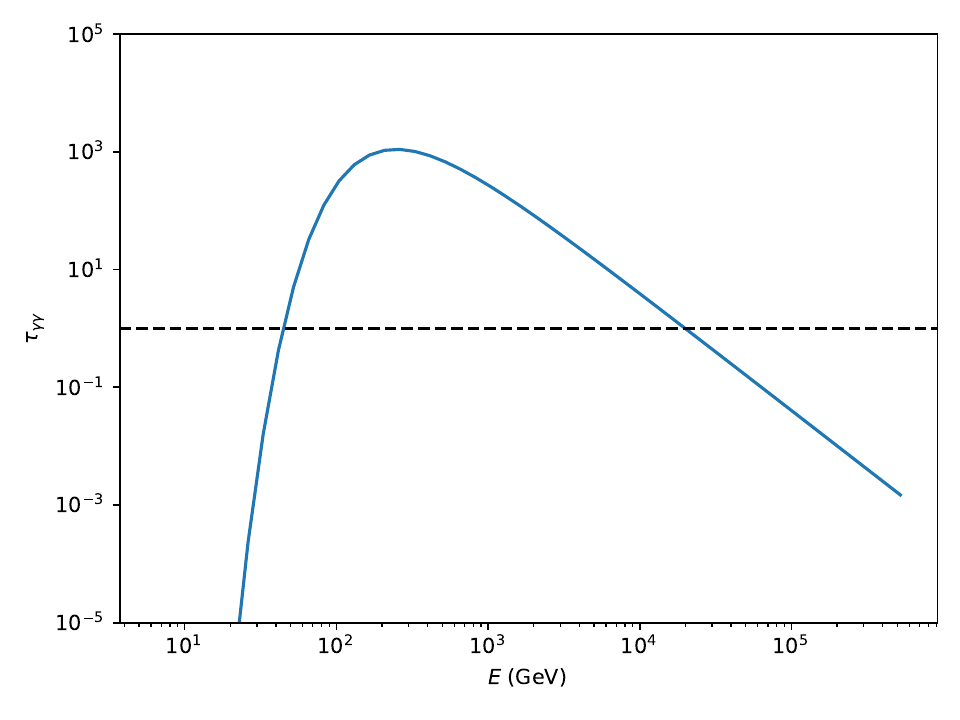} 
   
   \vspace{-0.3cm}
   \caption{Top: Temporal evolution of $\tau_{\gamma \gamma}$ for photons of 0.1, 1, and 10 TeV. Vertical gray lines represent the Bayesian block interval identifying the LAT emission, while the horizontal dashed line ($\tau_{\gamma \gamma}=1$) represents the transition to a transparent medium ($\tau_{\gamma \gamma}\lesssim 1$). Shaded areas represent the error propagation from temperature and luminosity measurements in \cite{Zhu23}. Bottom: Absorption above 10 GeV, averaged over the Bayesian block time interval in Fig. \ref{fig:B_tautime}. The overall shape is a result of the peaked SN blackbody spectrum and the cross section~$\sigma_{\gamma \gamma}$.}
    \label{fig:B_tautime}%

    \end{figure}

For a first estimate of the impact that $\gamma$-$\gamma$ absorption could have on the CSM model prediction for CTAO, we derived the expected $\tau_{\gamma \gamma}$ following Sect. 2.3 in \cite{Fang20}. Assuming isotropy in the blackbody component, we estimated the photon density at the shock ($n_{\rm opt}$) during the relevant evolution using the results from Table 3 in \cite{Zhu23}. The derived optical depth ($\tau_{\gamma \gamma} \sim n_{\rm opt}\sigma_{\gamma \gamma}R_s$) for photons of 0.1, 1, and 10 TeV is represented in Fig. \ref{fig:B_tautime}. It is apparent that the medium would be  optically thick at least until the end of the emission interval. To derive the absorption directly applicable to Fig. \ref{fig:cta}, we computed the weighted time average $\tau_{\gamma \gamma}$ for the Bayesian block interval (Fig. \ref{fig:B_tautime}, bottom). As was expected, $\gamma$-$\gamma$ absorption is not relevant for the LAT energy band but would prevent CTAO detection (i.e., at approximately tera-electronvolt energies). For the specific case of a SN 2017egm-like object, since no emission is expected for the CSM model after 250-300 days (see Fig. \ref{fig:ECRandflux}), a CTAO detection would support a magnetar-powered scenario.

 \end{appendix}

%
%
\end{document}